\documentclass[british]{timesjhep}
\usepackage{CJK}
\usepackage{listings}
\usepackage{multicol}
\newenvironment{block}[1]{\vspace{0.4\baselineskip}\hrule height 2\arrayrulewidth\vspace{\doublerulesep}\hrule\vspace{0.30\baselineskip}{\bfseries #1}\vspace{0.2\baselineskip}\hrule\vspace{0.3\baselineskip}}{\vspace{0.2\baselineskip}\hrule\vspace{\doublerulesep}\hrule height 2\arrayrulewidth\vspace{0.5\baselineskip}}
\newcommand{\Beta}{\mathrm{B}}

\newcommand{\Sp}{\mathrm{Sp}}
\newcommand{\SO}{\mathrm{SO}}
\newcommand{\SU}{\mathrm{SU}}
\newcommand{\rd}{\mathrm{d}}
\newcommand{\rO}{\mathrm{O}}
\newcommand{\rU}{\mathrm{U}}
\newcommand{\bnh}{\hat{\mathbf{n}}}
\newcommand{\bzh}{\hat{\mathbf{z}}}
\newcommand{\br}{\mathbf{r}}
\newcommand{\bq}{\mathbf{q}}
\newcommand{\bM}{\mathbf{M}}
\newcommand{\bX}{\mathbf{X}}
\newcommand{\cA}{\mathcal{A}}
\newcommand{\cH}{\mathcal{H}}
\newcommand{\cO}{\mathcal{O}}
\newcommand{\cP}{\mathcal{P}}
\newcommand{\cR}{\mathcal{R}}
\newcommand{\cZ}{\mathcal{Z}}
\newcommand{\BI}{\mathbb{I}}
\newcommand{\BR}{\mathbb{R}}
\newcommand{\BZ}{\mathbb{Z}}
\newcommand{\Yb}{\bar{Y}}
\newcommand{\el}{\text{el}}
\newcommand{\Int}{\text{int}}
\newcommand{\FS}{\text{(FS)}}

\begin{document}

\begin{CJK*}{UTF8}{bkai}
    
\title{FuzzifiED}
\subtitle{Julia Package for Numerics on the Fuzzy Sphere}
\date{1st December 2025}
\author{Zheng Zhou 周正}
\affiliation{Perimeter Institute for Theoretical Physics, Waterloo, Ontario N2L 2Y5, Canada}
\affiliation{Department of Physics and Astronomy, University of Waterloo, Waterloo, Ontario N2L 3G1, Canada}
\emailAdd{physics@zhengzhou.page}
\abstract{The Julia package \emph{FuzzifiED} aims at simplifying the numerical calculations on the fuzzy sphere. It supports exact diagonalisation (ED) and density matrix renormalisation group (DMRG) calculations. FuzzifiED can also apply to generic fermionic and bosonic models. This documentation provides a review of the fuzzy sphere regularisation and an instruction for using FuzzifiED for numerical calculations.}
\extra{\textsc{Documentation}~: \url{https://docs.fuzzified.world}\\\textsc{Source code}~: \url{https://github.com/FuzzifiED/FuzzifiED.jl}\\\textsc{Version}~: 1.2.1}
\arxivnumber{2503.00100}

\maketitle

\end{CJK*}

\section{Purpose and Outline}

Since its proposal, the fuzzy sphere regularisation has made significant contributions to the study of 3D CFTs~\cite{Zhu2022,Hu2023Mar,Han2023Jun,Zhou2023,Lao2023,Hu2023Aug,Hofmann2023,Han2023Dec,Zhou2024Jan,Hu2024,Cuomo2024,Zhou2024Jul,Dedushenko2024,Fardelli2024,Fan2024,Zhou2024Oct,Voinea2024,Yang2025Jan,Han2025,Laeuchli2025,Fan2025,ArguelloCruz2025,EliasMiro2025,He2025Jun,Taylor2025,Yang2025Jul,Zhou2025Jul,Dong2025,Zhou2025Sep,Voinea2025,Wiese2025,Dey2025,Guo2025,Huffman2025}. The Julia package FuzzifiED aims at simplifying the numerical calculations on the fuzzy sphere. It supports exact diagonalisation (ED) calculations, as well as the density matrix renormalisation group (DMRG) using the ITensor~\cite{ITensor} library. FuzzifiED can also apply to generic fermionic and bosonic models. This package offers the following features~:
\begin{enumerate}
    \item Flexibility~: FuzzifiED can reproduce nearly all the ED and DMRG results from fuzzy sphere research. Its flexible design also makes it straightforward to adapt to new models.
    \item Usability~: The expressive Julia interface~\cite{Julia} simplifies coding and comprehension. To help users get started, we provide a collection of examples.
    \item Efficiency~: FuzzifiED produces results on reasonable system sizes within minutes.
    \item Open source~: The FuzzifiED codebase is freely available under the MIT License, welcoming reviews and contributions from the wider community.
\end{enumerate}

This documentation provides a review of the fuzzy sphere regularisation and an instruction for numerical calculations with FuzzifiED. The rest of this documentation is organised as follows~: The Part~\ref{pt:intro} (Sections~\ref{sec:intro}--\ref{sec:numerics}) is devoted to a review of the fuzzy sphere regularisation and the numerical methods applied to it, mainly targeting at providing the necessary technical information for those who want to get started with the research on the fuzzy sphere. The Part~\ref{pt:numerics} (Sections~\ref{sec:usage}--\ref{sec:examples}) gives a detailed instruction for numerical calculation with the package FuzzifiED.
\begin{itemize}
    \item In Section~\ref{sec:intro}, we briefly introduce the conformal field theories in dimensions $d\ge 3$ and the fuzzy sphere regularisation.
    \item In Section~\ref{sec:review}, we review the existing works related to the fuzzy sphere.
    \item In Section~\ref{sec:construct}, we review the set-up of fuzzy sphere, the construction of interaction models and the extraction of CFT data.
    \item In Section~\ref{sec:numerics}, we review the numerical methods used in the package FuzzifiED, \textit{viz.}~ED and DMRG.
    \item In Section~\ref{sec:usage}, we give an instruction for installing and using FuzzifiED.
    \item In Section~\ref{sec:ed}, we give an instruction for performing ED calculation with FuzzifiED.
    \item In Section~\ref{sec:dmrg}, we give an instruction for performing DMRG calculation with FuzzifiED.
    \item In Section~\ref{sec:examples}, we present a collection of practical examples that reproduce many existing results.
    \item In Appendix~\ref{app:data}, we present the specialised data structures that we use in the ED calculation.
    \item In Appendix~\ref{app:manifolds}, we present the construction of the Landau level and pseudo-potentials on the torus.
    \item In Appendix~\ref{app:code}, we give a collection of the ED and DMRG tutorial codes.
    \item In Appendix~\ref{app:glossary}, we list a glossary for all the interfaces in FuzzifiED to facilitate searching.
\end{itemize}

\cleardoublepage
\part{Review of the Fuzzy Sphere}
\label{pt:intro}

\section{Introduction}
\label{sec:intro}

\subsection{Conformal Field Theory}

Conformal field theory (CFT) is one of the central topics of modern physics. It refers to a field theory that is invariant under conformal transformations that preserve the angles between vectors. In space-time dimension $d>2$, the global conformal symmetry transformations form a group $\SO(d+1,1)$, generated by translation, $\SO(d)$ rotation,\footnote{Here we work in Euclidean signature. In Lorentzian signature it is the Lorentz transformation $\SO(1,d-1)$} dilatation (scale transformation), and special conformal transformation (SCT)~\cite{Rychkov2016CFT,SimmonsDuffin2016CFT}. Each CFT operator must transform under irreducible representations of rotation and dilatation. The representations are labelled by the $\SO(d)$ spin $l$ and scaling dimension $\Delta$, respectively. A special kind of operators that are invariant under SCT, called ``primaries,'' deserve particular attention. By acting spatial derivatives on the primaries, their ``descendants'' are obtained. Each local operator in a CFT is a linear combination of primaries and descendants. The conformal symmetry is the maximal space-time symmetry (except supersymmetry) that a field theory can have. It gives powerful constraints on the property of the field theory. In particular, conformal symmetry uniquely determines the form of two-point (2-pt) and three-point (3-pt) correlation functions. The 3-pt correlator of three primary operators $\Phi_i,\Phi_j,\Phi_k$ contains a universal coefficient called the OPE coefficient $f_{\Phi_i\Phi_j\Phi_k}$. The collection of the scaling dimensions and the OPE coefficients of primaries $\{\Delta_{\Phi_i},f_{\Phi_i\Phi_j\Phi_k}\}$ is called the conformal data. Theoretically, with full knowledge of the CFT data, an arbitrary correlation function of a CFT can be obtained.\footnote{The full knowledge is not often possible in practice, as the number of primaries is often infinite.}

CFT has provided important insights into various aspects of theoretical physics. It has produced useful predictions about the critical phenomena~\cite{Polyakov1970Conformal,Cardy1996Scaling,Sachdev2011Quantum} in condensed-matter physics. Many classical and quantum phase transitions are conjectured to have emergent conformal symmetry in the infra-red (IR), \textit{i.~e.}~at long wave-length or low energy. The universal critical exponents are directly determined by the scaling dimensions of the primary operators. \textit{E.~g.}, in the 3D Ising transition that spontaneously breaks a $\BZ_2$ symmetry, most critical exponents are given by the scaling dimensions of the lowest $\BZ_2$-odd operator $\sigma$ and $\BZ_2$-even operator $\epsilon$, such as
\begin{equation}
    \eta=2\Delta_\sigma-1,\qquad\nu=\frac{1}{3-\Delta_\epsilon}.
\end{equation}
CFT is also closely related to string theory and quantum gravity in high-energy physics. In the string theory, CFT describes the 2D worldsheet~\cite{Polchinski1998String}~; in quantum gravity, there is a conjectured duality between the gravity theory in $(d+1)$-dimensional anti-de Sitter (AdS) space in the bulk and a $d$-dimensional CFT on the boundary~\cite{Maldacena1998AdSCFT}. Moreover, CFT plays an important role in understanding of quantum field theories in general. It describes many fixed points in the renormalisation-group (RG) flow, and many QFTs can be seen as CFTs with perturbations. It also helps us understand how physics changes under a change of scale and reveals some fundamental structures of the RG flow~\cite{Zamolodchikov1986Irreversibility}.

In 2D CFTs, besides the global conformal symmetry $\SO(3,1)$, there also exists an infinite-dimensional local conformal symmetry~\cite{DiFrancesco1997CFT,Ginsparg1988CFT}. Altogether, they form the Virasoro algebra. The infinite-dimensional conformal algebra has made many theories exactly solvable, especially the rational theories such as the minimal models~\cite{Belavin1984BPZ} and, more generally, the Wess-Zumino-Witten (WZW) theories~\cite{Wess1971WZW,Witten1983WZW}. However, going to higher dimensions, the CFTs are less well-studied due to a much smaller conformal group. The existing methods include numerical conformal bootstrap and Monte Carlo lattice simulations. Numerical bootstrap bounds the conformal data by making use of consistency conditions such as reflection positivity and crossing symmetry, together with some information of the CFT such as the global symmetry and a certain amount of assumptions~\cite{Poland2018Bootstrap,Rychkov2023Bootstrap}. It has achieved great success in 3D Ising~\cite{ElShowk2012Ising,Kos2016Ising}, $\rO(N)$ Wilson-Fisher~\cite{Chester2019O2,Chester2020O3}, Gross-Neveu-Yukawa CFTs~\cite{Iliesiu2015GNY}, \textit{etc.} On the other hand, one can study a CFT by constructing a lattice model that goes through a phase transition in the corresponding universality class, and study the phase transition by Monte Carlo simulation. This method has achieved success in many phase transitions assuming conformal symmetry, \textit{e.~g.}~the 3D Ising model~\cite{Ferrenberg2018IsingMC}. However, the extraction of universal data usually involves complicated and computation-intensive finite-size scaling~\cite{Cardy1996Scaling,Sandvik2010FSS,Fisher1972FSS}, and only the lowest few CFT operators can be accessed.

Among these higher dimensional CFTs, we mainly focus on three space-time dimensions.

\subsection{Fuzzy Sphere}

In addition to these existing approaches, the \emph{``fuzzy sphere regularisation''} has recently emerged as a new powerful method to study 3D CFTs. It involves studying interacting electrons moving on a sphere
under the influence of a magnetic monopole at its centre.

The idea begins with putting an interacting quantum Hamiltonian on a 2-sphere $S^2$. This geometry preserves the full rotation symmetry (on the contrary, lattice models often only preserve a discrete subgroup). Moreover, when the system is tuned to a critical point or critical phase, combined with the time evolution direction, the system is described by a quantum field theory living on a generalised cylinder $S^2\times\BR$, a manifold that is conformally equivalent to flat space-time through the Weyl transformation
\begin{equation}
    (\br,\tau)\in S^2\times\BR\ \longmapsto\ e^{\tau/R}\bnh\in\BR^3,
\end{equation}
where $R$ is the radius of the sphere and $\bnh$ is the unit vector corresponding to $\br$. This conformal transformation maps each time slice of the cylinder to a co-centric sphere in the flat space-time.

Thanks to the conformal flatness that is not owned by other manifolds (\textit{e.~g.}, a lattice model with periodic boundary condition lives on the torus $T^2$, which is not conformally flat), we can make use of some nice properties of conformal field theories. The most important one is the state-operator correspondance~\cite{Pappadopulo2012Radial,Rychkov2016CFT,SimmonsDuffin2016CFT}. Specifically, there is a one-to-one correspondence between the eigenstates of the critical Hamiltonian on the sphere and the CFT operators. One can colloquially understand the state $|\Phi\rangle$ as the insertion of the corresponding operator $\Phi(0)$ at the origin point into the vacuum $|0\rangle$~: $|\Phi\rangle=\hat{\Phi}(0)|0\rangle$. The state and its corresponding operator have the same $\SO(3)$ spin and representation under global symmetry. More importantly, as the Weyl transformation maps the Hamiltonian $H$ that generates the time translation on the cylinder to the dilatation $D$ on the flat space-time, the excitation energy of a state $|\Phi\rangle$ is proportional to the scaling dimension of the corresponding operator $\Delta_\Phi$
\begin{equation}
    E_\Phi-E_0=\frac{v}{R}\Delta_\Phi,
\end{equation}
where $E_0$ is the ground state energy, $R$ is the radius of the sphere, and $v$ is the speed of light that is dependent on the microscopic model and is the same for every state. With this property, one can calculate the scaling dimensions simply by solving the energy spectrum of the quantum Hamiltonian without doing complicated finite-size scalings, and one can obtain the OPE coefficients simply by taking the inner product of a local operator.

Although the quantum Hamiltonians on a sphere enjoy the full rotation symmetry and the property of state-operator correspondence, it is difficult to put a lattice on the sphere due to the curvature (in particular the non-zero Euler characteristic), especially to recover an $\SO(3)$-symmetric thermodynamic limit~\cite{Brower2024Sphere}. An alternative way is to fuzzify the sphere~\cite{Madore1991Fuzzy}. We consider charged free particles moving on a sphere with a magnetic monopole with a flux $4\pi s$ ($s\in\BZ/2$) placed at its centre. The monopole exerts a uniform magnetic field on the sphere, which modifies the single-particle Hamiltonian and the single-particle eigenstates. Now, the single particle eigenstates form highly degenerate spherical Landau levels~\cite{Haldane1983LLL,Wu1976LLL,Greiter2011LLL,Hasebe2010LLL}. The lowest Landau level has a degeneracy $(2s+1)$. By setting the single-particle gap to be the leading energy scale, and projecting onto the lowest Landau level, we obtain a finite Hilbert space. For the purpose of numerical simulation, the system is analogous to a length-$(2s+1)$ chain with long-range interaction, where different Landau level orbitals behave like the lattice sites. The difference is that the $(2s+1)$ orbitals form a spin-$s$ representation of the $\SO(3)$ rotation group, and in this way, the continuous rotation symmetry is preserved. By putting multiple flavours on an orbital and adding interactions, various 3D CFTs can be realised. The interaction Hamiltonians are designed through matching the global symmetry and phase diagram. The word ``fuzzy'' means the non-commutativity, in our case, due to the presence of magnetic field~\cite{Madore1991Fuzzy,Hasebe2010LLL}. The non-commutativity provides a natural length scale which serves as a UV regulator of the quantum field theory. The radius of the sphere scales as $R\sim\sqrt s$. The thermodynamic limit can be taken as $s\to\infty$, and we then recover a regular sphere without non-commutativity.

The power of this approach has been first demonstrated in the context of the 3D Ising transition~\cite{Zhu2022}, where the presence of emergent conformal symmetry has been convincingly established, and a wealth of conformal data has been accurately computed. The study has then been extended to accessing various conformal data such as the OPE coefficients~\cite{Hu2023Mar}, correlation functions~\cite{Han2023Jun}, entropic $F$-function~\cite{Hu2024}, conformal generators~\cite{Fardelli2024,Fan2024} and the cross-cap coefficients~\cite{Dong2025}, developing techniques to improve the numerical precision, such as quantum Monte Carlo~\cite{Hofmann2023}, conformal perturbation~\cite{Laeuchli2025} and the finite-size scaling of the ground-state energy~\cite{Wiese2025}, realising various 3D CFTs such as the free scalar~\cite{He2025Jun,Taylor2025}, Wilson-Fisher CFTs~\cite{Han2023Dec,Dey2025,Guo2025}, $\SO(5)$~\cite{Zhou2023} and $\rO(4)$~\cite{Yang2025Jul} deconfined criticality, $\Sp(N)$-symmetric CFTs~\cite{Zhou2024Oct}, multi-flavour $\SU(2)$ QCD~\cite{Huffman2025}, 3-state Potts model~\cite{Yang2025Jan} and Yang-Lee non-unitary CFT~\cite{Fan2025,ArguelloCruz2025,EliasMiro2025}, exploring fractional quantum Hall transitions, such as the Ising CFT on the FQHE states~\cite{Voinea2024}, the confinement transition of the $\nu=1/2$ bosonic Laughlin state~\cite{Zhou2025Jul}, the free Marjorana fermion theory~\cite{Zhou2025Sep}, and the transition between bosonic Pfaffian and the Halperin 220 state~\cite{Voinea2025}, studying conformal defects and boundaries such as the pinning-field defect~\cite{Hu2023Aug,Zhou2024Jan}, various conformal boundaries in the 3D Ising CFT~\cite{Zhou2024Jul,Dedushenko2024}, and lower-dimensional CFTs like 2D Ising on the fuzzy circle~\cite{Han2025}. In the following sections, we shall review the existing works, technical details and numerical methods. 

\section{Review of Existing Works}
\label{sec:review}

In this section, we review the existing works related to the fuzzy sphere.

\paragraph{The pioneering work~\cite{Zhu2022}}

This work first proposes the idea of the fuzzy sphere and applies it to a pedagogical example of the 3D Ising CFT. Zhu \textit{et al.}~construct a model with two flavours of fermions that resemble the spin-up and spin-down in the lattice transverse-field Ising model. At half-filling, one can colloquially think that a spin degree of freedom lives on each orbital. The Hamiltonian contains a density-density interaction\footnote{Here the density operator refers to a local fermion bilinear.} that resembles the Ising ferromagnetic interaction and a polarising term that resembles the transverse field. By tuning the ratio between the two terms, a transition between quantum Hall ferromagnet~\cite{Pasquier2000HallFM,Girvin2010HallFM} (a two-fold degenerate state where either of the two flavours is completely occupied) and paramagnet (a one-fold degenerate state where the superposition of the two flavours at each orbital is occupied) occurs. This transition spontaneously breaks a $\BZ_2$ symmetry and falls into the Ising criticality. They then make use of a unique feature of spherical models described by CFT --- state-operator correspondence --- at the critical point to extract the scaling dimensions of the scaling local operators. They find evidence for the conformal symmetry, including that (1) there exists a conserved stress tensor with $\Delta=3$ (which is used as the calibrator), and (2) all the levels can be classified into conformal multiplet where the spacings between operators' scaling dimensions are very close to integer. This is one of the first numerical evidence that the 3D Ising transition has emergent conformal symmetry. More remarkably, the scaling dimensions of primaries such as $\sigma,\epsilon,\epsilon'$ are already very close to the most accurate known value by numerical bootstrap with an error within $1.2\%$ at a small system size with the number of orbitals $N_m=16$, for which the computational cost is comparable to a $4\times4$ lattice system. The structure of the Ising CFT operator spectrum already starts to show up at an even smaller system size $N_m=4$. All these clues point towards a curious observation that the fuzzy sphere suffers from a remarkably small finite-size effect. The detail for the construction of models is presented in Sections~\ref{sec:construct_denint} and \ref{sec:construct_pspot}, and the detail for the analysis of the spectrum is presented in Section~\ref{sec:construct_spec}.

This seminal work opens a new avenue for studying 3D conformal field theories. After that, most of the research on the fuzzy sphere can roughly be categorised into the following directions~:

\begin{enumerate}
    \item Accessing various conformal data,
    \item Developing techniques to improve the numerical precision,
    \item Realising various 3D CFTs, 
    \item Exploring fractional quantum Hall transitions and
    \item Studying conformal defects and boundaries.
\end{enumerate}

\subsection{Accessing Various Conformal Data}

The first direction is to develop methods to calculate various universal data of 3D CFTs on the fuzzy sphere. Typically, these methods are tested with the simplest example of the 3D Ising CFT. For many of those CFT data, the fuzzy sphere is the first non-perturbative method to access them~; for the others, the fuzzy sphere has achieved great consistency with previous methods such as quantum Monte Carlo and conformal bootstrap. So far, the accessible CFT data include operator spectrum, OPE coefficients~\cite{Hu2023Mar}, correlation functions~\cite{Han2023Jun}, entropic $F$-function~\cite{Hu2024}, conformal generators~\cite{Fardelli2024,Fan2024}, the cross-cap coefficients~\cite{Dong2025}. 

\paragraph{OPE coefficients~\cite{Hu2023Mar}}

Apart from the operator spectrum, a wealth of CFT data can be obtained from the local operators. This work studies the local observables on the fuzzy sphere, including the density operators and certain four-fermion operators. These observables can be expressed as linear combinations of local scaling operators in the CFT. After a finite-size scaling with the data from different system sizes, the sub-leading contribution can be subtracted, and only the leading contribution is left. In this way, the lowest primaries in the Ising CFT in each symmetry sector, \textit{viz.}~the $\BZ_2$-odd $\sigma$ and the $\BZ_2$-even $\epsilon$, can be realised. The OPE coefficients are then evaluated by taking the inner product of a fuzzy sphere local observable with two CFT states $\langle\Phi_1|\Phi_2(\br)|\Phi_3\rangle$. Hu \textit{et al.}~compute 17 OPE coefficients of low-lying CFT primary fields with high accuracy, including four that have not been reported before. The rest are consistent with numerical bootstrap results. It is also worth noting that this work starts to apply DMRG to the fuzzy sphere. The maximal system size is increased from $N_m=18$ by ED to $N_m=48$ by DMRG. The detail for calculating the OPE coefficient is presented in Section~\ref{sec:construct_obs}.

\paragraph{Correlation functions~\cite{Han2023Jun}}

In addition to the OPE coefficients, the local observables can also be used to calculate correlation functions. By taking the inner product of two local observables (density operators) at a time displacement $\langle\Phi_1|\Phi_2(\br_0)\Phi_3(\br,\tau)|\Phi_4\rangle$ with two CFT states, a general 4-pt function can be calculated. In practice, this piece of CFT data cannot be derived from the scaling dimensions and the OPEs due to the existence of infinitely many primaries. Han \textit{et al.}~calculate the 4-pt functions in the 3D Ising CFT with DMRG. A non-trivial check of conformality, the crossing symmetry, is verified for the correlator $\langle\sigma\sigma\sigma\sigma\rangle$. The special case --- 2-pt functions by taking $\Phi_1=\Phi_4=\BI$ --- is also studied and compared with the expected results by conformal symmetry. The detail for calculating the correlation functions is presented in Section~\ref{sec:construct_obs}.

\paragraph{Entropic $F$-function~\cite{Hu2024}}

Beyond the correlators of local operators, a wealth of information can be learnt from the entanglement entropy and entanglement spectrum. A remarkable quantity is called the $F$-function, which is defined through the scaling behaviour of the entanglement entropy~\cite{Myers2010Fthm,Casini2011Fthm,Jafferis2011Fthm,Klebanov2011Fthm,Casini2012Fthm}. Specifically, consider a quantum system that lives on $\BR^2$. A circle with radius $R_d$ divides the system into inner part $A$ and outer part $B$. The entanglement entropy is defined and expected to scale with $R_d$ as 
\begin{equation}
    S_A(R_d)=-\operatorname{tr}_A\rho\log\rho=\alpha R_d/\delta-F,
\end{equation}
where $\delta$ is a UV-regulator. The constant piece is known as the $F$-function of a 3D CFT. The $F$-function is proved to be RG-monotonic, \textit{i.~e.}, along a RG flow from UV to IR, the value of $F$-function is non-increasing, analogous to the central charge in 2D CFTs. Despite its importance, it has never been calculated before through non-perturbative approaches in interacting 3D CFTs. This work has performed the first non-perturbative computation of $F$ function for the 3D Ising CFT on the fuzzy sphere. The sphere is cut in the real space into two crowns along a latitude circle $\theta$, and the entanglement entropy $S_A(\theta)$ as a function of $\theta$ is calculated~\cite{Sterdyniak2011RealEnt,Dubail2011RealEnt,Zaletel2012RealEnt,Rodriguez2011RealEnt}. The $F$-function is extracted from the $S_A(\theta)$ in the vicinity of the equator, and the result yields $F=0.0612(5)$ after a finite-size scaling.

\paragraph{Conformal generators~\cite{Fardelli2024,Fan2024}}

Within the generators of conformal symmetry, the $\SO(3)$ rotation and the dilatation are manifest and act as rotation and time translation on the fuzzy sphere. The rest two, \textit{viz.}~translation $P^\mu$ and special conformal transformation (SCT) $K^\mu$, need to be emergent in the IR at the conformal point. It is worthwhile to construct these IR generators by the UV operators on the fuzzy sphere. These works investigate such construction with the help of stress tensor $T^{\mu\nu}$. The time component $T^{\tau\tau}$ of stress tensor equals the Hamiltonian density $\cH$ and it integrates into the generator $\Lambda^\mu=P^\mu+K^\mu=\int\rd^2\br\,2n^\mu\cH$. The action of this generator sends a scaling operator to other operators in the same multiplet, with the number of spatial derivatives increased or decreased by one. Fardelli \textit{et al.}~and Fan calculate the matrix elements of the generators $\Lambda^\mu$ and find good agreement with the theoretical values in the CFT, which is another non-trivial verification of conformal symmetry. Furthermore, the separate generators $P^\mu$ and $K^\mu$ can be obtained by considering the commutator $[H,\Lambda^\mu]$, which is useful in determining the primaries. The detail for constructing the conformal generators is presented in Section~\ref{sec:construct_gen}.

\paragraph{The cross-cap coefficients~\cite{Dong2025}}

This work focuses on the 3D CFTs on the space-time manifold of the the real projective space $\BR\mathrm{P}^3$ obtained by identifying antipodal points on $S^2$. On $\BR\mathrm{P}^3$, one-point functions of scalar primary fields are generally non-vanishing and encodes the ``cross-cap coefficients.'' Dong \textit{et al.}~extracts the cross-cap coefficients of the 3D Ising CFT through simulating microscopic models on the lattice models on polyhedrons and continuum models in Landau levels and entangle the degrees of freedom at anti-podal points in Bell-type states.

\subsection{Developing Techniques to Improve the Numerical Precision}

There have also been techniques to improve the precision through accessing larger system size and detailed analysis of finite-size data, such as quantum Monte Carlo simulation~\cite{Hofmann2023}, conformal perturbation~\cite{Laeuchli2025}, and finite-size scaling the ground-state energy~\cite{Wiese2025}.

\paragraph{Quantum Monte Carlo on the fuzzy sphere~\cite{Hofmann2023}}

Until this work, the numerical methods that have been applied to the fuzzy sphere include exact diagonalisation (ED) and density matrix renormalisation group (DMRG). Hofmann \textit{et al.}~further present the numerical studies of the fuzzy sphere with quantum Monte Carlo (QMC) simulation, which is known for its potential for studying criticalities in $(2+1)$ dimensions at larger system size. Specifically, they make use of the determinant quantum Monte Carlo (DQMC) method that converts the simulation of fermions into the simulation of bosonic auxiliary fields. To overcome the sign problem, they consider two copies of the original model and construct the Ising CFT on a 4-flavour model. They determine the lowest energy spectra within each symmetry sector by calculating the imaginary-time-displaced correlation functions. They also calculate the equal-time correlation functions and compare them with the 2-pt functions of CFT.

\paragraph{Conformal perturbation~\cite{Laeuchli2025}}

The energy spectrum calculated numerically at a finite size does not coincide with that of the CFT. Part of the finite-size correction comes from the higher irrelevant operators that are not precisely tuned to zero (\textit{e.~g.}, in the Ising CFT, the irrelevant operators include $\epsilon', C_{\mu\nu\rho\sigma}, T'_{\mu\nu}$, \textit{etc.}, and the lowest singlets $\epsilon$ and $\epsilon'$ are tuned away through the two parameters). These irrelevant operators exert perturbations on the states and their energies. This work captures this kind of correction using the conformal perturbation theory. By making use of the fact that the corrections from an irrelevant operator on the energy of the primary and its descendants are not independent, the coefficients of the irrelevant operators can be fitted and their corrections can be removed.

The conformal perturbation theory is first studied on the 2D CFTs~\cite{Reinicke1987Perturbation1,Reinicke1987Perturbation2} and the icosahedron~\cite{Lao2023} and then applied to the fuzzy sphere. This work opens up a new route to improving the precision of scaling dimensions on the fuzzy sphere by making better use of the existing data. The method to partly remove the finite-size correction through conformal perturbation theory is widely used by the following works.

\paragraph{Finite-size scaling of the ground-state energy~\cite{Wiese2025}}

This work proposes a new approach to locate the phase-transition line from a finite-size scaling analysis of its ground-state energy with the example of the 3D Ising CFT. Wiese performs a finite-size scaling with the ansatz $E_{\text{GS}}=E_0 R^2+E_1+E_{3/2}R^{-1}$ and identify the minima of $E_{3/2}/E_0$ as the critical curve and the ``sweet spot.''

\subsection{Realising Various 3D CFTs}

The third direction is to study various other CFTs beyond 3D Ising. The fuzzy sphere has revealed lots of new information about these theories~; the previously known results are also consistent with the fuzzy sphere. So far, the accessible CFTs include the 3-state Potts model~\cite{Yang2025Jan}, the Yang-Lee non-unitary CFT~\cite{Fan2025,ArguelloCruz2025,EliasMiro2025} and three widely studied classes~:
\begin{enumerate}
    \item the free-scalar~\cite{He2025Jun,Taylor2025} and Wilson-Fisher CFTs realised as Heisenberg bilinear~\cite{Han2023Dec} and truncated quantum rotor model~\cite{Dey2025,Guo2025}, 
    \item CFTs with $\Sp(N)$ global symmetry related to the non-linear $\sigma$ models (NLSM) with a Wess-Zumino-Witten (WZW) topological term, including $\SO(5)$ deconfined criticality~\cite{Zhou2023} and $\rO(4)$ deconfined criticality~\cite{Yang2025Jul} through a symmetry-breaking perturbation, a series of new theories with $\Sp(N)$ symmetry~\cite{Zhou2024Oct}, and the multi-flavour $\SU(2)$ QCD~\cite{Huffman2025}.
\end{enumerate}
While the CFTs above are all built from quantum Hall ferromagnets at integer filling, another class is 
\begin{enumerate}[resume]
    \item fractional quantum Hall transitions, which will be discussed in more detail in Section~\ref{sec:review_fqh}.
\end{enumerate}

\paragraph{The 3-state Potts model~\cite{Yang2025Jan}}

The Potts models describe transitions that spontaneously break $S_Q$ symmetries where $Q\in\BZ$ is known as the number of states. In 2D, the transitions with $Q\leq Q_c=4$ are continuous and captured by CFTs~\cite{Dotsenko1984Potts}, while $Q>Q_c$ are first order. Specifically, 2D 5-state Potts transition is pseudo-critical and described by a pair of complex CFTs in its vicinity~\cite{Tang2024Potts} in a similar manner with the conjectured $\SO(5)$ DQCP. In 3D, the 3-state Potts model is found to be first-order~\cite{Barkema1991Potts,Chester2022Potts}. This work constructs a 3-flavour model on the fuzzy sphere with $S_3$ permutation symmetry among flavours. The interacting Hamiltonian resembles the Ising model. Yang \textit{et al.}~find out that the transition point of the 3D 3-state Potts model, despite being probably first-order, exhibits approximate conformal symmetry, indicating that there might be an underlying CFT describing it. However, it is difficult to determine the nature of the transition from the operator spectrum (specifically, from the relevance of the second singlet $\epsilon'$) due to the complicated finite-size effect.

\paragraph{Yang-Lee non-unitary CFT~\cite{Fan2025,ArguelloCruz2025,EliasMiro2025}} 

The Yang-Lee singularity~\cite{Yang1952Singularity,Lee1952Singularity,Cardy2023Singularity} is one of the simplest non-unitary CFTs. It is a natural extension of the Ising CFT triggered by the $i\sigma$ deformation, and plays an important role in understanding order-disorder transformation. While the 2D Yang-Lee theory is described by $M(2,5)$ minimal model and 3D Yang-Lee CFT can be solved perturbatively up to five loops, Fan \textit{et~al.}, Arguello Cruz \textit{et~al.}, and Elias Miro \textit{et~al.}~present the non-perturbative results of the Yang-Lee CFT on the fuzzy sphere. They have studied the operator spectrum, the OPE coefficients, and the RG flow from the Ising CFT to the Yang-Lee CFT. They have also shown how the finite-size effects can be controlled by finite-size scaling and conformal perturbation.

\subsubsection{Free-Scalar and Wilson-Fisher CFTs}

The $\rO(N)$ vector model is arguably the most well studied theory that describes the critical behaviours. It can be formulated as a quantum field theory (QFT) of $N$ interacting scalar fields. This QFT has an ultra-violet fixed point of $N$ free scalars. A quartic interaction term induces a RG flow to the infra-red fixed point described by the Wilson-Fisher (WF) CFT~\cite{Wilson1971WF,Sachdev2011Quantum,Cardy1996Scaling}. It describes a transition between an ordered phase with $\rO(N)$ spontaneous symmetry breaking and an $\rO(N)$-symmetric disordered phase. This theory has been extensively studied using perturbative calculations, quantum Monte Carlo and conformal bootstrap. The $\rO(2)$ WF has been experimentally measured with high precision in the normal-superfluid transition of Helium-4~\cite{Lipa2003SF}. Its fuzzy-sphere realisation paves the way for studying operators with certain quantum numbers, multi-scalar CFTs, defects and boundaries, \textit{etc.}

\paragraph{The free scalar CFT~\cite{He2025Jun,Taylor2025}}

This work introduces a simple model to realise the free real scalar CFT on the fuzzy sphere that is structurally similar to the 3D Ising CFT. A weakly-broken $\rU(1)$ symmetry in the fuzzy-sphere model realises the $\BR$-shift symmetry of the free scalar in the thermodynamic limit, so that the fix point can be accessed with only a single tuning parameter. He numerically demonstrates that the model correctly reproduces the operator spectrum, correlation functions and the harmonic oscillator algebra of the real-scalar CFT. He generalises the Girvin-MacDonald-Platzman algebra~\cite{Girvin1986GMP} to the fuzzy-sphere algebra of the density operators, which is potentially useful for defining quantum field theories on non-commutative geometries. He proposes a wave-function ansatz for the ground states which exhibit remarkable agreement with the CFT ground state wave-functions of the fuzzy-sphere model.

\paragraph{The bilayer Heisenberg transition~\cite{Han2023Dec}}

This work studies the $\rO(3)$ WF CFT on the set-up of a bilayer Heisenberg model. The construction involves two copies of $\SU(2)$ ferromagnet with altogether four flavours. Briefly speaking, the model contains two competing terms~: (1) a $\SU(2)$ ferromagnetic interaction which favours a Heisenberg ferromagnetic phase where each of the two copies is half-filled and the symmetry-breaking order parameter lives on a $S^2$ manifold, (2) a polarising term which favours one of the two copies being completely filled, corresponding to a Heisenberg paramagnet. The transition between these two phases falls into the $\rO(3)$ Wilson-Fisher universality. Through the energy spectrum at the transition, Han \textit{et al.}~provide evidence that $\rO(3)$ Wilson-Fisher fixed point exhibits conformal symmetry, as well as revealing a wealth of information about the CFT, \textit{e.~g.}~the instability to cubic anisotropy. They also calculate several OPE coefficients.

\paragraph{The general model for $\rO(N)$ free-scalar and Wilson-Fisher CFTs~\cite{Guo2025}}

This work proposes a general model that realises the $\rO(N)$ free-scalar and Wilson-Fisher CFTs. The set-up contains altogether $(N+1)$ flavours with total filling $\nu=1$, where one flavour $c_0$ is $\rO(N)$ singlet and the rest $N$ flavours $c_i$ transform as $\rO(N)$ vector. The scalar field is realised as the bilinear $\phi_i=c_i^\dagger c_0+c_0^\dagger c_i$, and the Hamiltonian for the WF CFT contains (1) a particle density interaction, (2) a $\phi$ density interaction, and (3) a relative chemical potential of the singlet flavour. Tuning the chemical potential realises a phase transition between a paramagnetic phase with $c_0$ fully filled and an $\rO(N)$ symmetry-breaking phase where electrons are in a superposed state between a vector flavour and the singlet flavour. Turning to the $\rO(N)$ free-scalar theory, it corresponds to a pseudo-Goldstone phase of a $\mathrm{PO}(N)$ SSB.

Guo \textit{et al.}~provide numerical evidence for $N=2,3,4$ such as the operator spectra and two-point conformal correlators. They discuss several quantities that may not have a direct CFT interpretation, but imply that the CFTs become to some extent semi-classical once regularised on the fuzzy sphere. 

\paragraph{The detailed study of the $\rO(3)$ Wilson-Fisher CFT~\cite{Dey2025}}

This work focuses on the $\rO(3)$ WF and fomulate the four-flavour model as a truncated rotor model. Dey \textit{et al.}~study the model numerically using ED and DMRG. They locate the critical point through a careful conformal perturbation analysis. They obtain scaling dimensions from finite-size spectra and OPE coefficients through conformal perturbation. The results are benchmarked with conformal bootstrap and perturbative calculations.

\subsubsection{NLSM-WZW with Symplectic Symmetry}
\newcommand{\rG}{\mathrm{G}}

One route to realise many CFTs is through the non-linear sigma model (NLSM) with a Wess-Zumino-Witten (WZW) term. The NLSM on a Grassmannian 
\begin{equation*}
    \frac{\rG(N)}{\rG(M)\times\rG(N-M)},\qquad\rG(N)=\Sp(\tfrac{N}{2}),\rO(N),\rU(N),\dots
\end{equation*}
captures a symmetry breaking pattern of fully filling $M$ out of $N$ flavours of fermions with a global symmetry $\rG(N)$. The NLSM often allows a topological WZW term with a quantised level. On the Landau level, a WZW term naturally appear~\cite{Lee2014WZW,Huffman2025}. While the 3D NLSM in its renormalisable region can only flow to the SSB phase, the symmetry and anomaly of the NLSM-WZW can be matched with critical gauge theories with critical scalars or fermions coupled to a dynamic gauge field~\cite{Komargodski2017QCD}. These gauge theories can either serve as the UV completion of the NLSM, or has the same UV completion with the NLSM. The simplest ones to construct on the fuzzy sphere are the NLSM on symplectic Grassmannians (\textit{i.~e.}~with $\rG(N)=\Sp(\tfrac{N}{2})$) with a level-1 WZW term.\footnote{The construction for $\rU$ or $\rO$ faces the difficulty that their centres $\rU(1)$ or $\BZ_2$, as the electric charge conservation or its subgroup, must be decoupled~; the construction for $\SU$ or $\SO$ faces the difficulty in realising the topological $\rU(1)$ or $\BZ_2$ symmetry.} They can be constructed by considering $N$ flavours of fermions and break the flavour symmetry to $\Sp(\tfrac{N}{2})$. This construction can be used to realise various 3D QCDs. 

\paragraph{The $\SO(5)$ deconfined criticality~\cite{Zhou2023}}

Deconfined quantum critical point (DQCP) is one of the pioneering examples of phase transitions beyond Landau paradigm~\cite{Senthil2003DQCP,Senthil2004DQCP,Senthil2023DQCP}. It has led to numerous theoretical surprises, including the emergent $\SO(5)$ symmetry~\cite{Nahum2015DQCP} and the duality between interacting theories~\cite{Wang2017DQCP}. Despite extensive studies over the past two decades, its nature remains controversial. Numerical simulations have shown no signal of discontinuity, but abnormal scaling behaviours have been observed~\cite{Senthil2023DQCP}. A plausible proposal to reconcile the tension is that DQCP is pseudo-critical, \textit{i.~e.}~a weakly first-order phase transition that has approximate critical behaviour, and is controlled by a pair of complex fixed points very close to the pseudo-critical region~\cite{Wang2017DQCP,Gorbenko2018Complex}.

The DQCP can be conveniently studied on the fuzzy sphere by constructing a NLSM on $S^4$ with WZW level-1, which serves as a dual description of the DQCP with an exact $\SO(5)$ symmetry~\cite{Nahum2015DQCP,Wang2017DQCP}. The idea is to construct a 4-flavour model with symmetry $\Sp(2)/\BZ_2=\SO(5)$ ($\BZ_2$ means to project the pseudo-real representations). At half-filling, it can be described by a NLSM on the Grassmannian $\tfrac{\Sp(2)}{\Sp(1)\times\Sp(1)}\cong S^4$ and the WZW level can be matched~\cite{Ippoliti2018DQCP,Wang2020DQCP}. This work provides evidence that the DQCP exhibits approximate conformal symmetry. Zhou \textit{et al.}~have identified 19 conformal primaries and their 82 descendants. Furthermore, by examining the RG flow of the lowest symmetry singlet, they demonstrate that the DQCP is more likely pseudo-critical, with the approximate conformal symmetry plausibly emerging from nearby complex fixed points. Several works~\cite{Chen2023WZW,Chen2024WZW} appear later to follow up.

\paragraph{The $\rO(4)$ deconfined criticality~\cite{Yang2025Jul}}

Apart from the $\SO(5)$ deconfined criticality, one can reach the $\rO(4)$, or easy-plane DQCP by adding a perturbation in the $\SO(5)$ symmetric tensor representation that breaks the global symmetry to $\rO(4)$. The $\rO(4)$ DQCP appears in several lattice models and has several gauge theory descriptions, \textit{e.~g.}~QED with two flavours of Dirac fermions~\cite{Wang2017DQCP}. This work studies the RG flow from the $\SO(5)$ to the $\rO(4)$ DQCP and traces the $\rO(4)$ operators back to the $\SO(5)$ theory.

\paragraph{A series of new $\Sp(N)$-symmetric CFTs~\cite{Zhou2024Oct}}

The quest to discover new 3D CFTs has been intriguing for physicists. A virgin land on this quest is the parity-breaking CFTs. In 3D, the Chern-Simons-matter theories stand out as the most well-known and possibly the only known type of parity-breaking CFTs. The fuzzy sphere is a promising platform for studying these theories. This work makes a concrete construction by generalising the DQCP to the WZW-NLSM on the target space of a general symplectic Grassmannian $\frac{\Sp(N)}{\Sp(M)\times\Sp(N-M)}$. Several candidate Chern-Simons-matter theories are known to exist on its phase diagram with $N$ flavour of gapless bosons or fermions coupled to a non-Abelian (\textit{viz.}~$\Sp(1)$, $\Sp(2)$, \textit{etc.}) Chern-Simons gauge field~\cite{Komargodski2017QCD}. On the fuzzy sphere, this WZW-NLSM can be realised by a $2N$ layer model with $\Sp(N)$ flavour symmetry, and $2M$ out of the $2N$ layers are filled. Zhou \textit{et al.}~numerically verify the emergent conformal symmetry by observing the integer-spaced conformal multiplets and studying the finite-size scaling of the conformality.

\paragraph{Multi-flavour $\SU(2)$ QCD~\cite{Huffman2025}}

The conformal window is an outstanding problem for 3D gauge theories. For a given gauge field, when the number of flavours coupled to it exceeds some critical value $N>N_c$, the field theory flows to a conformal fixed point~; otherwise the gauge field is confined and the field theory flows to SSB. The conformal windows of QED\textsubscript{3} and $\SU(2)$ QCD\textsubscript{3} are directly related to the nature of deconfined criticality and the Dirac spin liquid. However, results beyond perturbative calculations for certain gauge groups are still hard to access.

On the fuzzy sphere, the $\SU(2)$ QCD\textsubscript{3} with $N$ flavours of fermions can be realised with the $\Sp(N)$ model at half-filling $M=N/2$. The model is free of sign problem and thus accessible through quantum Monte Carlo. Notably, $N$ is treated as a parameter instead of the number of fermions simulated, so the calculation can reach arbitrarily large number of flavours. 

This work shows that the phase diagram at $N\geq 4$ has a stable conformal fixed point corresponding to the QCD\textsubscript{3} through examining the RG flow. Huffmann \textit{et al.}~extracts the various scaling dimensions from the equal-time and imaginary-time two-point correlation functions and finds agreement with conformal symmetry and large-$N$ expansion result.

\subsection{Exploring Fractional Quantum Hall Transitions}
\label{sec:review_fqh}

The models above are all built from quantum Hall ferromagnets at integer filling. Specifically, in the absence of the interaction, an integer number of lowest Landau levels are fully occupied. The lowest Landau level at fractional filling realises various topological orders (TO) with long-range entanglement and anyon excitations. \textit{E.~g.}, the Laughlin states~\cite{Laughlin1983Anomalous} with filling $\nu=1/k$ realise the Abelian topological orders captured by Chern-Simons theories $\rU(1)_{-k}$, the Jain sequence~\cite{Jain1989Composite} with $\nu=p/(mp+1)$ realises the Abelian topological orders with composite fermion descriptions, and the Read-Rezayi sequence with $\nu=p/(mp+2)$~\cite{Read1998Rezayi}, as a natural extension of the Moore-Read states~\cite{Moore1991Pfaffian}, realises in general a series of non-Abelian topological orders. These states have finite charge gaps as well that guarantees that the gapless spin degree of freedom can decouple from the strongly non-commutative charge degree of freedom.

The fuzzy sphere provides a suitable platforms to study the transitions between or out of these fractional quantum Hall (FQH) states. Theoretically, these transitions are often described by Chern-Simons theories coupled to matter, many of which are conjectured to be conformal with enhanced symmetries and field theory dualities~\cite{Dasgupta1981Duality,Metlitski2015Duality,Seiberg2016Duality}. The Moir\'e materials~\cite{Cai2023Moire,Zeng2023Moire,Park2023Moire} provide an experimental platform to realise these transitions. The critical points can be conveniently reached by tuning many knobs like potential amplitude.

The exploration starts at realising the Ising CFT on the background of a FQHE state~\cite{Voinea2024}, for which the topological order stays the same across the transition. The transition between $\nu=1/2$ bosonic Laughlin state and a $\nu=2$ fermionic Laughlin state, realising the confinement transition of $\nu=1/2$ bosonic Laughlin state~\cite{Zhou2025Jul}, is the first fractional quantum Hall transition on the fuzzy sphere. Umsing similar strategy, the phase diagram between fermionic integer quantum Hall and bosonic Laughlin state realising two transitions of free Majorana fermion and gauged Ising CFT~\cite{Zhou2025Sep}, and the transition between bosonic Pfaffian and Halperin 220 state~\cite{Voinea2025}, are also studied.

\paragraph{Ising CFT from FQHE state~\cite{Voinea2024}}

Voinea \textit{et al.}~explore the possibility of constructing CFTs on the Haldane-Laughlin states that capture the fractional quantum Hall effect (FQHE)~\cite{Haldane1983LLL,Laughlin1983FQHE}. Specifically, they study the fermionic LLL at fillings of $\nu=1/3$ and $1/5$. The model Hamiltonian contains (1) a dominant projection term that puts the ground state on the Haldane-Laughlin state, and (2) an interaction term as a perturbation that drives the Ising-type phase transition. They show that the energy spectra at the critical point exhibit conformal symmetry. Notably, they also make the construction with respect to the bosonic LLL at a filling of $\nu=1/2$.

\paragraph{The confinement transition of $\nu=1/2$ bosonic Laughlin state~\cite{Zhou2025Jul}}

This work studies one of the simplest instances of Chern-Simons-matter theories~: one complex critical scalar coupled to a $\rU(1)_2$ Chern-Simons gauge field and its three other dual Lagrangian descriptions~\cite{Seiberg2016Duality,Hsin2016LevelRank,Benini2017Duality}. They describe the transition between a $\nu=1/2$ bosonic Laughlin state and a trivially gapped phase. This transition appears in various contexts in condensed-matter physics, \textit{viz.}~Kalmeyer-Laughlin chiral spin liquid~\cite{Kalmeyer1987CSL,Kalmeyer1989CSL}, anyon superconductivity~\cite{Divic2025AnyonSC,Kim2024AnyonSC,Shi2024AnyonSC}, Feshbach resonance~\cite{Yang2008Feshbach,Liou2018Feshbach}, \textit{etc.}

Zhou \textit{et al.}~realise this theory on a set-up with fermion-boson mixture containing two flavours of fermion and one flavour of bosons carrying electric charge $Q_b=2Q_f$. Tuning the relative chemical potential induces a transition between a $\nu_f=2$ fermionic integer quantum Hall state and a $\nu_b=1/2$ bosonic fractional quantum Hall state. They show that the transition is continuous and has emergent conformal symmetry. The operator spectrum has exactly one relevant singlet with scaling dimension $\Delta_S=1.52(18)$, signature of a critical point. This work opens up the route of using the fuzzy sphere to study transitions between distinct topological order. 

\paragraph{The free Majorana fermion theory~\cite{Zhou2025Sep}} 

The CFTs realised on the fuzzy sphere until this work contains only bosonic CFT operators. The microscopic fermions on the fuzzy sphere have strong non-commutativity, transform under projective representations of the sphere rotation symmetry, and thus cannot flow to CFT operators. This work overcomes this challenge by constructing a boson-fermion mixed set-up with both microscopic bosons $b$ and fermions $f$ with the same electric charge and let their angular momenta differ by $1/2$. Their bilinears $b^\dagger f$ or $f^\dagger b$ can be used to realise fermionic local operators in a CFTs.

Zhou \textit{et al.}~allow conversion between two bosons and two fermions and set the total filling $\nu=1$. On the phase diagram, there are three gapped phases, \textit{viz.}~a fermionic integer quantum Hall phase, an $f$-wave chiral topological superconductor, and a bosonic Pfaffian phase~\cite{Moore1991Pfaffian}. They are separated by two continuous transitions described respectively by a massless free Majorana fermion and a gauged Ising CFT. Notably, the Hamiltonian that realises the free Majorana fermion is analogous with the field theory Hamiltonian in addition to a density-density interaction.

\paragraph{The transition between bosonic Pfaffian and Halperin 220~\cite{Voinea2025}}

The fuzzy sphere can be used to answer long-standing questions on fractional quantum Hall transitions. An example is the transition between the bosonic Moore-Read Pfaffian~\cite{Moore1991Pfaffian} and the Halperin 220 state~\cite{Halperin1983} on a bilayer bosonic system, which has been predicted to be described by a massless free Majorana fermion with the fermion parity $\BZ_2$ symmetry gauged~\cite{Wen2000Majorana}, but lacks simulations. This work identifies the low-energy spectrum with the gauged Majorana fermion. 

\subsection{Studying Conformal Defects and Boundaries}

Apart from the bulk CFTs, the fuzzy sphere can also be used to study their conformal defects and boundaries. Deforming a CFT with interactions living on a sub-dimensional defect may trigger a RG flow towards a non-trivial interacting IR fixed point. A defect IR theory with a smaller conformal symmetry is called a defect CFT (dCFT)~\cite{Billo2013Defect,Billo2016Defect}. The dCFTs have rich physical structures, such as defect operators and bulk-to-defect correlation functions. Moreover, a bulk CFT can flow to several different dCFTs. Similarly, deformation on the boundary may trigger a flow towards a boundary CFT (bCFT). So far, the accessible defects/boundaries include the pinning-field defect of the 3D Ising CFT, in particular, its defect operator spectrum, correlators~\cite{Hu2023Aug}, $g$-function, defect-changing operators~\cite{Zhou2024Jan}, its cusp~\cite{Cuomo2024}, and the conformal boundaries of the 3D Ising CFT~\cite{Zhou2024Jul,Dedushenko2024}. Besides the defects and boundaries of 3D CFTs, one can also studying lower dimensional bulk CFTs on a $(2+1)$D set-up, like the on a fuzzy thin torus~\cite{Han2025}.

\paragraph{Conformal pinning-field defect~\cite{Hu2023Aug}}

This is the first work that studies conformal defects with the fuzzy sphere. The simplest example of conformal defect is the pinning-field defect of the 3D Ising CFT~\cite{Andreas2000MagLine,Allais2014MagLine,Allais2013MagLine,Pannell2023MagLine}, where the defect line is completely polarised and the $\BZ_2$ symmetry is explicitly broken. A defect line along $z$-direction that passes the origin point, after the radial quantisation, corresponds to the north and south poles of the sphere being polarised. Hence, to realise the pinning-field defect on the fuzzy sphere, one only needs to apply a pinning magnetic field to the north and south poles (Since only the $m=+s$ orbital has non-zero amplitude at the north pole and $m=-s$ at the south pole due to the locality, one only need to pin the $m=\pm s$ orbitals).

This work demonstrates that the defect IR fixed point has emergent conformal symmetry $\SO(2,1)\times\rO(2)$~: in the operator spectrum, there exists a displacement operator as the non-conservation of stress tensor at exactly $\Delta_\rd=2$, and the defect primaries and descendants have integer spacing~; the bulk-to-defect 1-pt and 2-pt correlation functions follow a power law. Hu \textit{et al.}~have identified six low-lying defect primary operators, extracted their scaling dimensions, and computed the 1-pt function of bulk primaries and 2-pt bulk-to-defect correlators.

\paragraph{The $g$-function and defect-changing operators~\cite{Zhou2024Jan}}

This work studies the $g$-function of conformal defects and the defect-creation and changing operators. Similar to the central charge and the $F$-function in bulk CFTs, there exists a RG-monotonic quantity called the $g$-function for the line defects that is non-increasing along the flow~\cite{Cuomo2021gfn,Casini2022gfn}. It is defined as the ratio between the partition functions of the defect CFT and the bulk CFT. On a different note, consider two semi-infinite pinning-field defects pinned towards opposite directions joint at one point, a defect-changing operator (also known as the domain wall operator) lives at the joining point~\cite{Affleck1994DefCh,Affleck1996DefCh}. Similarly, a defect-creation operator lives at the end-point of a semi-infinite line defect. The relevance of the defect-changing operator is related to the stability of spontaneous symmetry breaking (SSB) on the line defect.

Zhou \textit{et al.}~realise the defect-creation and changing operators for the Ising pinning-field defect by acting a pinning field at the north pole, and opposite pinning fields at the north and south poles, respectively. The scaling dimensions are calculated through state-operator correspondence $\Delta_{\textrm{creation}}=0.108(5),\Delta_{\textrm{changing}}=0.84(5)$, indicating the instability of SSB on the Ising pinning-field defect. Moreover, they show that the $g$-function and many other CFT data can be calculated by taking the overlaps between the eigenstates of different defect configurations. Notably, this work has given the first non-perturbative result for the $g$-function $g=0.602(2)$.

\paragraph{Cusp~\cite{Cuomo2024}}

A cusp is two semi-infinite defect lines joined at one point at an angle. This can be realised on the fuzzy sphere through pinning fields at two points at an angle. Cuomo \textit{et al.}~study the cusps through various theoretical and numerical approaches. In particular, on the fuzzy sphere, they calculate the cusp anomalous dimension as a function of the angle for the Ising pinning-field defect, and verify its relation with the Zamolodchikov norm of the displacement operator.

\paragraph{Conformal boundaries of 3D Ising CFT~\cite{Zhou2024Jul,Dedushenko2024}}

Apart from line defects, boundaries are also important extended objects in CFT. For the Ising CFT, there exist several conformal boundaries~: normal bCFT with explicitly broken $\BZ_2$ symmetry, ordinary bCFT that is stable and has preserved $\BZ_2$ symmetry, extraordinary bCFT with spontaneously broken $\BZ_2$ symmetry, and special bCFT as the transition between ordinary and extraordinary bCFTs~\cite{Metlitski2020IsingBd,Krishnan2023IsingBd,Giombi2023IsingBd}. These works focus on the normal and ordinary bCFTs and show that they can be realised by acting a polarising field on a hemisphere. By noting that the LLL orbitals are localised along latitude circles, the bCFTs can equivalently be realised by pinning the orbitals with $m<0$. By studying the operator spectrum, Zhou \textit{et al.}~and Dedushenko \textit{et al.}~show numerical evidence for conformal symmetry and estimate the scaling dimensions of the conformal primaries. They also calculate the bulk-to-boundary 1-pt and 2-pt functions and extract the corresponding OPE coefficients. Interestingly, Zhou \textit{et al.}~notice a certain correspondence between the boundary energy spectrum and bulk entanglement spectrum through the orbital cut.

\paragraph{Fuzzy circle~\cite{Han2025}}

Besides the fuzzy sphere $S^2$, the regularisation with the lowest Landau level can also be used for other manifolds or dimensions. This work studies the 2D CFT on a ``fuzzy circle.'' Although Landau levels can only be defined on even space dimensions~\cite{Hasebe2020Landau}, one can reach odd space dimensions by compactifying one of the even dimensions. Specifically, Han \textit{et al.}~construct the LLL on a thin torus $T^2$ and sends one of the lengths to infinity while fixing the other. In this way, a circle $S^1$ is recovered in the thermodynamic limit. They construct the 2D Ising and 3-state Potts CFTs on the fuzzy circle and compare the operator spectrum and OPE coefficients with the Virasoro multiplet structure and the known results in the exactly solvable minimal models.

\section{Model Construction}
\label{sec:construct}

In this section, we review the process of constructing a model on the fuzzy sphere and extracting the conformal data. We aim to provide the necessary technical information for those who want to get started with the research on the fuzzy sphere, especially the aspects rarely covered by other literature.

\subsection{Projection onto the Lowest Landau Level}
\label{sec:setup}

To build the set-up of the fuzzy sphere, we consider a sphere with radius $R$ and put a $4\pi s$-monopole at its centre. Consider free electrons moving on the sphere. The monopole modifies the single particle Hamiltonian~\cite{Haldane1983LLL,Wu1976LLL,Greiter2011LLL,Hasebe2010LLL}
\begin{equation}
    H_0=\frac{1}{2MR^2}(\partial^\mu+iA^\mu)^2,
\end{equation} 
where $\mu=\theta,\phi$ and the gauge connexion is taken as
\begin{equation}
    A_\theta=0,\quad A_\phi=-\frac{s}{R}\operatorname{ctg}\theta.
\end{equation}
The eigenstates of the Hamiltonian are the monopole spherical harmonics
\begin{equation}
    \Phi_{nm}(\br)=\frac{1}{R}Y_{lm}^{(s)}(\bnh),\qquad n=0,1,\dots,\quad l=n+s,\quad m=-l,\dots,l-1,l,
\end{equation}
where $\bnh$ is the unit vector of the point on the sphere specified by angular co-ordinates $\theta$ and $\phi$, and the energies are
\begin{equation}
    E_n=\frac{1}{2MR^2}n(2s+n+1).
\end{equation}
Each level, known as a Landau level, has a degeneracy $(2l+1)$. Specifically, the wave-functions on the lowest Landau level (LLL) $n=0,l=s$ are easy to write out~:
\begin{equation}
    \Phi_{0m}(\br)=\frac{1}{R}Y_{sm}^{(s)}(\bnh),\qquad Y_{sm}^{(s)}(\bnh)=C_me^{im\phi}\cos^{s+m}\frac{\theta}{2}\sin^{s-m}\frac{\theta}{2}
\end{equation}
where $C_m=1/\sqrt{4\pi\Beta(s+m+1,s-m+1)}$ is the normalising factor, and $\Beta$ is the Euler's beta function. The LLL has a degeneracy $N_m=2s+1$.

We now consider $N_f$ flavours of fermions moving on the sphere, characterised by the second-quantised fermion operator $\psi_f(\br)$, with a flavour index $f=1,\dots,N_f$. We partially fill the lowest Landau level and set the single energy gap to be much larger than the scale of interaction $H_0\gg H_\Int$, so that the quantum fluctuation can be constrained on the lowest Landau level. In practice, we often fill integer number of flavours $N_e=kN_m$ ($k\in\BZ$) so that a quantum Hall ferromagnet (\textit{i.~e.}~the state where integer number of LLLs are fully filled) is preferred in the absence of interaction, for which the charge degree of freedom is gapped and decouples from the gapless CFT degree of freedom when the interactions are introduced.\footnote{The charge degree of freedom cannot flow to any gapless CFT local operator because the charged operators carry projective representations of the rotation group due to the non-commutativity while the CFT operators carry linear prepresentations.}

We then project the system onto the LLL. Technically, this can be done by writing the fermion operators in terms of the annihilation operators of the LLL orbitals\footnote{In the old version, we used a different convension
\begin{equation*}
    \psi_f(\bnh)_\text{(old)}=\sum_{m=-s}^s Y^{(s)}_{sm}(\bnh)c_{mf}.
\end{equation*}
In the old convension, the components of density operator $n_{M,lm}$ are extensive and their action decreases the angular momentum $L^z$ by $m$~; in the current convension, $n_{M,lm}$ is intensive and its action increases the angular momentum $L^z$ by $m$. In the code, two parameters \href{https://docs.fuzzified.world/models/\#FuzzifiED.ObsNormRadSq}{\lstinline[basicstyle=\ttfamily\scriptsize]|ObsNormRadSq|} and \href{https://docs.fuzzified.world/models/\#FuzzifiED.ObsMomIncr}{\lstinline[basicstyle=\ttfamily\scriptsize]|ObsMomIncr|} control the convension.}
\begin{equation}
    \psi^\dagger_f(\br)=\sum_{m=-s}^s \Phi_{0m}(\br)c^\dagger_{mf}=\frac{1}{R}\sum_{m=-s}^s Y^{(s)}_{sm}(\bnh)c^\dagger_{mf},
\end{equation}
where $c^{(\dagger)}_{mf}$ annihilates/creates an electron with $L^z$-quantum number $m$ at the $f$-th flavour of the lowest Landau level. Hereafter, we will omit the hats on the operators.

After the projection, we obtain a finite Hilbert space on which numerical simulations can be carried out. For this purpose, the system is analogous to a length-$(2s+1)$ spin chain with long-range interaction, where different Landau level orbitals behave like the lattice sites. The difference is that the $(2s+1)$ orbitals form a spin-$s$ representation of the $\SO(3)$ rotation group, and in this way the continuous rotation symmetry is preserved. The exact rotation symmetry shortens the RG flow from the UV to the IR and reduces the finite-size effect, so that the numerical results are considerably accurate even at a small system size.

The word ``\emph{fuzzy}'' means non-commutativity~\cite{Madore1991Fuzzy}. Here, the magnetic field results in the non-commutativity of the co-ordinates. More concretely, we write the co-ordinate operators as a matrix on the lowest Landau level
\begin{equation}
    X^\mu_{m_1m_2}=\int\rd^2\br\,x^\mu \frac{Y_{sm_1}^{(s)}(\br)}{R}\frac{\Yb_{sm_2}^{(s)}(\br)}{R}.
\end{equation}
These matrices $\bX^\mu$ ($\mu=x,y,z$) satisfy relation~\cite{Hasebe2010LLL,Zhou2023}
\begin{equation}
    \bX_\mu\bX^\mu=\frac{s}{s+1}R^2\BI,\qquad [\bX^\mu,\bX^\nu]=\frac{1}{s+1}i\epsilon^{\mu\nu\rho}R\bX_\rho.
\end{equation}
The first equation involves a renormalised radius $\tilde{R}$ of the sphere, and the second equation involves the magnetic length $l_B$ that determines the non-commutativity.
\begin{equation*}
    \bX_\mu\bX^\mu=\tilde{R}^2\BI,\qquad [\bX^\mu,\bX^\nu]=l_B^2\,i\epsilon^{\mu\nu\rho}(\bX_\rho/R).
\end{equation*}
Their ratio scales with the square root of the number of orbitals
\begin{equation}
    \tilde{R}/l_B=[s(s+1)]^{1/4}\sim\sqrt{N_m}.
\end{equation}
We can take $l_B=1$ as the unit length. In this way, the thermodynamic limit can be taken as $N_m\to\infty$, where a regular sphere is recovered. Hereafter we take the radius of the sphere $R=\sqrt{N_m}$

\subsection{Density Operator}
\label{sec:construct_den}

Having constructed the single-particle states, we then consider the interacting many-body Hamiltonian. The simplest building block is the density operator, \textit{i.~e.}, local fermion bilinear
\begin{equation}
    n_M(\br)=\psi_{f'}^\dagger(\br)M_{f'f}\psi_f(\br).
    \label{eq:den_def}
\end{equation}
Here, the matrix insertion $M$ puts the density operators in a certain representation of the flavour symmetry. \textit{E.~g.}, for a 2-flavour system, $M$ can be taken as the Pauli matrices $\BI,\sigma^x,\sigma^y,\sigma^z$~; for a system with $N_f$ flavours in the fundamental representation of $\SU(N_f)$ flavour symmetry, one can put $n_M$ in the singlet or adjoint representation
\begin{align}
    n_S(\br)&=\psi_{c}^\dagger(\br)\psi^c(\br)\nonumber\\
    n_a{}^b(\br)&=\psi_{a}^\dagger(\br)\psi^b(\br)-\tfrac{1}{N}\delta_{a}{}^b\psi_c^\dagger(\br)\psi^c(\br).
\end{align}
Like the fermion operator, the density operator can also be expressed in the orbital space
\begin{equation}
    n_M(\br)=\sum_{lm}Y_{lm}(\bnh)n_{M,lm}.
    \label{eq:den_decomp}
\end{equation}
Conversely,
\begin{align}
    n_{M,lm}&=\frac{1}{R^2}\int\rd^2\br\,\Yb_{lm}(\bnh)n_M(\br)\nonumber\\
    &=\int\rd^2\bnh\,\Yb_{lm}(\bnh)\left(\frac{1}{R}\sum_{m_1}Y^{(s)}_{sm_1}(\bnh)c^\dagger_{m_1f_1}\right)M_{f_1f_2}\left(\frac{1}{R}\sum_{m_2}\Yb^{(s)}_{sm_2}(\bnh)c_{m_1f_2}\right)\nonumber\\
    &=\frac{1}{R^2}\sum_{m_1m_2}c^\dagger_{m_1f_1}M_{f_1f_2}c_{m_1f_2}\int\rd^2\bnh\,\Yb_{lm}(\bnh)Y^{(s)}_{sm_1}(\bnh)\Yb^{(s)}_{sm_2}(\bnh)\nonumber\\
    &=\frac{1}{R^2}\sum_{m_1}c^\dagger_{m_1f_1}M_{f_1f_2}c_{m-m_1,f_2}\nonumber\\
    &\qquad\qquad\times(-1)^{s+2m-m_1}(2s+1)\sqrt{\frac{2l+1}{4\pi}}\begin{pmatrix}l&s&s\\-m&m_1&-m_1+m\end{pmatrix}\begin{pmatrix}l&s&s\\0&-s&s\end{pmatrix}.
    \label{eq:den_mod}
\end{align}
Here, we have used the properties of the monopole spherical harmonics~\cite{Wu1977Monopole,Shnir2005Monopole}
\begin{subequations}
\begin{align}
    \Yb_{lm}^{s}&=(-1)^{s+m}Y_{l,-m}^{(-s)}\\
    \int\rd^2\bnh\,Y_{lm}^{(s)}\Yb_{lm}^{(s)}&=\delta_{ll'}\delta_{mm'}\\
    \int\rd^2\bnh\,Y_{l_1m_1}^{(s_1)}Y_{l_2m_2}^{(s_2)}Y_{l_3m_3}^{(s_3)}&=\sqrt{\frac{(2l_1+1)(2l_2+1)(2l_3+1)}{4\pi}}\begin{pmatrix}l_1&l_2&l_3\\m_1&m_2&m_3\end{pmatrix}\begin{pmatrix}l_1&l_2&l_3\\-s_1&-s_2&-s_3\end{pmatrix},
\end{align}
\end{subequations}
where $\left(\begin{smallmatrix}\bullet&\bullet&\bullet\\\bullet&\bullet&\bullet\end{smallmatrix}\right)$ is the $3j$-symbol~\cite{Biedenharn1984Angular}, and we denote the common spherical harmonics by $Y_{lm}^{(0)}=Y_{lm}$. In this way, we have fully expressed the density operator in terms of the operators in the orbital space $c^{(\dagger)}_{mf}$.

\subsection{Density-Density Interaction}
\label{sec:construct_denint}

The most straightforward way to construct an interaction term is to add a density-density interaction with a potential function\footnote{Computationally, this is not the simplest construction and we will present the simpler construction in terms of pseudo-potentials in the next section.}
\begin{equation}
    H_\Int=\int\rd^2\br_1\,\rd^2\br_2\,U(|\br_1-\br_2|)n_M(\br_1)n_M(\br_2).
\end{equation}
The interacting potentials can be expanded in terms of the Legendre polynomials
\begin{equation}
    U(|\br_{12}|)=\sum_l\tilde{U}_lP_l(\cos\theta_{12})=\sum_{lm}\frac{4\pi\tilde{U}_l}{2l+1}\Yb_{lm}(\bnh_1)Y_{lm}(\bnh_2),
\end{equation}
where the cordial distance $\br_{12}=\br_1-\br_2$ and $|\br_{12}|=2R\sin\theta_{12}/2$. Conversely
\begin{equation}
    \tilde{U}_l=\int\sin\theta_{12}\rd\theta_{12}\,\frac{2l+1}{2}U(|\br_{12}|)P_l(\cos\theta_{12}).
\end{equation}
Specifically, for local and super-local interactions
\begin{align}
    U(|\br_{12}|)&=g_0\delta(\br_{12}),&\tilde{U}_l&=\frac{g_0}{R^2}(2l+1)\nonumber\\
    U(|\br_{12}|)&=g_1\nabla^2\delta(\br_{12}),&\tilde{U}_l&=-\frac{g_1}{R^4}l(l+1)(2l+1).
\end{align}
Here we make use of the conversion relations 
\begin{equation*}
    \delta(\br)=\frac{1}{R^2}\delta(\bnh),\qquad \nabla^2_\br=\frac{1}{R^2}\nabla^2_{\bnh}
\end{equation*}
for $\br=R\bnh$. By expanding the density operators into the orbital space and completing the integrals,
\begin{equation}
    H_\Int=\sum_{lm}\frac{4\pi R^4\tilde{U}_l }{2l+1}n^\dagger_{M,lm}n_{M,lm}.
\end{equation}

With these ingredients, we can now consider how to construct models. This comes down to matching the symmetry and phase diagram. \textit{E.~g.}, for the Ising model~\cite{Zhu2022}, the $\BZ_2$ global symmetry is realised as the exchange of the two flavours $\psi_\uparrow(\br)\leftrightarrow\psi_\downarrow(\br)$. We need a phase diagram with a paramagnetic (PM) phase where the $\BZ_2$ symmetry is conserved and a ferromagnetic (FM) phase where the $\BZ_2$ symmetry is spontaneously broken. The PM phase is favoured by a polarising term that resembles a transverse field
\begin{equation*}
    -h\int\rd^2\br\,n_x(\br)
\end{equation*}
and the FM phase where either of the two flavours is fully filled is favoured by a repulsion between the two flavours
\begin{equation*}
    \int\rd^2\br_1\,\rd^2\br_2\,U(|\br_{12}|)n_\uparrow(\br_1)n_\downarrow(\br_2),
\end{equation*}
where the density operators are defined as
\begin{equation*}
    n_x(\br)=\psi^\dagger_\downarrow(\br)\psi_\uparrow(\br)+\psi^\dagger_\uparrow(\br)\psi_\downarrow(\br),\quad n_\uparrow(\br)=\psi^\dagger_\uparrow(\br)\psi_\uparrow(\br),\quad n_\downarrow(\br)=\psi^\dagger_\downarrow(\br)\psi_\downarrow(\br),
\end{equation*}
and the potentials can be most conveniently taken as a combination of local and super-local interactions. Altogether, the model Hamiltonian reads
\begin{equation}
    H_\Int=\int\rd^2\br_1\,\rd^2\br_2\,U(|\br_{12}|)n_\uparrow(\br_1)n_\downarrow(\br_2)-h\int\rd^2\br\,n_x(\br).
    \label{eq:ising_hmt}
\end{equation}
By tuning the ratio between $h$ and $U(|\br_{12}|)$, a phase transition described by the Ising CFT is realised.

\subsection{Interaction in Terms of Pseudo-Potentials}
\label{sec:construct_pspot}

Another way that is computationally more convenient to construct the four-fermion interaction terms is through Haldane pseudo-potential~\cite{Haldane1983LLL,Trugman1985Pseudo}. To explain the idea, we take the 3D Ising model as an example. We first classify all the fermion bilinears $\lambda_{m_1f_1m_2f_2}c_{m_1f_1}c_{m_2f_2}$. To simplify the discussion, we can take a specific flavour index $\lambda_{m_1m_2}c_{m_1\uparrow}c_{m_2\downarrow}$. The fermion bilinears can be classified into irreducible representations (irreps) of $\SO(3)$ rotation symmetry. Since $c_{mf}$ carries the spin-$s$ representation, the spin of its bilinear ranges from $0$ to $2s$ and takes integer values. The spin-$(2s-l)$ combination reads
\begin{equation}
    \Delta_{lm}=\sum_{m_1}\langle sm_1,s(m-m_1)|(2s-l)m\rangle c_{m_1\uparrow}c_{m-m_1\downarrow},
    \label{eq:ising_pair}
\end{equation}
where $m=-(2s-l),\dots,(2s-l)$, and the Clebshbar-Gordan coefficients~\cite{Biedenharn1984Angular} is related to the $3j$-symbol by
\begin{equation}
    \langle l_1m_1,l_2m_2|lm\rangle=(-1)^{-l_1+l_2-m}\sqrt{2l+1}\begin{pmatrix}l_1&l_2&l\\m_1&m_2&-m\end{pmatrix}
\end{equation}
A four-fermion interaction term is formed by contracting these pairing operators with its conjugate
\begin{equation}
    H=\sum_lU_lH_l,\quad H_l=\sum_m\Delta_{lm}^\dagger\Delta_{lm}.
\end{equation}
Putting these together, the interaction Hamiltonian can be expressed as
\begin{equation}
    H=\sum_{l,m_1m_2m_3m_4}U_lC^l_{m_1m_2m_3m_4}c^\dagger_{m_1\uparrow}c^\dagger_{m_2\downarrow}c_{m_3\downarrow}c_{m_4\uparrow}-h\sum_m(c_{m\uparrow}^\dagger c_{m\downarrow}+\text{h.~c.}),
\end{equation}
where the matrix elements are
\begin{equation}
    C^l_{m_1m_2m_3m_4}=\delta_{m_1+m_2,m_3+m_4}\langle sm_1,sm_2|(2s-l)(m_1+m_2)\rangle\langle (2s-l)(m_3+m_4)|sm_3,sm_4\rangle.
    \label{eq:ps_pot_coeff}
\end{equation}
The coupling strengths $U_l$ of the spin-$(2s-l)$ channel are called the Haldane pseudo-potentials.

We also need to consider the constraint that the two fermions must be anti-symmetrised~: for even $l$, the orbital index is symmetrised, so the spin index must be anti-symmetrised, so the two fermions form a spin-singlet which is invariant under the $\SU(2)$ transformation~; for odd $l$, the orbital index is anti-symmetrised, so the spin index is symmetrised, breaking the flavour symmetry from $\SU(2)$ to $\BZ_2$. Hence, an odd-$l$ pseudo-potential must be added (This fact escapes the construction by density-density interaction).

The fermion bilinears with other flavour configurations $\lambda_{m_1m_2,\pm}(c_{m_1\uparrow}c_{m_2\uparrow}\pm c_{m_1\downarrow}c_{m_2\downarrow})$ can be analysed in a similar way. After that, we have enumerated all possible four-fermion interaction terms.

Each pseudo-potential corresponds to a profile of interaction potential functions. The conversion between the pseudo-potentials $U_l$ and the Legendre expansion coefficients of the potential function $\tilde{U}_l$
\begin{equation}
    U(|\br_{12}|)=\sum_l\tilde{U}_lP_l(\cos\theta_{12})
\end{equation}
is
\begin{equation}
    U_l=\sum_k \tilde{U}_k(-1)^l(2s+1)^2\begin{Bmatrix}2s-l&s&s\\k&s&s\end{Bmatrix}\begin{pmatrix}s&k&s\\-s&0&s\end{pmatrix}^2,
\end{equation}
where $\left\{\begin{smallmatrix}\bullet&\bullet&\bullet\\\bullet&\bullet&\bullet\end{smallmatrix}\right\}$ is the $6j$-symbol. Specifically, a local interaction $\delta(\br_{12})$ contains only pseudo-potential $U_0$~; a superlocal interaction of form $(\nabla^2)^l\delta(\br_{12})$ contains $U_0,U_1,\dots,U_l$. Here we explicitly give the expressions for the lowest pseudo-potentials
\begin{align}
    U(|\br_{12}|)&=\delta(\br_{12}),&U_0&=\frac{(2s+1)^2}{4s+1}\nonumber\\
    U(|\br_{12}|)&=\nabla^2\delta(\br_{12}),&U_0&=-\frac{s(2s+1)^2}{4s+1},&U_1&=\frac{s(2s+1)^2}{4s-1}.
\end{align}
More details are given in Ref.~\cite{Fan2024}. We also note that the construction for interaction Hamiltonian for LLL with pseudo-potentials is not restricted to sphere. \textit{E.~g.}, on a torus, the interaction can also be parametrise by pseudo-potentials~\cite{Yoshioka1983Torus,Haldane1985Torus,Haldane1987Torus}. More details are presented in Appendix~\ref{app:manifolds}.

For systems with more complicated continuous flavour symmetries, classification in terms of representation of flavour symmetry must also be considered, and the indices must be overall anti-symmetrised. We explain that through the example of a $2N$-flavour system with $\Sp(N)$ symmetry~\cite{Zhou2024Oct}. The maximal flavour symmetry is $\SU(2N)$, so interactions must be added to break the symmetry from $\SU(2N)$ to $\Sp(N)$. The fermion operators $c_{ma}$ live in the $\Sp(N)$ fundamental representation, where $a$ is the $\Sp(N)$ index. We shall show that all the allowed terms are
\begin{multline}
    H=\sum_{\substack{l\in\BZ\\m_1m_2m_3m_4}}U_lC^l_{m_1m_2m_3m_4}c^\dagger_{m_1a}c^\dagger_{m_2b}c_{m_3b}c_{m_4a}\\
    -\frac{1}{2}\sum_{\substack{l\in 2\BZ\\m_1m_2m_3m_4}}V_lC^l_{m_1m_2m_3m_4}\Omega_{aa'}\Omega_{bb'}c^\dagger_{m_1a}c^\dagger_{m_2a'}c_{m_3b'}c_{m_4b}.
\end{multline}
where $\Omega=\begin{pmatrix}0&\BI_N\\-\BI_N&0\end{pmatrix}$.

To find out all the four-fermion interactions allowed by the rotation symmetry $\SO(3)$ and flavour symmetry $\Sp(N)$, we classify all the fermion bilinears $c_{m_1a}c_{m_2b}$ into irreps of $\SO(3)\times\Sp(N)$. For each irrep, by contracting the bilinear with its Hermitian conjugate, we obtain an allowed four-fermion interaction term. Each fermion carries $\SO(3)$ spin-$s$ and $\Sp(N)$ fundamental. For the rotation symmetry $\SO(3)$, the bilinears can carry spin-$(2s-l)(l=0,\dots,2s)$ represetation~; for even $l$, the orbital indices are symmetrised~; for odd $l$, the orbital indices are anti-symmetrised. For the flavour symmetry $\Sp(N)$, the bilinears can carry singlet $S$, traceless anti-symmetric rank-2 tensor $A$ and symmetric rank-2 tensor $T$ representation~; for $S$ and $A$, the flavour indices are anti-symmetrised~; for $T$, the flavour indices are symmetrised. As the two fermions altogether should be anti-symmetrised, the allowed combinations are

\begin{enumerate}
    \item $\Sp(N)$ singlet and $\SO(3)$ spin-$(2s-l)$ with even $l$, the bilinears are
    \begin{equation}
        \Delta_{lm}=\sum_{m_1m_2}\langle sm_1,sm_2|(2s-l)m\rangle\Omega_{cc'}c_{m_1c}c_{m_2c'}\delta_{m,m_1+m_2}.
    \end{equation}
    The corresponding interaction term $H_{S,l}=\sum_m\Delta_{lm}^\dagger\Delta_{lm}$ is the even-$l$ pseudo-potential for the $V$-term.
    \item $\Sp(N)$ anti-symmetric and $\SO(3)$ spin-$(2s-l)$ with even $l$, the bilinears are
    \begin{equation}
        \Delta_{lm,[ab]}=\sum_{m_1m_2}\langle sm_1,sm_2|(2s-l)m\rangle(c_{m_1a}c_{m_2b}-c_{m_1b}c_{m_2a}-\tfrac{1}{N}\Omega_{ab}\Omega_{cc'}c_{m_1c'}c_{m_2c})\delta_{m,m_1+m_2}.
    \end{equation}
    The corresponding interaction term $H_{A,l}=\sum_m\Delta_{lm,[ab]}^\dagger\Delta_{lm,[ab]}$ is the even-$l$ pseudo-potential for the $U$-term.
    \item $\Sp(N)$ symmetric and $\SO(3)$ spin-$(2s-l)$ with odd $l$, the bilinears are
    \begin{equation}
        \Delta_{lm,(ab)}=\sum_{m_1m_2}\langle sm_1,sm_2|(2s-l)m\rangle(c_{m_1a}c_{m_2b}+c_{m_1b}c_{m_2a})\delta_{m,m_1+m_2}.
    \end{equation}
    The corresponding interaction term $H_{T,l}=\sum_m\Delta_{lm,(ab)}^\dagger\Delta_{lm,(ab)}$ is the odd-$l$ pseudo-potential for the $U$-term.
\end{enumerate}

In summary, by classifying all the bilinears, we show that all allowed interactions are the $U_l$ terms with both even and odd $l$, and the $V_l$ terms with only even $l$.

\subsection{Operator Spectrum and Search for Conformal Point}
\label{sec:construct_spec}

Having introduced the construction of an interacting model on the fuzzy sphere, we now turn to the verification of the conformal symmetry and the extraction of the CFT data. The most straightforward approach is to extract the scaling dimensions from the energy spectrum through the state-operator correspondence. Specifically, there is a one-to-one correspondence between the eigenstates of the Hamiltonian and the CFT operators. The state and its corresponding operator has the same $\SO(3)$ spin and representation under flavour symmetry, and the excitation energy of a state $|\Phi\rangle$ is proportional to the scaling dimension of the corresponding operator $\Delta_\Phi$
\begin{equation}
    E_\Phi-E_0=\frac{v}{R}\Delta_\Phi,
\end{equation}
where $E_0$ is the ground state energy, $R$ is the radius of the sphere (here we take $R=\sqrt{N_m}$), and $v$ is a model-dependent speed of light. The constant $v/R$ can be determined through a calibration process, \textit{i.~e.}~comparing the spectrum to some known properties of a CFT spectrum. The criteria to determine the conformal symmetry include

\begin{enumerate}
    \item The existence of a conserved stress tensor $T^{\mu\nu}$. The stress tensor is the symmetry current of the translation transformation. It is known to be a singlet under the flavour symmetry, have spin-2 under $\SO(3)$ rotation and scaling dimension exactly $\Delta_{T^{\mu\nu}}=3$.
    \item The existence of a conserved flavour symmetry current $J^\mu$ if there is a continuous flavour symmetry. The symmetry current typically lives in the anti-symmetric rank-2 tensor representation of the flavour symmetry. \textit{E.~g.}, if the flavour symmetry is $\rU(1)$, then the symmetry current has charge-0~; if the flavour symmetry is $\rO(3)$, then the symmetry current has spin-1 and is odd under the improper $\BZ_2$ transformation~; if the flavour symmetry is $\rO(n)$ ($n\ge 4$) or $\SU(n)$ ($n\ge 3$), then the symmetry current lives in the anti-symmetric rank-$2$ tensor representation.
    \item The organisation of the operator spectrum into conformal multiplets. All the levels in the spectrum of a CFT can be organised into the conformal primaries and their descendants. The descendants live in the same representation under the flavour symmetry as the primary, and the difference between the scaling dimensions of a primary and its descendant is an integer. Specifically, for a scalar primary $\Phi$, its descendants have the form\footnote{Hereafter, we will presume the subtraction of trace and omit the terms.}
    \begin{equation*}
        \Box^n\partial^{\mu_1}\partial^{\mu_2}\dots\partial^{\mu_l}\Phi-\textrm{(trace)}\qquad (n,l=0,1,2,\dots)
    \end{equation*}
    with $\SO(3)$ spin-$l$ and scaling dimension 
    \begin{equation*}
        \Delta=\Delta_\Phi+2n+l,
    \end{equation*}
    where $\Box=\partial_\mu\partial^\mu$. For a spinning primary $\Phi^{\mu_1\dots\mu_s}$, its descendants have two forms~:
    \begin{equation*}
        \Box^n\partial^{\nu_1}\dots\partial^{\nu_m}\partial_{\rho_1}\dots\partial_{\rho_k}\Phi^{\rho_1\dots\rho_{k}\mu_1\dots \mu_{s-k}}\qquad (k=0,\dots,s,\quad n,m=0,1,\dots)
    \end{equation*}
    with scaling dimension and $\SO(3)$ spin
    \begin{equation*}
        \Delta=\Delta_\Phi+k+m+2n,\qquad l=s-k+m,
    \end{equation*}
    and
    \begin{equation*}
        \Box^n\partial^{\nu_1}\dots\partial^{\nu_m}\partial_{\rho_1}\dots\partial_{\rho_k}\epsilon^{\sigma}{}_{\tilde{\mu}\tilde{\nu}}\partial^{\tilde{\nu}}\Phi^{\rho_1\dots\rho_{k}\tilde{\mu}\mu_1\dots \mu_{s-k-1}}\qquad (k=0,\dots,s-1,\quad n,m=0,1,\dots)
    \end{equation*}
    with
    \begin{equation*}
        \Delta=\Delta_\Phi+k+m+2n+1,\qquad l=s-k+m.
    \end{equation*}
    For the second form, the fully anti-symmetric tensor $\epsilon$ alters the parity. For conserved currents only $m=0$ descendants exist due to the conservation $\partial_{\mu_1}\Phi^{\mu_1\dots\mu_l}=0$. 
\end{enumerate}

The most convenient way of determining the coefficient $v/R$ is by utilising criteria 1 or 2~:
\begin{equation}
    \frac{v}{R}=\frac{E_{T^{\mu\nu}}-E_0}{3}\quad\textrm{or}\quad\frac{E_{J^\mu}-E_0}{2}.
\end{equation}
Alternatively, one can define a cost function that depends on the tuning parameter and the speed of light and compares the scaling dimensions obtained from the fuzzy sphere and the prediction by conformal symmetry. \textit{E.~g.}, for the Ising CFT, the tuning parameters are the pseudo-potentials $\{U_i\}$ and the transverse field $h$. The criteria for conformal symmetry we use include the stress tensor $T^{\mu\nu}$ and the descendants $\partial^\mu\sigma$, $\partial^\mu\partial^\nu\sigma$, $\Box\sigma$, $\partial^\mu\epsilon$. The cost function is the root-mean-square of the deviations of these criteria from the expectation of the conformal symmetry
\begin{multline}
    Q^2(\{U_i\},h,v;N_m)=\frac{1}{N_s}\left[(\Delta_{T^{\mu\nu}}^\FS-3)^2+(\Delta_{\partial^\mu\sigma}^\FS-\Delta_\sigma^\FS-1)^2\right.\\
    \left.+(\Delta_{\partial^\mu\partial^\nu\sigma}^\FS-\Delta_\sigma^\FS-1)^2+(\Delta_{\Box\sigma}^\FS-\Delta_\sigma^\FS-1)^2+(\Delta_{\partial^\mu\epsilon}^\FS-\Delta_\epsilon^\FS-1)^2\right]
\end{multline}
where $N_s=5$ is the number of criteria, the scaling dimension of an operator $\Phi$ on the fuzzy sphere is determined as
\begin{equation}
    \Delta_\Phi^\FS(\{U_i\},h,v;N_m)=\frac{E_\Phi-E_0}{v/R}.
\end{equation}
The optimal conformal point and calibrator are determined by minimising this cost function for each system size $N_m$. Note that this optimal point depends on the system size. In order to do finite-size scaling, if the CFT describes a phase transition, one could fix all but one parameters at the optimal point in the largest accessible system size and tune the last parameter to determine the critical point through a finite-size scaling.

\subsection{Local Observables}
\label{sec:construct_obs}

\newcommand{\Flat}{\text{(flat)}}
\newcommand{\cyl}{\text{(cyl.)}}

We have introduced how to determine the scaling dimensions from the energy spectrum. Beyond that, evaluating other CFT quantities requires realising local CFT operators on the fuzzy sphere. Any gapless local observables $\cO(\br)$ on the fuzzy sphere can be written as the linear combination of CFT operators that live in the same representation of flavour symmetry and parity\footnote{The realisation of CFT operators in the microscopic model has been investigated in more detail in 2D in Ref.~\cite{Zou2019Operator2d}.}~\cite{Hu2023Mar}
\begin{equation}
    \cO(\br,\tau)=\sum_\alpha \lambda_\alpha\Phi^\cyl_\alpha(\br,\tau).
\end{equation}
Here special care should be taken for the CFT operator $\Phi^\cyl_\alpha(\br,\tau)$ on the cylinder. A conformal transformation produces a scale factor $\Lambda(\br)^{\Delta_\Phi}$ to a primary operator $\Phi$. The scale factor is $\Lambda(\br)=r/R$ for the Weyl transformation from the flat space-time to the cylinder. Hence,\footnote{We need to clarify some of the notations~: $\cO$ represents an operator in the microscopic model, and $\Phi$ represents a CFT operator~; the arguments $\Phi(\br)$ or $\Phi(\br,\tau)$ by default mean the operator is defined on a cylinder, and $\Phi(x)$ by default on flat space-time.}
\begin{equation}
    \Phi^\cyl_\alpha(\br,\tau)=\left(\frac{e^{\tau/R}}{R}\right)^{\Delta_{\Phi_\alpha}}\Phi_\alpha^\Flat(x).
\end{equation}
For descendants, certain other factors may be produced, but the conversion factors still scale with the radius of the sphere as $R^{-\Delta}$ where $\Delta$ is the scaling dimension of the descendants. For simplicity, hereafter we focus on the equal-time correlators with $\tau=0$, for which $\Phi^\cyl_\alpha(\br)=R^{-\Delta_{\Phi_\alpha}}\Phi_\alpha^\Flat(x)$. The operator with larger system size decays faster when increasing system size.

The simplest local observable is the density operator defined in Eq.~\eqref{eq:den_def}, and its decomposition into angular modes is given in Eqs.~\eqref{eq:den_decomp} and \eqref{eq:den_mod}. From the CFT perspective, the density operators are the superpositions of scaling operators with corresponding quantum numbers, \textit{i.~e.}~with the same representation under flavour symmetry and parity.

Take the Ising model as an example. Consider the density operators $n^x$ and $n^z$ with matrix insertion $M=\sigma^x,\sigma^z$. In the leading order, they can be used as UV realisations of CFT operators $\sigma$ and $\epsilon$.
\begin{align}
    n^x(\br)&=\lambda_0+\lambda_\epsilon\epsilon(\br)+\lambda_{\partial^\mu\epsilon}\partial^\mu\epsilon(\br)+\lambda_{T^{\mu\nu}}T^{\mu\nu}(\br)+\dots&\epsilon_\textrm{FS}&=\frac{n^x-\lambda_0}{\lambda_\epsilon}+\dots\nonumber\\
    n^z(\br)&=\lambda_\sigma\sigma(\br)+\lambda_{\partial^\mu\sigma}\partial^\mu\epsilon(\br)+\lambda_{\partial^\mu\partial^\nu\sigma}\partial^\mu\partial^\nu\sigma(\br)+\dots&\sigma_\textrm{FS}&=\frac{n^z}{\lambda_\sigma}+\dots
\end{align}
where the coefficients $\lambda_0,\lambda_\epsilon,\lambda_\sigma,\dots$ are model-dependent and need to be determined, and all the operators on the right-hand side are defined on the cylinder.

We first consider the insertion of a single operator $\langle\Phi_1|\Phi_2(\br)|\Phi_3\rangle$. It helps us produce the OPE coefficients. For the simplest example of three scalars, \begin{equation} f_{\Phi_1\Phi_2\Phi_3}=\lim_{r_\infty\to\infty}r_\infty^{-2\Delta_{\Phi_1}}\langle \Phi_1(x_\infty)\Phi_2(x)\Phi_3(0)\rangle_\Flat=\langle\Phi_1|\Phi_2^\Flat(x)|\Phi_3\rangle \end{equation} where $x_\infty$ is a point on the sphere with radius $r_\infty$, $x$ is a point on the unit sphere, the states are obtained from acting the operator at the origin point on the vacuum state
\begin{equation}
    |\Phi_3\rangle=\Phi_3(0)|0\rangle
\end{equation}
and its Hermitian conjugate is defined as
\begin{equation}
    \Phi_1^\dagger(\infty)=(\Phi_1(0))^\dagger=\lim_{r_\infty\to\infty}r_\infty^{2\Delta_{\Phi_1}}\Phi_1(x_\infty),\qquad\langle\Phi_1|=\langle0|\Phi_1^\dagger(\infty).
\end{equation}
After the Weyl transformation from the flat space-time to the cylinder, we obtain the expression on the fuzzy sphere
\begin{equation}
    f_{\Phi_1\Phi_2\Phi_3}=R^{\Delta_{\Phi_2}}\langle\Phi_1|\Phi^{(\cyl)}_2(\br)|\Phi_3\rangle.
\end{equation}
The UV realisation of $\Phi_2$ contains many other operators with different spins. By integrating the correlation function against different spherical harmonics, \textit{i.~e.}~take the angular modes of the operator inserted
\begin{equation}
    \int\rd^2\br\,\Yb_{lm}(\bnh)\langle\Phi_1|\Phi_2(\br)|\Phi_3\rangle=\langle\Phi_1|\Phi_{2,lm}|\Phi_3\rangle,
\end{equation}
we can filter out the sub-leading contributions with different spins. For the spinning operators, this also tells us about different OPE structures. By taking $\Phi_3=\BI$, we can recover the 2-pt functions
\begin{align}
    \langle\Phi_2|\Phi_{2,00}|0\rangle&=R^{-\Phi_2}\nonumber\\
    \Phi_2(\br)|0\rangle&=R^{-\Phi_2}\left[|\Phi_2\rangle+\lambda'_\mu(\br)|\partial^\mu\Phi_2\rangle+\lambda''(\br)|\Box\Phi_2\rangle+\lambda''_{\mu\nu}(\br)|\partial^\mu\partial^\nu\Phi_2\rangle\right].
\end{align}
It is worth noting that acting a primary $\Phi_2(\br)$ on the vacuum also produces various descendants in the multiplet.

In the example of Ising CFT, we first use the insertion of a single operator to determine the coefficients $\lambda_0,\lambda_\epsilon,\lambda_\sigma$
\begin{equation}
    \lambda_0=\frac{1}{\sqrt{4\pi}}\langle 0|n^x_{00}|0\rangle,\quad\lambda_\epsilon=\frac{R^{\Delta_\epsilon}}{\sqrt{4\pi}}\langle \epsilon|n^x_{00}|0\rangle,\quad\lambda_\sigma=\frac{R^{\Delta_\sigma}}{\sqrt{4\pi}}\langle \sigma|n^z_{00}|0\rangle.
\end{equation}
Take the OPE coefficient $f_{\sigma\sigma\epsilon}$ as an example. It can be expressed either as an inner product of $\sigma$ or $\epsilon$
\begin{align}
    f_{\sigma\sigma\epsilon}&=R^{\Delta_\sigma}\langle\epsilon|\sigma(\br)|\sigma\rangle=\frac{\langle\epsilon|n_{00}^z|\sigma\rangle}{\langle 0|n_{00}^z|\sigma\rangle}+\cO(R^{-2})\nonumber\\
    &=R^{\Delta_\epsilon}\langle\sigma|\epsilon(\br)|\sigma\rangle=\frac{\langle\sigma|n^x_{00}|\sigma\rangle-\langle0|n^x_{00}|0\rangle}{\langle\epsilon|n^x_{00}|0\rangle}+\cO(R^{-(3-\Delta_\epsilon)}).
\end{align}
For the first line, the sub-leading contribution comes from the contribution of the descendant $\Box\sigma$ to $n_{00}^z$. As $\sigma(\br)$ scales as $R^{-\Delta_\sigma}$ and $\Box\sigma(\br)$ as $R^{-\Delta_\sigma-2}$,
\begin{align}
    \langle\epsilon|n_{00}^z|\sigma\rangle&=f_{\sigma\sigma\epsilon}\lambda_\sigma R^{-\Delta_\sigma}(1+c_1R^{-2}+\dots)\nonumber\\
    \langle\epsilon|n_{00}^z|\sigma\rangle&=\lambda_\sigma R^{-\Delta_\sigma}(1+c'_1R^{-2}+\dots)\nonumber\\
    \frac{\langle\epsilon|n_{00}^z|\sigma\rangle}{\langle 0|n_{00}^z|\sigma\rangle}&=f_{\sigma\sigma\epsilon}+\cO(R^{-2}).
\end{align}
Here $c_1$ and $c'_1$ are constant factors that represent the contribution of $\Box\sigma$ and do not scale with system size. Hence, the sub-leading contribution scales as $R^{-2}$. For the second line, the sub-leading contribution comes from the stress tensor $T^{\mu\nu}$. Similarly, the power of the scaling is the difference of the scaling dimension $R^{-(\Delta_{T^{\mu\nu}}-\Delta_\epsilon)}=R^{-(3-\Delta_\epsilon)}$.

We then proceed to the insertion of two operators. This can help us determine up to a 4-pt function~\cite{Han2023Jun}. Through conformal transformation, any 4-pt function can be expressed in the form of
\begin{equation}
    \langle\Phi_1|\Phi^{\textrm{(cyl.)}}_2(\br,\tau)\Phi^{\textrm{(cyl.)}}_3(\bzh)|\Phi_4\rangle=\frac{e^{\Delta_{\Phi_2}\tau/R}}{R^{\Delta_{\Phi_2}+\Delta_{\Phi_3}}}\langle\Phi_1^\dagger(\infty)\Phi_2(x)\Phi_3(\bzh)\Phi_4(0)\rangle,
\end{equation}
where the time-displaced operator can be defined as
\begin{equation}
    \Phi_2(\br,\tau)=e^{-H\tau}\Phi_2(\br)e^{H\tau}.
\end{equation}
As a sanity check, By taking $\Phi_1=\Phi_4=\BI$, $\Phi_2=\Phi_3$ and $\tau=0$, the 2-pt function on the unit sphere is recovered
\begin{multline}
    \langle0|\Phi^{\textrm{(cyl.)}}_2(\br)\Phi^{\textrm{(cyl.)}}_2(\bzh)|0\rangle=R^{-2\Delta_{\Phi_2}}\langle\Phi_2(\br)\Phi_2(\bzh)\rangle\\=\frac{1}{R^{2\Delta_{\Phi_2}}|\br-\bzh|^{2\Delta_{\Phi_2}}}=\frac{1}{R^{2\Delta_{\Phi_2}}(1-\cos\theta)^{\Delta_{\Phi_2}}}.
\end{multline}

\subsection{Conformal Generators}
\label{sec:construct_gen}

So far, in the conformal group, we know that the rotation and the dilatation are manifest on the fuzzy sphere. The rest, \textit{viz.}~translation and SCT, are emergent. In this section, we consider how to express the generators of these emergent symmetries in terms of the microscopic operators~\cite{Fardelli2024,Fan2024}.

A general Noether current and corresponding generator of the infinitesimal space-time transformation $x^\mu\mapsto x^\mu+\epsilon^\mu(x)$ can be expressed in terms of the stress tensor
\begin{equation}
    j_\epsilon^\mu(x)=\epsilon^\nu(x)T^\mu{}_\nu(x),\quad Q_\epsilon=\int_\Sigma\rd^{d-1}x\,\sqrt{g}j_\epsilon^0(x),
\end{equation}
where for the second equation, the integral is evaluated on a closed surface $\Sigma$. Specifically, for the generators $P^\mu,K^\mu$ of translation and SCT in the embedded sphere
\begin{align}
    P^\mu&=\int\rd^2\br\,(r^\mu T^0{}_0+iT^{0\mu}),\nonumber\\ K^\mu&=\int\rd^2\br\,(r^\mu T^0{}_0-iT^{0\mu}).
\end{align}
Hence, the conformal generator $\Lambda^\mu=P^\mu+K^\mu$ is the $l=1$ component of the Hamiltonian density $\cH=T^0{}_0$\footnote{Here the indices $\mu$ and $m$ are two equivalent way to express the components}
\begin{equation}
    \Lambda_m=P_m+K_m=\sqrt\frac{16\pi}{3}\int\rd^2\br\,\Yb_{1m}(\bnh)\cH(\br).
\end{equation}
By acting it on the states, the number of derivatives is increased or decreased by $1$, \textit{e.~g.}, for a primary $\Phi$
\begin{align}
    \Lambda^\mu|\Phi\rangle&=\textrm{const.}\times|\partial^\mu\Phi\rangle\nonumber\\     \Lambda^\mu|\partial_\mu\Phi\rangle&=\textrm{const.}\times|\Phi\rangle+\textrm{const.}\times|\partial^\mu\partial^\nu\Phi\rangle+\textrm{const.}\times|\Box\Phi\rangle.
\end{align}
The derivation of the expression and the constant factors are calculated and given in Refs.~\cite{Fardelli2024,Fan2024}.

We then need to find the expression for the Hamiltonian density. \textit{E.~g.}, for the Ising model, it is the local density operator and density-density interactions with some full derivatives
\begin{equation}
    \cH(\br)=n_z\left(g_0+g_1\nabla^2\right)n_z-hn_x+g_{D,1}\nabla^2n_x+g_{D,2}\nabla^2n_z^2+\dots,
\end{equation}
where $g_{D,i}$ are undetermined constants that does not affect the Hamiltonian $H=\int\rd^2\br\,\cH$. We have only listed a few examples of the allowed full derivatives.

To determine those constants, we consider another strategy by considering all the possible two-fermion and four-fermion operators that are singlet under flavour symmetry and spin-1 under $\SO(3)$. We consider the example of Ising CFT. The two-fermion terms include the density operators
\begin{equation*}
    n^x_{1m}\quad\textrm{and}\quad n^0_{1m}.
\end{equation*}
Similar to what we have done for Hamiltonian, the four-fermion operators can be obtained by combining the fermion bilinears $\Delta_{lm}$ defined in Eq.~\eqref{eq:ising_pair}
\begin{equation}
    \sum_{\substack{l_1l_2m_1m_2}}\tilde{U}_{l_1l_2}\Delta^\dagger_{l_1m_1}\Delta_{l_2m_2}\langle (2s-l_1)m_1,(2s-l_2)(-m_2)|1m\rangle
\end{equation}
For $l_1\in2\BZ$, the spin index in the pairing operator is anti-symmetrised~; For $l_1\in2\BZ+1$, the spin index in the pairing operator is symmetrised. Therefore, $l_1-l_2\in2\BZ$ for non-zero results. And since $|l_1-l_2|\leq 1$, we conclude $l_1=l_2$. so

\begin{equation}
    \Lambda_m=\sum_{\substack{lm_1m_2}}\tilde{U}_{l}\Delta^\dagger_{lm_1}\Delta_{lm_2}\begin{pmatrix}2s-l&2s-l&1\\-m_1&m_2&m     \end{pmatrix}+\tilde{h}n^x_{1m}+\tilde{\mu}n^0_{1m}
\end{equation}
Here, $\tilde{U}_l,\tilde{h},\tilde{\mu}$ are tuning parameters.

After obtaining $\Lambda^\mu=P^\mu+K^\mu$, the separate $P^\mu$ and $K^\mu$ can be obtained by considering the commutator with the dilatation generator $D$, which is proportional to the Hamiltonian. As $[D,P^\mu]=P^\mu$ and $[D,K^\mu]=-K^\mu$,
\begin{align}
    P^\mu&=\tfrac{1}{2}\Lambda^\mu+\tfrac{1}{2}[D,\Lambda^\mu]\nonumber\\
    K^\mu&=\tfrac{1}{2}\Lambda^\mu-\tfrac{1}{2}[D,\Lambda^\mu].
\end{align}

\section{Numerical Methods}
\label{sec:numerics}

In this section, we briefly review the numerical methods supported in FuzzifiED. The numerical methods that have been applied to the fuzzy sphere include exact diagonalisation (ED), density matrix renormalisation group (DMRG) and determinant quantum Monte Carlo (DQMC). Among these, ED and DMRG have been implemented in FuzzifiED.

\subsection{Exact Diagonalisation (ED)}

Exact diagonalisation (ED) might be the most straightforward method for solving a quantum many-body Hamiltonian. In ED, one constructs a many-body basis and writes down all the non-zero elements of the Hamiltonian as a sparse matrix on this basis. The eigenstates of the Hamiltonian with the lowest energy can be solved without finding the full eigensystem by Arnoldi or Lanczos algorithm.

Briefly speaking, the Arnoldi algorithm~\cite{Arnoldi1951} is an iterative method. Each iteration constructs an orthonormal basis of the Krylov subspace from an initial vector and finds an approximation to the eigenvector on that basis. This approximate eigenvector is then used as the initial vector for the next iteration. An example of Krylov subspace is spanned by acting the matrix $H$ repeatedly on the initial vector $|i\rangle$
\begin{equation}
    \mathcal{K}_r(H,|i\rangle)=\operatorname{span}\left\{|i\rangle,H|i\rangle,H^2|i\rangle,\dots,H^{r-1}|i\rangle\right\}.
\end{equation}

The ED calculation can be optimised in several ways. The storage of the Hamiltonian matrix may be compressed by data structure tailored for sparse matrix such as compressed sparse column (CSC). The Hamiltonian matrix is usually block diagonal due to the symmetry of the Hamiltonian. The Hilbert space is divided into several sectors that carry different representations under the symmetry, and acting the Hamiltonian on a state in a sector results in a state in the same sector. \textit{E.~g.}, in the ED calculation for the Ising model on the fuzzy sphere, the symmetries we can use include two $\rU(1)$ symmetries, \textit{viz.}~the conservation of particle number and the angular momentum in the $z$-direction, and three $\BZ_2$ symmetries, \textit{viz.}~the Ising $\BZ_2$ flavour symmetry, the particle-hole symmetry and the $\pi$-rotation along the $y$-axis\footnote{So far, FuzzifiED only supports $\rU(1)$ and $\BZ_p$ symmetries.}

The ED method enjoys several advantages, including (1) the full knowledge of the eigenstate wave-function and (2) the ability to access relatively high excited states. However, despite these optimisations, the dimension of the Hilbert space scales exponentially with the number of orbitals. This results in exponentially growing space and time complexity. \textit{E.~g.}, for the Ising model on the fuzzy sphere, for $N_m=14$, the dimension of Hilbert space $\dim\cH=1.8\times10^5$ and the number of elements in the Hamiltonian is $N_
\el=1.1\times 10^7$~; for $N_m=16$, the numbers have already grown to $\dim\cH=2.2\times10^6$ and $N_\el=2.1\times 10^8$, which translates to a memory demand of $3.1$ gigabytes.

In FuzzifiED, we use the Fortran library Arpack~\cite{Arpack1998} to perform the Arnoldi algorithm.

\subsection{Density Matrix Renormalisation Group (DMRG)}

To overcome the size limit of ED, the density matrix renormalisation group (DMRG) is a powerful method to calculate the ground state of a quasi-one-dimensional system. It was first invented by White~\cite{White1992DMRG} as an improvement to the numerical renormalisation group (NRG) used in the Kondo problem. Since its proposal, it has been proven potent in various problems in condensed-matter physics, such as the static and dynamic properties of one-dimensional models such as the Heisenberg, $t$--$J$ and Hubbard models~\cite{Schollwock2005DMRG}. Later, Schollw\"ock has discovered a new point of view that implements the DMRG in the language of matrix product states (MPS)~\cite{Schollwoeck2010DMRG}.

Briefly speaking, in this language, DMRG is a variational method that optimises the fidelity between the exact ground state and the variational MPS. During each ``sweep,'' DMRG solves a local eigenvalue problem for the active tensors to improve the approximation. To find the excited states, one needs to add projection $|0\rangle\langle 0|$ of the ground state $|0\rangle$ to the Hamiltonian by hand.

Although the fuzzy sphere deals with $(2+1)$-dimensional quantum systems, the basis of the lowest Landau level provides a natural way to express it as a quasi-1D problem. Therefore, DMRG has been a powerful numerical method for the fuzzy sphere. However, like other $(2+1)$D models, the DMRG on the fuzzy sphere also suffers from the divergence of the required maximal bond dimension with system size. One should thus be careful with the convergence of the results when doing DMRG.

In FuzzifiED, we use the ITensor library~\cite{ITensor} in Julia to perform the DMRG calculations.

\cleardoublepage
\part{Numerical Calculation with FuzzifiED}
\label{pt:numerics}

\section{Installation and Usage}
\label{sec:usage}

The package FuzzifiED is implemented in the Julia language~\cite{Julia}.\footnote{The most intensive core functionalities are written in Fortran and wrapped by Julia interfaces, but the users are not required to have any interaction with or knowledge of the Fortran part of the source code.} Some useful links are given in Table~\ref{tbl:link}. To install the package, enter \lstinline|julia| to enter the Julia REPL (read-eval-print loop), and then run the following command.
\begin{lstlisting}
    using Pkg ; Pkg.add("FuzzifiED")
\end{lstlisting}
To use the package, include at the head of the Julia script.
\begin{lstlisting}
    using FuzzifiED
\end{lstlisting}
To obtain the documentation for an interface, type ``\lstinline|?|'' followed by the keyword in the Julia REPL, \textit{e.~g.}~``\lstinline|?Confs|.''

\begin{table}[htbp]
    \centering
    \begin{tabular}{l|l}
        \hline\hline
        Installation for Julia&\url{https://julialang.org/downloads}\\
        Homepage&\url{https://www.fuzzified.world}\\
        Documentation&\url{https://docs.fuzzified.world}\\
        Julia source code&\url{https://github.com/FuzzifiED/FuzzifiED.jl}\\
        JLL wrapper&\url{https://github.com/FuzzifiED/FuzzifiED_jll.jl}\\
        Fortran source code&\url{https://github.com/FuzzifiED/FuzzifiED_Fortran}\\
        Registry of the package&\url{https://juliahub.com/ui/Packages/General/FuzzifiED}\\
        \hline\hline
    \end{tabular}
    \caption{Some useful links.}
    \label{tbl:link}
\end{table}

\section{Exact Diagonalisation}
\label{sec:ed}

In this section, we briefly describe the procedure for exact diagonalisation (ED) calculation and give an instruction for using FuzzifiED for ED.

Practically, the ED calculation can be divided into four steps, which will be described in detail in the following sections~:
\begin{enumerate}
    \item To construct a many-body basis that respects a given set of quantum numbers (Sections~\ref{sec:ed_confs} and \ref{sec:ed_basis}). Specifically, in FuzzifiED we support quantum numbers of commuting $\rU(1)$ or discrete $\BZ_p$ symmetries.
    \item To construct the sparse matrix corresponding to the Hamiltonian on the basis above~(Sections~\ref{sec:ed_term} and \ref{sec:ed_opmat}).
    \item To find the lowest eigenstates of the sparse matrix and their corresponding eigenenergies~(Section~\ref{sec:ed_diag}).
    \item To make measurements on the eigenstates~(Sections~\ref{sec:ed_inner}, \ref{sec:ed_obs} and \ref{sec:ed_ent}). This includes the total angular momentum, density operators, entanglement, \textit{etc.}
\end{enumerate}

To demonstrate the usage of FuzzifiED interfaces, in this section, we use an example that calculates the eigenstates for the Ising model on the fuzzy sphere. Specifically, it
\begin{enumerate}
    \item calculates the lowest eigenstates in the symmetry sector $L^z=0$ and $(\cP,\cZ,\cR)=(+,+,+)$,
    \item measures their total angular momenta, and
    \item calcultes the OPE coefficient $f_{\sigma\sigma\epsilon}=\langle \sigma|n^z_{00}|\epsilon\rangle/\langle \sigma|n^z_{00}|0\rangle$.
\end{enumerate}
The full code is collected in Appendices~\ref{app:code_ed1} and \ref{app:code_ed2}.

\subsection{Set-up}
\label{sec:ed_setup}

Before starting the calculation, we need to input the set-up for the system, including the number of flavours $N_f$, orbitals $N_m$ and sites $N_o$. Here we specify some of our notations.
\begin{itemize}
    \item A ``flavour'' is labelled by $f$. The number of flavours is $N_f$.
    \item An ``orbital'' is specified by the magnetic quantum number labelled by $m$. The number of orbitals is $N_m=2s+1$.
    \item A ``site'' is specified by both the flavour and the orbital index $o=(f,m)$. The number of sites is $N_o=N_mN_f$. In practice, we label the sites with an integer from $1$ to $N_o$. We store the sites in an ascending order of first $m$ and then $f$~: $o=(m+s)N_f+f$.
\end{itemize}

In the example of the Ising model with $s=5.5$,
\begin{lstlisting}
    nm = 12
    nf = 2
    no = nm * nf
\end{lstlisting}

FuzzifiED also provides three environment parameters~:
\begin{itemize}
    \item \href{https://docs.fuzzified.world/core/\#FuzzifiED.ElementType}{\lstinline|FuzzifiED.ElementType|} --- the type of the matrix elements by default, either \lstinline|Float64| or \lstinline|ComplexF64|.
    \item \href{https://docs.fuzzified.world/core/\#FuzzifiED.NumThreads}{\lstinline|FuzzifiED.NumThreads|} --- an integer to define how many threads OpenMP uses by default.
    \item \href{https://docs.fuzzified.world/core/\#FuzzifiED.SilentStd}{\lstinline|FuzzifiED.SilentStd|} --- a bool that determines whether logs of the FuzzifiED should be turned off by default.
\end{itemize}

\subsection{Constructing the Configurations}
\label{sec:ed_confs}

The first step for the ED calculation is to construct the basis that respects the symmetries of the Hamiltonian. This is divided into two procedures~: (1) to generate the ``configurations'' that carry the diagonal quantum numbers, and (2) to generate the ``basis'' that also carries the off-diagonal quantum numbers (under discrete transformations). The ``\emph{configurations}'' are the collection of states that can be written as direct products of occupied $|1\rangle$ or empty $|0\rangle$ on each site and carry certain diagonal quantum numbers (QNDiag).

The QNDiags supported by FuzzifiED are the charges of $\rU(1)$ or $\BZ_p$ symmetry in the form of
\begin{align}
    Q&=\sum_oq_on_o,&\rU(1)&\textrm{ symmetry}\nonumber\\
    Q&=\sum_oq_on_o\mod p,&\BZ_p&\textrm{ symmetry},
\end{align}
where $n_o=c^\dagger_oc_o$ is the particle number on each site, and $q_o$ is the charge that each site carries. FuzzifiED restricts $q_o$ to be integer-valued. In FuzzifiED, the QNDiags are recorded in the mutable type \lstinline|QNDiag|.

\begin{block}{\href{https://docs.fuzzified.world/core/\#FuzzifiED.QNDiag}{\lstinline|QNDiag|} --- Type}
The type contains the fields
\begin{itemize}
    \item \lstinline|name :: String| --- the name of the diagonal quantum number.\footnote{The name is only needed for conversion into quantum numbers in ITensor.}
    \item \lstinline|charge :: Vector{Int64}| --- the symmetry charge $q_o$ of each site.
    \item \lstinline|modul :: Vector{Int64}| --- the modulus $p$, set to $1$ for $\rU(1)$ QNDiags.
\end{itemize}
and can be initialised by the method
\begin{lstlisting}
    QNDiag([name :: String, ]charge :: Vector{Int64}[, modul :: Int64])
\end{lstlisting}
where the arguments in the brackets are facultative.
\end{block}

Several useful QNDiags are built-in\footnote{For a more detailed description of the interfaces, refer to the documentation at \url{https://docs.fuzzified.world}.}
\begin{itemize}
    \item \href{https://docs.fuzzified.world/models/\#FuzzifiED.GetNeQNDiag-Tuple{Int64}}{\lstinline|GetNeQNDiag(no)|} --- the number of electrons ;
    \item \href{https://docs.fuzzified.world/models/\#FuzzifiED.GetLz2QNDiag-Tuple{Int64,\%20Int64}}{\lstinline|GetLz2QNDiag(nm, nf)|} --- twice the angular momentum $2L_z$ ;
    \item \href{https://docs.fuzzified.world/models/\#FuzzifiED.GetFlavQNDiag}{\lstinline|GetFlavQNDiag(nm, nf, qf[, id, modul])|} --- a linear combination of number of electrons in each flavour $Q=\sum_fq_fn_f$, where $\{q_f\}$ is stored in \lstinline|qf| in the format of either an array or a dictionary.\footnote{\textit{E.~g.}, for $Q=n_{f=1}-n_{f=3}$ in a 4-flavour system, both \lstinline[basicstyle=\ttfamily\scriptsize]|qf = [1, 0, -1, 0]| and \lstinline[basicstyle=\ttfamily\scriptsize]|qf = Dict(1 => 1, 3 => -1)| are acceptable.}
    \item \href{https://docs.fuzzified.world/models/\#FuzzifiED.GetZnfChargeQNDiag-Tuple{Int64,\%20Int64}}{\lstinline|GetZnfChargeQNDiag(nm, nf)|} --- a $\BZ_{N_f}$-charge $Q=\sum_{f=1}^{N_f}(f-1)n_f\mod N_f$.
    \item \href{https://docs.fuzzified.world/models/\#FuzzifiED.GetPinOrbQNDiag}{\lstinline|GetPinOrbQNDiag(no, pin_o[, id])|} --- the number of electrons in the subset of sites \lstinline|pin_o|. This QNDiag is useful for defects and boundaries.
\end{itemize}

The collection of configurations is generated from the QNDiags. It is recorded in the mutable type \lstinline|Confs|.

\begin{block}{\href{https://docs.fuzzified.world/core/\#FuzzifiED.Confs}{\lstinline|Confs|} --- Type}
The mutable type contains the fields
\begin{itemize}
    \item \lstinline|no :: Int64| --- the number of sites.
    \item \lstinline|ncf :: Int64| --- the number of configurations.
    \item \lstinline|conf :: Vector{Int64}| --- an array of length \lstinline|ncf| containing all the configurations.
    \item \lstinline|nor :: Int64|, \lstinline|lid :: Vector{Int64}| and \lstinline|rid :: Vector{Int64}| contain the information of Lin table that is used to inversely look up the index from the configuration.
\end{itemize}
and can be constructed by the method
\begin{lstlisting}
    Confs(no :: Int64, secd :: Vector{Int64}, qnd :: Vector{QNDiag})
\end{lstlisting}
where \lstinline|qnd| is the array of QNDiags, and \lstinline|secd| is the array of charges $Q$ of each QNDiag.\footnote{In general, many methods in FuzzifiED admits keyword arguments \lstinline[basicstyle=\ttfamily\scriptsize]|num_th :: Int64| that specifies the number of threads and \lstinline[basicstyle=\ttfamily\scriptsize]|disp_std :: Bool| that specifies whether or not the log shall be displayed. Hereafter, we will omit these two arguments.}
\end{block}

Here, each configuration is stored as a binary number with $N_o$ bits. If the $o$-th site in the configuration is occupied, the $(o-1)$-th bit of the configuration is $1$; if the site is empty, then the bit is $0$. Besides the storage of the configuration, we also need a reverse look-up process that returns the index from the binary string. This is realised by a Lin table stored in \lstinline|lid| and \lstinline|rid|. The details are given in Appendix~\ref{app:data_lin}.

In the example of Ising model Eq.~\eqref{eq:ising_hmt}, there are two QNDiags, \textit{viz.}~the particle number and the angular momentum
\begin{align}
    Q_1&=N_e,& q_{1,mf}&=1\nonumber\\
    Q_2&=2L_z,&q_{2,mf}&=2m.
\end{align}
The code for generating the configurations in the $L_z=0$ sector is
\begin{lstlisting}
    qnd = [
        QNDiag(fill(1, no)),
        QNDiag([ 2 * m - nm - 1 for m = 1 : nm for f = 1 : nf ])
    ]
    cfs = Confs(no, [nm, 0], qnd)
\end{lstlisting}
Alternatively, using the built-in models,
\begin{lstlisting}
    qnd = [
        GetNeQNDiag(no),
        GetLz2QNDiag(nm, nf)
    ]
    cfs = Confs(no, [nm, 0], qnd)
\end{lstlisting}

\subsection{Constructing the Basis}
\label{sec:ed_basis}

Having constructed the configurations, we now construct the basis of the Hilbert space. The ``\emph{basis}'' is the collection of states that are linear combinations of the configurations carrying certain diagonal and $\BZ_p$ off-diagonal quantum numbers (QNOffd).

The QNOffds supported by FuzzifiED are the $\BZ_p$ symmetries in the form of
\begin{equation}
    \cZ:\ c_o\to \alpha_o^* c^{(p_o)}_{\pi_o},\quad c_o^\dagger\to \alpha_o c^{(1-p_o)}_{\pi_o},
\end{equation}
where we use a notation $c^{(1)}=c^\dagger$ and $c^{(0)}=c$ for convenience, $\pi_o$ is a permutation of the sites $1,\dots N_o$, $\alpha_o$ is a coefficient, and $p_o$ specified whether or not particle-hole transformation is performed for the site. Note that one must guarantee that all these transformations commute with each other and also commute with the diagonal quantum numbers. In FuzzifiED, the QNOffds are recorded in the mutable type \lstinline|QNOffd|.

\begin{block}{\href{https://docs.fuzzified.world/core/\#FuzzifiED.QNOffd}{\lstinline|QNOffd|} --- Type}
The mutable type contains the fields
\begin{itemize}
    \item \lstinline|perm :: Vector{Int64}| --- a length-$N_o$ array that records the permutation $\pi_o$.
    \item \lstinline|ph :: Vector{Int64}| --- a length-$N_o$ array that records $p_o$ to determine whether or not to perform a particle-hole transformation on a site.
    \item \lstinline|fac :: Vector{ComplexF64}| --- a length-$N_o$ array that records the factor $\alpha_o$ in the transformation.
    \item \lstinline|cyc :: Int64| --- the cycle $p$.
\end{itemize}
It can be initialised by the methods
\begin{lstlisting}
    QNOffd(perm :: Vector{Int64}[, ph :: Vector{Int64}][, fac :: Vector{ComplexF64}][, cyc :: Int64])
    QNOffd(perm :: Vector{Int64}, ph_q :: Bool[, fac :: Vector{ComplexF64}])
\end{lstlisting}
By default, \lstinline|ph| is set to all 0, \lstinline|fac| is set to all 1 and \lstinline|cyc| is set to 2. If \lstinline|ph_q| is set to be true, \lstinline|ph| is set to all 1.
\end{block}

Several useful QNOffds are built-in
\begin{itemize}
    \item \href{https://docs.fuzzified.world/models/\#FuzzifiED.GetParityQNOffd}{\lstinline|GetParityQNOffd(nm, nf[, permf, fac])|} --- the particle-hole transformation $\cP:c^\dagger_{mf}\allowbreak\mapsto\alpha_fc_{m\pi_f}$, with the permutation of flavours $\pi_f$ and the factors $\alpha_f$ stored in \lstinline|permf| and \lstinline|fac| as either an array or a dictionary.
    \item \href{https://docs.fuzzified.world/models/\#FuzzifiED.GetFlavPermQNOffd}{\lstinline|GetFlavPermQNOffd(nm, nf, permf[, fac][, cyc])|} --- the flavour permutation transformation $\cZ:c^\dagger_{mf}\mapsto\alpha_fc_{m\pi_f}^\dagger$, with the permutation of flavours $\pi_f$ and the factors $\alpha_f$ stored in \lstinline|permf| and \lstinline|fac| as either an array or a dictionary, and the cycle stored in \lstinline|cyc|.
    \item \href{https://docs.fuzzified.world/models/\#FuzzifiED.GetRotyQNOffd-Tuple{Int64,\%20Int64}}{\lstinline|GetRotyQNOffd(nm, nf)|} --- the $\pi$-rotation with respect to the $y$-axis $\cR_y:c^\dagger_{mf}\mapsto(-1)^{m+s}c_{(-m)f}^\dagger$
\end{itemize}

After implementing the QNOffds, a state on the new basis should look like
\begin{equation}
    |I\rangle=\lambda_{i_{I1}}|i_{I1}\rangle+\lambda_{i_{I2}}|i_{I2}\rangle+\cdots+\lambda_{i_{Im_I}}|i_{Im_I}\rangle,
\end{equation}
where the $|i\rangle$'s are configurations, and $|I\rangle$ is a linear combination of them. In other words, the configurations are organised into groups of size $m_I$. In FuzzifiED, the basis $\{|I\rangle\}$ is recorded in the mutable type \lstinline|Basis|.

\begin{block}{\href{https://docs.fuzzified.world/core/\#FuzzifiED.Basis}{\lstinline|Basis|} --- Type}
The mutable type contains the fields
\begin{itemize}
    \item \lstinline|cfs :: Confs| --- the configurations $\{|i\rangle\}$ that respect the QNDiags.
    \item \lstinline|dim :: Int64| --- the dimension of the basis.
    \item \lstinline|szz :: Int64| --- the maximum size $\max m_I$ of groups.
    \item \lstinline|cfgr :: Vector{Int64}| --- an array of length \lstinline|cfs.ncf| and records which group $|I\rangle$ each configuration $|i\rangle$ belong to.
    \item \lstinline|cffac :: Vector{ComplexF64}| --- an array of length \lstinline|cfs.ncf| and records the coefficients $\lambda_i$ of each configuration.
    \item \lstinline|grel :: Matrix{Int64}| --- a \lstinline|szz|$\times$\lstinline|dim| matrix that records the configurations in each group $|i_{Ik}\rangle$ ($k=1,\dots,m_I$).
    \item \lstinline|grsz :: Vector{Int64}| --- an array of length \lstinline|dim| that records the size $m_I$ of each group.
\end{itemize}
It can be constructed by the methods
\begin{lstlisting}
    Basis(cfs :: Confs, secf :: Vector{ComplexF64}, qnf :: Vector{QNOffd})
    Basis(cfs :: Confs)
\end{lstlisting}
where \lstinline|secf| records the eigenvalue of each transformation, typically in the form $e^{i2\pi q/p}$ where $p$ is the cycle and $q$ is the $\BZ_p$ charge.
\end{block}

In the example of Ising model Eq.~\eqref{eq:ising_hmt}, There are three $\BZ_2$ symmetries, \textit{viz.}~the particle-hole transformation $\cP$, the $\pi$-rotation along the $y$-axis $\cR_y$, and the flavour (Ising) symmetry $\cZ$
\begin{align}
    \cP:\ c^\dagger_{\sigma m}&\mapsto\sigma c_{-\sigma,m}\nonumber\\
    \cZ:\ c^\dagger_{\sigma m}&\mapsto c^\dagger_{-\sigma,m}\nonumber\\
    \cR_y:\ c^\dagger_{\sigma m}&\mapsto c^\dagger_{\sigma,-m}
\end{align}
The code to generate the basis in the all-positive sector is
\begin{lstlisting}
    qnf = [
        QNOffd([ (m - 1) * nf + [2, 1][f] for m = 1 : nm for f = 1 : nf ], true,
            ComplexF64[ [-1, 1][f] for m = 1 : nm for f = 1 : nf ]),
        QNOffd([ (m - 1) * nf + [2, 1][f] for m = 1 : nm for f = 1 : nf ]),
        QNOffd([ (nm - m) * nf + f for m = 1 : nm for f = 1 : nf],
            ComplexF64[ iseven(m) ? 1 : -1 for m = 1 : nm for f = 1 : nf ])
    ]
    bs = Basis(cfs, [1, 1, 1], qnf)
\end{lstlisting}
Alternatively, using the built-in functions
\begin{lstlisting}
    qnf = [
        GetParityQNOffd(nm, 2, [2, 1], [-1, 1]),
        GetFlavPermQNOffd(nm, 2, [2, 1]),
        GetRotyQNOffd(nm, 2)
    ]
    bs = Basis(cfs, [1, 1, 1], qnf)
\end{lstlisting}

\subsection{Recording the Many-Body Operator Terms}
\label{sec:ed_term}

Having constructed the basis, we now construct the many-body operators. A general many-body operator can be written as
\begin{equation}
    \cO =\sum_{t=1}^{N_t}U_tc^{(p_{t1})}_{o_{t1}}c^{(p_{t2})}_{o_{t2}}\dots c^{(p_{tl_t})}_{o_{tl_t}},
\end{equation}
where $c^{(0)}=c$ and $c^{(1)}=c^\dagger$. In FuzzifiED, this is recorded as an array of \lstinline|Term|, and each \lstinline|Term| records the building block $Uc^{(p_{1})}_{o_{1}}c^{(p_{2})}_{o_{2}}\dots c^{(p_{l})}_{o_{l}}$.

\begin{block}{\href{https://docs.fuzzified.world/core/\#FuzzifiED.Term}{\lstinline|Term|} --- Type}
The mutable type contains the fields
\begin{itemize}
    \item \lstinline|coeff :: ComplexF64| --- the coefficient $U$.
    \item \lstinline|cstr :: Vector{Int64}| --- a length-$2l$ array $\{p_1,o_1,p_2,o_2,\dots p_l,o_l\}$ recording the operator string.
\end{itemize}
It can be initialised by the method
\begin{lstlisting}
    Term(coeff :: ComplexF64, cstr :: Vector{Int64})
\end{lstlisting}
The addition and multiplication of terms are supported, and the terms can be simplified by the method
\begin{lstlisting}
    SimplifyTerms(tms :: Vector{Term})
\end{lstlisting}
After the simplification, the resulting terms satisfy
\begin{enumerate}
    \item Each term is normal-ordered --- the creation operators are in front of the annihilation operators~; the site index of the creation operators are in ascending order and the annihilation operators in descending order.
    \item Like terms are combined, and terms with zero coefficients are removed.
\end{enumerate}
\end{block}

In FuzzifiED, several useful operator terms are built-in~:
\begin{itemize}
    \item \href{https://docs.fuzzified.world/models/\#FuzzifiED.GetDenIntTerms}{\lstinline|GetDenIntTerms(nm, nf[, ps_pot][, mat_a[, mat_b]])|} --- the normal-ordered den\-sity-density interaction term in the Hamiltonian
    \begin{equation}
        \sum_{l\{m_if_i\}}U_lC^l_{m_1m_2m_3m_4}M^A_{f_1f_4}M^B_{f_2f_3}c^\dagger_{m_1f_1}c^\dagger_{m_2f_2}c_{m_3f_3}c_{m_4f_4}
    \end{equation}
    where $C^l_{m_1m_2m_3m_4}$ is given in Eq.~\eqref{eq:ps_pot_coeff}.
    \item \href{https://docs.fuzzified.world/models/\#FuzzifiED.GetPairIntTerms}{\lstinline|GetPairIntTerms(nm, nf, ps_pot, mat_a[, mat_b])|} --- the normal-ordered pair-pair interaction term in the Hamiltonian
    \begin{equation}
        \sum_{l\{m_if_i\}}U_lC^l_{m_1m_2m_3m_4}M^A_{f_1f_2}M^B_{f_3f_4}c^\dagger_{m_1f_1}c^\dagger_{m_2f_2}c_{m_3f_3}c_{m_4f_4}.
    \end{equation}
    \item \href{https://docs.fuzzified.world/models/\#FuzzifiED.GetPolTerms-Tuple{Int64,\%20Int64,\%20Matrix{\%3C:Number}}}{\lstinline|GetPolTerms(nm, nf[, mat])|} --- the polarisation term in the Hamiltonian
    \begin{equation}
        \sum_{mf_1f_2}c^\dagger_{mf_1}M_{f_1f_2}c_{mf_2}.
    \end{equation}
    \item \href{https://docs.fuzzified.world/models/\#FuzzifiED.GetL2Terms-Tuple{Int64,\%20Int64}}{\lstinline|GetL2Terms(nm, nf)|} --- the total angular momentum.
    \item \href{https://docs.fuzzified.world/models/\#FuzzifiED.GetC2Terms-Tuple{Int64,\%20Int64,\%20Vector{\%3C:AbstractMatrix{\%3C:Number}}}}{\lstinline|GetC2Terms(nm, nf, mat_gen[, mat_tr])|} --- the quadratic Casimir
    \begin{equation}
        C_2=\sum_{imm'\{f_i\}}\frac{(c^\dagger_{mf_1}G^i_{f_1f_2}c_{mf_2})(c^\dagger_{m'f_3}(G^i_{f_3f_4})^\dagger c_{m'f_4})}{\operatorname{tr}G_i^\dagger G_i}-\textrm{(trace)}
    \end{equation}
    where $G^i$ are the generator matrices.
\end{itemize}

In the example of the Ising model, the code that records the Hamiltonian Eq.~\eqref{eq:ising_hmt} is
\begin{lstlisting}
    using WignerSymbols
    ps_pot = [ 4.75, 1.0 ] * 2.
    h = 3.16
    tms_hmt = Term[]
    m = zeros(Int64, 4)
    for m[1] = 0 : nm - 1, m[2] = 0 : nm - 1, m[3] = 0 : nm - 1
        m[4] = m[1] + m[2] - m[3]
        (m[4] < 0 || m[4] >= nm) && continue
        f = [0, 1, 1, 0]
        o = m .* nf .+ f .+ 1
        mr = m .- s

        val = ComplexF64(0)
        for l in eachindex(ps_pot)
            (abs(mr[1] + mr[2]) > nm - l || abs(mr[3] + mr[4]) > nm - l) && break
            val += ps_pot[l] * (2 * nm - 2 * l + 1) * wigner3j(s, s, nm - l, mr[1], mr[2], -mr[1] - mr[2]) * wigner3j(s, s, nm - l, mr[4], mr[3], -mr[3] - mr[4])
        end
        tms_hmt += Terms(val, [1, o[1], 1, o[2], 0, o[3], 0, o[4]])
    end
    for m = 0 : nm - 1
        o = m * nf .+ [1, 2]
        tms_hmt += Terms(-h, [1, o[1], 0, o[2]])
        tms_hmt += Terms(-h, [1, o[2], 0, o[1]])
    end
\end{lstlisting}
Alternatively, using the built-in functions
\begin{lstlisting}
    sg1 = [  1  0 ;  0  0 ]
    sg2 = [  0  0 ;  0  1 ]
    sgx = [  0  1 ;  1  0 ]
    sgz = [  1  0 ;  0 -1 ]
    ps_pot = [ 4.75, 1.0 ] * 2.0
    fld_h = 3.16
    tms_hmt = SimplifyTerms(
        GetDenIntTerms(nm, nf, ps_pot, sg1, sg2)
        - fld_h * GetPolTerms(nm, 2, sgx)
    )
\end{lstlisting}

We also need to construct the total angular momentum. It is defined as
\begin{equation}
    L^2=L^+L^-+(L^z)^2-L^z.
\end{equation}
As $c_m$ carries the $\SO(3)$ spin-$s$ representation,
\begin{equation}
    L^z=\sum_{mf}mc_m^\dagger c_m,\quad L^\pm=\sum_{mf}\sqrt{(s\mp m)(s\pm m+1)}c^\dagger_{m\pm 1}c_m.
\end{equation}
We can first construct its building blocks and use the addition and multiplication of the terms
\begin{lstlisting}
    tms_lz = [ Term(m - s - 1, [1, (m - 1) * nf + f, 0, (m - 1) * nf + f]) for m = 1 : nm for f = 1 : nf ]
    tms_lp = [ Term(sqrt((nm - m) * m), [1, m * nf + f, 0, (m - 1) * nf + f]) for m = 1 : nm - 1 for f = 1 : nf ]
    tms_lm = tms_lp'
    tms_l2 = SimplifyTerms(tms_lz * tms_lz - tms_lz + tms_lp * tms_lm)
\end{lstlisting}
Alternatively, using the built-in functions,
\begin{lstlisting}
    tms_l2 = GetL2Terms(nm, nf)
\end{lstlisting}

\subsection{Generating Sparse Matrix}
\label{sec:ed_opmat}

Having obtained the terms in the many-body operator, we need to generate the matrix elements given the initial and final basis and find their eigenstates.

In FuzzifiED, the mutable type \lstinline|Operator| records the terms together with information about its symmetry, the basis of the state it acts on, and the basis of the resulting state.

\begin{block}{\href{https://docs.fuzzified.world/core/\#FuzzifiED.Operator}{\lstinline|Operator|} --- Type}
The type can be initialised with the method
\begin{lstlisting}
    Operator(bsd :: Basis[, bsf :: Basis], terms :: Vector{Term} ; red_q :: Int64, sym_q :: Int64)
\end{lstlisting}
where the arguments
\begin{itemize}
    \item \lstinline|bsd :: Basis| --- the basis of the initial state.
    \item \lstinline|bsf :: Basis| --- the basis of the final state. Facultative, the same as \lstinline|bsd| by default.
    \item \lstinline|terms :: Vector{Term}| --- the terms.
    \item \lstinline|red_q :: Int64| --- a flag that records whether or not the conversion to a sparse matrix can be simplified~: if \lstinline|bsd| and \lstinline|bsf| have exactly the same set of quantum numbers, and the operator fully respects the symmetries, then \lstinline|red_q = 1| ; otherwise \lstinline|red_q = 0| ; Facultative, if \lstinline|bsf| is omitted, 1 by default, otherwise 0 by default.
    \item \lstinline|sym_q :: Int64| --- the symmetry of the operator~: if its corresponding matrix is Hermitian, then \lstinline|sym_q = 1| ; if it is symmetric, then \lstinline|sym_q = 2| ; otherwise \lstinline|sym_q = 0|. Facultative, if \lstinline|bsf| is omitted, 1 by default, otherwise 0 by default.
\end{itemize}
\end{block}

The sparse matrix is recorded in the compressed sparse column (CSC) format, which is described in detail in Appendix~\ref{app:data_csc}. In FuzzifiED, the sparse matrix is stored in the mutable type \lstinline|OpMat{T}|

\begin{block}{\href{https://docs.fuzzified.world/core/\#FuzzifiED.OpMat}{\lstinline|OpMat{T}|} --- Type}
\noindent where \lstinline|T| is the type of the elements, it can either be \lstinline|ComplexF64| or \lstinline|Float64|. It contains the fields
\begin{itemize}
    \item \lstinline|dimd :: Int64|, \lstinline|dimf :: Int64|, \lstinline|nel :: Int64|, \lstinline|symq :: Int64| --- Parameters of the sparse matrix, the number of columns, rows, elements and the symmetry of matrix, respectively.
    \item \lstinline|colptr :: Vector{Int64}| with length \lstinline|dimd :: Int64 + 1|.
    \item \lstinline|rowid :: Vector{Int64}| with length \lstinline|nel|.
    \item \lstinline|elval :: Vector{T}| with length \lstinline|nel|.
\end{itemize}
It can be generated from the method
\begin{lstlisting}
    OpMat[{T}](op :: Operator)
\end{lstlisting}
\end{block}

\subsection{Finding Eigenstates}
\label{sec:ed_diag}

After generating the sparse matrix, the method \lstinline|GetEigensystem| uses the Fortran Arpack package to calculate its lowest eigenstates.

\begin{block}{\href{https://docs.fuzzified.world/core/\#FuzzifiED.GetEigensystem-Tuple{OpMat{ComplexF64},\%20Int64}}{\lstinline|GetEigensystem|} --- Method}
\begin{lstlisting}
    GetEigensystem(
        mat :: OpMat{T}, nst :: Int64 ;
        tol :: Float64, ncv :: Int64, initvec :: Vector{T}
    ) :: Tuple{Vector{T}, Matrix{T}}
\end{lstlisting}
The arguments are
\begin{itemize}
    \item \lstinline|mat :: OpMat{T}| --- the matrix.
    \item \lstinline|nst :: Int64| --- the number of eigenstates to be calculated.
    \item \lstinline|tol :: Float64| --- the tolerence for the Arpack process. The default value is $10^{-8}$.
    \item \lstinline|ncv :: Int64| --- an auxiliary parameter needed in the Arpack process. The default value is \lstinline|max(2 * nst, nst + 10)|.
    \item \lstinline|initvec :: Vector{T}| --- the initial vector. If empty, a random initialisation shall be used. Facultative, empty by default.
\end{itemize}
The output is a tuple that includes two elements~:
\begin{itemize}
    \item A length-\lstinline|nst| array recording the eigenvalues, and
    \item A \lstinline|dimd|$\times$\lstinline|nst| matrix where every column records an eigenstate.
\end{itemize}

\end{block}

In the example of the Ising model, the code to calculate the lowest $N_\textrm{st}=10$ eigenstates from the basis and the Hamiltonian terms is
\begin{lstlisting}
    nst = 10
    hmt = Operator(bs, tms_hmt)
    hmt_mat = OpMat(hmt)
    enrg, st = GetEigensystem(hmt_mat, nst)
\end{lstlisting}

\subsection{Inner Product of States, Operators and Transformations}
\label{sec:ed_inner}

Having obtained the eigenstates, we need to make measurements on them. The simplest kind of measurement is the inner product of a many-body operator with two states $\langle j|\cO|i\rangle$. FuzzifiED supports the inner product and vector product of \lstinline|Operator| and \lstinline|OpMat{T}| with vectors that represent the state
\begin{lstlisting}
    (op :: Operator) * (st_d :: Vector{T}) :: Vector{T}
    (mat :: OpMat{T}) * (st_d :: Vector{T}) :: Vector{T}
    (st_f :: Vector{T}) * (op :: Operator) * (st_d :: Vector{T}) :: T
    (st_f :: Vector{T}) * (mat :: OpMat{T}) * (st_d :: Vector{T}) :: T
\end{lstlisting}
where \lstinline|T| is the type of the elements. \textit{E.~g.}, the code to measure the angular momenta of each state is
\begin{lstlisting}
    tms_l2 = GetL2Terms(nm, 2)
    l2 = Operator(bs, tms_l2)
    l2_mat = OpMat(l2)
    l2_val = [ st[:, i]' * l2_mat * st[:, i] for i in eachindex(enrg)]
\end{lstlisting}

One might also need to act transformations on the state $\cZ|i\rangle$. In FuzzifiED, the mutable type \lstinline|Transf| records the transformation together with the basis of the initial and final states

\begin{block}{\href{https://docs.fuzzified.world/core/\#FuzzifiED.Transf}{\lstinline|Transf|} --- Type}
The type can be initialised from a \lstinline|QNOffd| by
\begin{lstlisting}
    Transf(bsd :: Basis[, bsf :: Basis], qnf :: QNOffd)
\end{lstlisting}
where the arguments
\begin{itemize}
    \item \lstinline|bsd :: Basis| --- the basis of the initial state.
    \item \lstinline|bsf :: Basis| --- the basis of the final state. Facultative, the same as \lstinline|bsd| by default.
    \item \lstinline|qnf :: QNOffd| --- records the transformation $c_o\to \alpha_o^* c^{(p_o)}_{\pi_o}$.
\end{itemize}
It can act on a state by
\begin{lstlisting}
    (trs :: Transf) * (st_d :: Vector{T}) :: Vector{T}
\end{lstlisting}
\end{block}

\subsection{Measuring Local Observables}
\label{sec:ed_obs}

Local observables are a particularly useful kind of operators on the fuzzy sphere. Their value at a point on the sphere can be decomposed into spherical components, and the multiplication of the components follows the triple integral formula of monopole spherical harmonics
\begin{align}
    \cO(\br)&=\sum_{lm}Y^{(s)}_{lm}(\bnh)\cO_{lm}\nonumber\\
    (\cO_1\cO_2)_{lm}&=\sum_{l_1l_2m_1m_2}(\cO_1)_{l_1m_1}(\cO_2)_{l_2m_2}\nonumber\\
    &\qquad\qquad\times(-1)^{s+m}\sqrt{\frac{(2l_1+1)(2l_2+1)(2l_3+1)}{4\pi}}\begin{pmatrix}l_1&l_2&l\\m_1&m_2&-m\end{pmatrix}\begin{pmatrix}l_1&l_2&l\\-s_1&-s_2&s\end{pmatrix}
\end{align}
In FuzzifiED, they are stored in the type \lstinline|SphereObs|.

\begin{block}{\href{https://docs.fuzzified.world/models/\#FuzzifiED.SphereObs}{\lstinline|SphereObs|} --- Type}
The type contains the fields
\begin{itemize}
    \item \lstinline|s2 :: Int64| and  \lstinline|l2m :: Int64| --- is twice the spin $2s$ and twice the maximal angular momentum $2l_{\max}$ of the observable.
    \item \lstinline|get_comp :: Function| --- a function that sends the component specified by a tuple of integers $(2l,2m)$ to a list of terms that specifies the expression of the component.
    \item \lstinline|stored_q :: Bool| --- a boolean that specifies whether or not the components of the observable are stored.
    \item \lstinline|comps :: Dict{Tuple{Int64, Int64}, Terms}| --- each component of the observable stores in the format of a dictionary whose keys are the pairs of integers $(2l,2m)$ and values are the lists of terms that specifies the expression of the component.
\end{itemize}
and can be initialised by the methods
\begin{lstlisting}
    SphereObs(s2 :: Int64, l2m :: Int64, get_comp :: Function)
    SphereObs(s2 :: Int64, l2m :: Int64, comps :: Dict)
\end{lstlisting}
Their adjoint, addition and multiplication are supported. The related functions are
\begin{itemize}
    \item \href{https://docs.fuzzified.world/models/\#FuzzifiED.StoreComps-Tuple{SphereObs}}{\lstinline|StoreComps(obs)|} calculates and stores each component of the observable.
    \item \href{https://docs.fuzzified.world/models/\#FuzzifiED.Laplacian-Tuple{SphereObs}}{\lstinline|Laplacian(obs)|} takes the Laplacian of an observable.
    \item \href{https://docs.fuzzified.world/models/\#FuzzifiED.GetComponent-Tuple{SphereObs,\%20Number,\%20Number}}{\lstinline|GetComponent(obs, l, m)|} returns a spherical component of the observable $\cO_{lm}$.
    \item \href{https://docs.fuzzified.world/models/\#FuzzifiED.GetPointValue-Tuple{SphereObs,\%20Float64,\%20Float64}}{\lstinline|GetPointValue(obs, theta, phi)|} returns an observable at one point $\cO(\br)$.
    \item \href{https://docs.fuzzified.world/models/\#FuzzifiED.GetIntegral-Tuple{SphereObs}}{\lstinline|GetIntegral(obs)|} returns the uniform spatial integral of an observable. 
    \item \href{https://docs.fuzzified.world/models/\#FuzzifiED.FilterComponent-Tuple{SphereObs,\%20Any}}{\lstinline|FilterComponent(obs, flt)|} filters a certain set of modes ofan observable.
\end{itemize}
\end{block}

Several important types of spherical observables are built-in in FuzzifiED\footnote{Two parameters \href{https://docs.fuzzified.world/models/\#FuzzifiED.ObsNormRadSq}{\lstinline[basicstyle=\ttfamily\scriptsize]|ObsNormRadSq|} and \href{https://docs.fuzzified.world/models/\#FuzzifiED.ObsMomIncr}{\lstinline[basicstyle=\ttfamily\scriptsize]|ObsMomIncr|} control how the components are normalised and whether they increase or decrease $L^z$.}
\begin{itemize}
    \item \href{https://docs.fuzzified.world/models/\#FuzzifiED.GetElectronObs-Tuple{Int64,\%20Int64,\%20Int64}}{\lstinline|GetElectronObs(nm, nf, f)|} --- electron annihilation operator $\psi_f(\br)$.
    \item \href{https://docs.fuzzified.world/models/\#FuzzifiED.GetDensityObs-Tuple{Int64,\%20Int64,\%20Matrix{\%3C:Number}}}{\lstinline|GetDensityObs(nm, nf[, mat])|} --- density operator $n_M(\br)=\sum_{ff'}\psi^\dagger_{f}(\br)M_{ff'}\psi_{f'}(\br)$.
    \item \href{https://docs.fuzzified.world/models/\#FuzzifiED.GetPairingObs-Tuple{Int64,\%20Int64,\%20Matrix{\%3C:Number}}}{\lstinline|GetPairingObs(nm, nf, mat)|} --- pair operator $\Delta_M(\br)=\sum_{ff'}\psi_{f}(\br)M_{ff'}\psi_{f'}$.
\end{itemize}

In the example of Ising model, to calculate the OPE coefficient $f_{\sigma\sigma\epsilon}=\langle \sigma|n^z_{00}|\epsilon\rangle/\langle \sigma|n^z_{00}|0\rangle$, one need to first calculate the eigenstates in the $\BZ_2$-odd sector
\begin{lstlisting}
    bs_m = Basis(cfs, [1, -1, 1], qnf)
    hmt_m = Operator(bs_m, bs_m, tms_hmt ; red_q = 1, sym_q = 1)
    hmt_mat_m = OpMat(hmt_m)
    enrg_m, st_m = GetEigensystem(hmt_mat_m, 10)
    st0 = st[:, 1]
    ste = st[:, 2]
    sts = st_m[:, 1]
\end{lstlisting}
and then construct the density operator
\begin{lstlisting}
    obs_nz = GetDensityObs(nm, 2, sgz)
    tms_nz00 = SimplifyTerms(GetComponent(obs_nz, 0.0, 0.0))
    nz00 = Operator(bs, bs_m, tms_nz00 ; red_q = 1)
    f_sse = abs((sts' * nz00 * ste) / (sts' * nz00 * st0))
\end{lstlisting}

Besides the spherical observable, we also provide a type \href{https://docs.fuzzified.world/models/\#FuzzifiED.AngModes}{\lstinline|AngModes|} that superposes under the rule of angular momentum superposition instead of spherical harmonics triple integral
\begin{equation}
    (\cA_1\cA_2)_{lm}=\sum_{l_1m_1l_2m_2}(\cA_1)_{l_1m_1}(\cA_2)_{l_2m_2}\langle l_1m_1,l_2m_2|lm\rangle.
\end{equation}
The interfaces are similar.

\subsection{Measuring the Entanglement}
\label{sec:ed_ent}

A non-local quantity that bears particular significance is the entanglement. To calculate the entanglement, we divide the sphere into two parts $A$ and $B$. The reduced density matrix of part $A$ is obtained by tracing the density matrix over the part $B$
\begin{equation}
    \rho_A(\Psi)=\operatorname{tr}_B|\Psi\rangle\langle\Psi|
\end{equation}
The entanglement entropy is $S=-\operatorname{tr}\rho_A\log\rho_A$ and the entanglement spectrum is the collection of eigenvalues of $\rho_A$ taken negative logarithm.

Here we sketch the process of the calculation. More detail is given in Ref.~\cite{Sterdyniak2011RealEnt}. The creation operator in each orbital is divided into the creation on $A$ part and the creation on $B$ part
\begin{equation}
    c^\dagger_o=\alpha_oc^\dagger_{o,A}+\beta_mc^\dagger_{o,B},
\end{equation}
where $|\alpha_o|^2+|\beta_o|^2=1$. For the cut in orbital space $m_c$,
\begin{equation*}
    \alpha_{mf}=\Theta(m_c-m)
\end{equation*}
where $\Theta$ is the Heaviside function~; for the cut in real space along latitude circle $\theta_c$,
\begin{equation*}
    \alpha_{mf}=\sqrt{\frac{\Beta_{\cos^2\theta_c/2}(s-m+1,s+m+1)}{\Beta(s-m+1,s+m+1)}}
\end{equation*}
where $\Beta_x(a,b)$ is the incomplete beta function.

To calculate the reduced density matrix, we decompose the state into the direct-product basis of two subsystems
\begin{equation}
    |\Psi\rangle=\sum_{K_0}v_{K_0}|K_0\rangle=\sum_{I_AJ_B}M_{I_AJ_B}|I_A\rangle|J_B\rangle,
\end{equation}
where the indices $K_0\in\cH,I_A\in\cH_A,J_B\in\cH_B$ are in the overall Hilbert space and the Hilbert space of subsystem $A$ and $B$. The density matrix is then
\begin{equation}
    \rho_A=\bM\bM^\dagger
\end{equation}
and the entanglement spectrum can be obtained from the SVD decomposition of the $\bM$ matrix. Like the Hamiltonian, the $\bM$ matrix is block diagonal, and each block carries different quantum numbers of the Hilbert spaces of $A$ and $B$ subsystem.\footnote{The $\alpha_o$ and $M_{IJ}$ in our convention is equivalent to $\mathcal{F}_{m,A}$ and $R_{\mu\nu}^A$ in the convention of Ref.~\cite{Sterdyniak2011RealEnt}, the conversions are $\alpha_{mf}=\sqrt{\mathcal{F}_{m,A}}$ and $M_{IJ}=R_{\mu\nu}^A/\sqrt{p}$.}

In FuzzifiED, the decomposition of state vector $v_{K_0}$ into matrix $M_{I_AJ_B}$ is done by the function \lstinline|StateDecompMat|, and the calculation of entanglement spectrum is done by the function \lstinline|GetEntSpec|

\begin{block}{\href{https://docs.fuzzified.world/core/\#FuzzifiED.GetEntSpec-Tuple{Vector{\%3C:Number},\%20Basis,\%20Vector{Vector{Vector{Int64}}},\%20Vector{Vector{Vector{Int64}}}}}{\lstinline|GetEntSpec|} --- Function}
\begin{lstlisting}
    GetEntSpec(
        st :: Vector{<:Number}, bs0 :: Basis,
        secd_lst :: Vector{Vector{Vector{Int64}}},
        secf_lst :: Vector{Vector{Vector{<:Number}}} ;
        qnd_a :: Vector{QNDiag}[, qnd_b :: Vector{QNDiag}],
        qnf_a :: Vector{QNOffd}[, qnf_b :: Vector{QNOffd}],
        amp_oa :: Vector{<:Number}[, amp_ob :: Vector{<:Number}]
    ) :: Dict{@NamedTuple{secd_a, secf_a, secd_b, secf_b}, Vector{Float64}}
\end{lstlisting}
The arguments are
\begin{itemize}
    \item \lstinline|st :: Vector{<:Number}| --- the state to be decomposed into the direct-product basis of two subsystems.
    \item \lstinline|bs0 :: Basis| --- the basis of the original state.
    \item \lstinline|secd_lst :: Vector{Vector{Vector{Int64}}}| --- the list of QNDiag sectors of subsystems to be calculated. Each of its elements is a two-element vector~; the first specifies the sector for subsystem $A$, and the second specifies the sector for subsystem $B$.
    \item \lstinline|secf_lst :: Vector{Vector{Vector{ComplexF64}}}| --- the list of QNOffd sectors of subsystems to be calculated. Each of its elements is a two-element vector~; the first specifies the sector for subsystem $A$, and the second specifies the sector for subsystem $B$.
    \item \lstinline|qnd_a :: Vector{QNDiag}, qnd_b :: Vector{QNDiag}, qnf_a :: Vector{QNOffd}| and \lstinline|qnf_b :: Vector{QNOffd}| --- the diagonal and off-diagonal quantum numbers of the subsystems $A$ and $B$. By default \lstinline|qnd_b = qnd_a| and \lstinline|qnf_b = qnf_a|.
    \item \lstinline|amp_oa :: Vector{ComplexF64}| and \lstinline|amp_ob :: Vector{ComplexF64}| --- the arrays that specify the amplitudes $\alpha_o$ and $\beta_o$. By default $\beta_o=\sqrt{1-\alpha_o^2}$.
\end{itemize}
and the output is a dictionary whose keys are named tuples that specify the sector containing entries \lstinline|secd_a|, \lstinline|secf_a|, \lstinline|secd_b|, \lstinline|secf_b| and values are lists of eigenvalues of the density matrix in those sectors.
\end{block}

In the example of the Ising model, to calculate the entanglement entropy cut from the equator, we first need to specify the quantum numbers of the subsystems~: the conservation of $N_e$, $L_z$ and the $\BZ_2$ symmetry.
\begin{lstlisting}
    qnd_a = [ GetNeQNDiag(no), GetLz2QNDiag(nm, nf) ]
    qnf_a = [ GetFlavPermQNOffd(nm, nf, [2, 1]) ]
\end{lstlisting}
we then specify the sectors to calculate~: The number of electrons in subsystem $A$ run from $0$ to $N_m$~; the angular momenta in subsystem $A$ can take all permitted values\footnote{The sectors with no allowed configurations are automatically ignored}~; for subsystem $B$, $N_{e,B}=N_m-N_{e,A}$, $L_{z,B}=-L_{z,A}$~; the $\BZ_2$ sectors of the two subsystems are the same.
\begin{lstlisting}
    secd_lst = Vector{Vector{Int64}}[]
    for nea = 0 : nm
        neb = nm - nea
        for lza = -min(nea, neb) * (nm - 1) : 2 : min(nea, neb) * (nm - 1)
            lzb = -lza
            push!(secd_lst, [[nea, lza], [neb, lzb]])
        end
    end
    secf_lst = [ [[1], [1]], [[-1], [-1]] ]
\end{lstlisting}
Finally, we specify the list of amplitudes $\alpha_o$.
\begin{lstlisting}
    amp_oa = [ sqrt(beta_inc(m, nm - m + 1, 0.5)) for m = 1 : nm for f = 1 : nf ]
\end{lstlisting}
To calculate the entanglement spectrum,
\begin{lstlisting}
    ent_spec = GetEntSpec(st_g, bs, secd_lst, secf_lst ; qnd_a, qnf_a, amp_oa)
\end{lstlisting}
The entanglement entropy can be calculated by collecting all the eigenvalues of the density matrix.
\begin{lstlisting}
    eig_rho = vcat(values(ent_spec)...)
    ent_entropy = -sum(eig_rho .* log.(eig_rho))
\end{lstlisting}

\subsection{Fuzzifino --- Module for Boson-Fermion Mixture}
\label{sec:ed_fuzzifino}

Fuzzifino is a module for ED calculation on the fuzzy sphere for systems with both bosons and fermions. The procedure is similar to FuzzifiED. We define several new types \href{https://docs.fuzzified.world/fuzzifino/\#FuzzifiED.Fuzzifino.SQNDiag}{\lstinline|SQNDiag|}, \href{https://docs.fuzzified.world/fuzzifino/\#FuzzifiED.Fuzzifino.SQNOffd}{\lstinline|SQNOffd|}, \href{https://docs.fuzzified.world/fuzzifino/\#FuzzifiED.Fuzzifino.SConfs}{\lstinline|SConfs|}, \href{https://docs.fuzzified.world/fuzzifino/\#FuzzifiED.Fuzzifino.SBasis}{\lstinline|SBasis|}, \href{https://docs.fuzzified.world/fuzzifino/\#FuzzifiED.Fuzzifino.STerm}{\lstinline|STerm|} and \href{https://docs.fuzzified.world/fuzzifino/\#FuzzifiED.Fuzzifino.SOperator}{\lstinline|SOperator|} that is parallel to the original versions \lstinline|QNDiag|, \lstinline|QNOffd|, \lstinline|Conf|, \lstinline|Basis|, \lstinline|Term| and \lstinline|Operator|. Here we note several points.

\begin{itemize}
    \item For each configuration, a boson part and a fermion part are stored. The boson configurations are indexed in ascending order from the last site to the first site. A maximal total occupation should be given. The details are given in Appendix~\ref{app:data_boson}.
    \item The operator string in a term is still stored in the form $\{p_1,o_1,p_2,o_2,\dots,p_l,o_l\}$, but now positive $o$ represents fermions, and negative $o$ represents boson with site number $|o|$. The bosonic creation and annihilation operator acts with an additional factor, \textit{e.~g.}, $b^\dagger|n\rangle=\sqrt{n+1}|n+1\rangle,b|n\rangle=\sqrt{n}|n-1\rangle$ for a single site.
    \item Several QNs and terms for purely bosonic models are built-in.
\end{itemize}

\subsection{FuzzyManifolds --- Module for Other Geometries}
\label{sec:ed_manifolds}

Apart from the sphere, FuzzifiED also supports the calculation for interacting particles on the lowest Landau level on a torus. The procedure is similar to sphere, except for two differences~:
\begin{itemize}
    \item The orbitals are now labelled by a $\BZ_{N_m}$ number $m=1,\dots,N_m$, and $m$ and $m+N_m$ are identified. Correspondingly, the conservation of angular momentum $L^z$ is now modified into a $\BZ_{N_m}$ translation symmetry with $\sum_{mf}mN_{mf}\mod N_m$ conserved. The \href{https://docs.fuzzified.world/manifolds/\#FuzzifiED.FuzzyManifolds.GetTorusLz2QNDiag-Tuple{Int64,\%20Int64}}{\lstinline|GetTorusLz2QNDiag|} function implemented the corresponding QNDiag. The translation in the other direction permutes $m$ and is implemented in \href{https://docs.fuzzified.world/manifolds/\#FuzzifiED.FuzzyManifolds.GetTorusTranslQNOffd-Tuple{Int64,\%20Int64}}{\lstinline|GetTorusTranslQNOffd|}.
    \item The interacting potentials are still parametrised by the pseudo-potentials in a similar sense, but the matrix $C_{m_1m_2m_3m_4}^l$ is different. The functions \href{https://docs.fuzzified.world/manifolds/\#FuzzifiED.FuzzyManifolds.GetTorusDenIntTerms}{\lstinline|GetTorusDenIntTerms|} and \href{https://docs.fuzzified.world/manifolds/\#FuzzifiED.FuzzyManifolds.GetTorusPairIntTerms}{\lstinline|GetTorusPairIntTerms|} generate the density-density interaction terms and the pair-pair interaction terms from the pseudo-potentials.
\end{itemize}

The detail of the set-up is given in Appendix~\ref{app:manifolds}.

\section{Density Matrix Renormalisation Group}
\label{sec:dmrg}

Having introduced ED, we now turn to density matrix renormalisation group (DMRG) that deals with larger systems. We briefly describe its procedure and give an instruction for using FuzzifiED for DMRG.

\subsection{DMRG with ITensor}

Practically, the \lstinline|dmrg| function in ITensor package automatically uses DMRG to optimise a matrix product state (MPS) to be the lowest eigenstate of a Hermitian Hamiltonian represented as a matrix product operator (MPO). To generate the input of the function, one needs to
\begin{enumerate}
    \item construct a set of sites that carries a certain set of QNDiags,
    \item construct a MPO representing the Hamiltonian on the sites from a set of terms (or OpSum in ITensor), and
    \item construct an initial MPS on the sites in the desired symmetry sector.
\end{enumerate}

In FuzzifiED, to use this extension for ITensor, one needs to use the packages \lstinline|ITensors| and \lstinline|ITensorMPS|. A new SiteType \lstinline|"FuzzyFermion"| is defined that behaves similarly to the built-in \lstinline|"Fermion"| type, and a single site can be generated from the QNDiags
\begin{lstlisting}
    siteind("FuzzyFermion" ; o :: Int64, qnd :: Vector{QNDiag})
\end{lstlisting}
The set of sites can be generated from the QNDiags (\textit{cf.}~Section \ref{sec:ed_confs}) by
\begin{lstlisting}
    GetSites(qnd :: Vector{QNDiag})
\end{lstlisting}

In the example of the Ising model, for convenience, we exchange the Pauli matrices $\sigma^x$ and $\sigma^z$ so that the two flavours carry $\BZ_2$-charge $0$ and $1$. The sites can be constructed by
\begin{lstlisting}
    nm = 12
    nf = 2
    no = nm * nf
    sites = GetSites([
        GetNeQNDiag(nm * nf),
        GetLz2QNDiag(nm, nf),
        GetZnfChargeQNDiag(nm, nf)
    ])
\end{lstlisting}

In ITensor, the MPO is generated from an OpSum and the sites. The OpSum can be directly converted from the terms (\textit{cf.}~Section \ref{sec:ed_term}). In the example of the Ising model,
\begin{lstlisting}
    sgx = [  0  1 ;  1  0 ]
    sgz = [  1  0 ;  0 -1 ]
    ps_pot = [4.75, 1.0] ./ 2
    tms_hmt = SimplifyTerms(
        GetDenIntTerms(nm, nf, ps_pot) -
        GetDenIntTerms(nm, nf, ps_pot, sgx) -0F
        3.16 * GetPolTerms(nm, nf, sgz)
    )
    os_hmt = OpSum(tms_hmt)
    hmt = MPO(os_hmt, sites)
\end{lstlisting}

To calculate the $\BZ_2$-even $L^z=0$ sector, the initial state can be taken as all the $\BZ_2$-even sites being filled and all the $\BZ_2$-odd sites being empty.\footnote{Note that ITensor takes the string \lstinline|"1"| instead of the number \lstinline|1| as occupied and \lstinline|"0"| instead of \lstinline|0| as filled.}
\begin{lstlisting}
    cfi_p = [ [1, 0][f] for m = 1 : nm for f = 1 : nf ]
    sti_p = MPS(sites, string.(cfi_p))
\end{lstlisting}

Having these ingredients ready, we can call the \lstinline|dmrg| function. To ensure performance, the maximal bond dimension should be increased gradually, and the noise decreased gradually to 0. An example that deals with maximal bond dimension $500$ is
\begin{lstlisting}
    E0, st0 = dmrg(hmt, sti_p ;
        nsweeps = 10,
        maxdim = [10,20,50,100,200,500],
        noise = [1E-4,3E-5,1E-5,3E-6,1E-6,3E-7],
        cutoff = [1E-8])
\end{lstlisting}
To generate a $\BZ_2$-odd initial state, we can simply flip the spin on the first orbital
\begin{lstlisting}
    cfi_m = [ m == 1 ? [0, 1][f] : [1, 0][f] for m = 1 : nm for f = 1 : nf ]
    sti_m = MPS(sites, string.(cfi_m))
    Es, sts = dmrg(hmt, sti_m ;
        nsweeps = 10,
        maxdim = [10,20,50,100,200,500],
        noise = [1E-4,3E-5,1E-5,3E-6,1E-6,3E-7],
        cutoff = [1E-8])
\end{lstlisting}
The first excited $\BZ_2$-even state can be generated by adding a projector $w|0\rangle\langle0|$ to the MPO
\begin{lstlisting}
    Ee, ste = dmrg(hmt, [st0], sti_p ;
        nsweeps = 10,
        maxdim = [10,20,50,100,200,500],
        noise = [1E-4,3E-5,1E-5,3E-6,1E-6,3E-7],
        cutoff = [1E-8],
        weight = 100)
\end{lstlisting}

The inner product can be measured by the ITensor function \lstinline|inner|. \textit{E.~g.}, to measure the angular momentum $L^2$ of the ground state,
\begin{lstlisting}
    tms_l2 = GetL2Terms(nm, 2)
    l2 = MPO(OpSum(tms_l2), sites)
    val_l20 = inner(st0', l2, st0)
\end{lstlisting}
To measure the OPE coefficient $f_{\sigma\sigma\epsilon}=\langle \sigma|n^x_{00}|\epsilon\rangle/\langle \sigma|n^x_{00}|0\rangle$\footnote{Note that the indices $x$ and $z$ have already been exchanged here.}
\begin{lstlisting}
    obs_nx = GetDensityObs(nm, 2, sgx)
    tms_nx00 = SimplifyTerms(GetComponent(obs_nx, 0.0, 0.0))
    nx00 = MPO(OpSum(tms_nx00), sites)
    f_sse = abs(inner(sts', nx00, ste) / inner(sts', nx00, st0))
\end{lstlisting}

\subsection{The Easy-Sweep Extension}

The extension Easy-Sweep facilitates the management of the DMRG process. It automatically records the intermediate results and recover the run from these intermediate results if a job is stopped on HPC. It also manages the gradual increase of maximal bond dimensions and the determination of convergence by the criteria of energy. This extension contains the following functions~:

\begin{itemize}
    \item \href{https://docs.fuzzified.world/itensors/\#FuzzifiED.GetMPOSites-Tuple{String,\%20Union{Sum{Scaled{ComplexF64,\%20Prod{Op}}},\%20Vector{Term}},\%20Vector{QNDiag}}}{\lstinline|GetMPOSites|} returns the MPO and sites for given operator terms and a Hilbert space with given quantum numbers. The function first checks if the MPO and sites are already stored in a specified file. If they are already stored, they are read and returned. Otherwise, the sites are generated from the quantum numbers, and the MPO is generated from the terms. The MPO and sites are then written into the file and returned.\footnote{We recommend the package \lstinline[basicstyle=\ttfamily\scriptsize]|ITensorMPOConstruction| stored at \url{https://github.com/ITensor/ITensorMPOConstruction.jl} to construct MPOs.}
    \item \href{https://docs.fuzzified.world/itensors/\#FuzzifiED.GetMPO-Tuple{String,\%20Union{Sum{Scaled{ComplexF64,\%20Prod{Op}}},\%20Vector{Term}},\%20Vector{\%3C:Index}}}{\lstinline|GetMPO|} returns the MPO for given operator terms and a given set of sites. The function first checks if the MPO is already stored in a specified file. If it is already stored, it is read and returned. Otherwise, MPO is generated from the terms and the sites. The MPO and sites are then written into the file and returned.
    \item \href{https://docs.fuzzified.world/itensors/\#FuzzifiED.SweepOne-Tuple{String,\%20MPO,\%20MPS,\%20Int64}}{\lstinline|SweepOne|} performs one round of DMRG sweeps for a given maximal bond dimensions $\chi_{\max}$ and returns the energy and the MPS. The function first checks if the calculation has already been done and the results are already stored in a specified file. If they are already stored, they are read and returned. Otherwise, the DMRG process is performed with a specified maximal bond dimension and the number of sweeps. The sweeps end if the energy difference is less than a tolerence or alternative criteria. The resulting energy and MPS are written into the file and returned.
    \item \href{https://docs.fuzzified.world/itensors/\#FuzzifiED.EasySweep-Tuple{String,\%20MPO,\%20MPS}}{\lstinline|EasySweep|} automatically performs several rounds of DMRG sweeps with increasing bond dimensions and returns energy and MPS. The function first checks whether the calculation has been partly done and the intermediate results have been stored in a specified file. The calculation is then picked up from the round that was previously stopped. The entire process is stopped if the energy difference between two rounds is less than a certain tolerence or the bond dimension of the result is less than $0.9$ times the maximal bond dimension. The resulting energy and MPS are written into the file and returned.
\end{itemize}

To use this extension for ITensor, one needs to use the packages \lstinline|ITensors|, \lstinline|ITensorMPS| and \lstinline|HDF5|. As the set-up, a path needs to be created to store the result HDF5 files, and a method that converts OpSum and sites to MPO needs to be defined.
\begin{lstlisting}
    function MyMPO(os, sites)
        operatorNames = [ "I", "C", "Cdag", "N" ]
        opCacheVec = [ [OpInfo(ITensors.Op(name, n), sites[n]) for name in operatorNames] for n in eachindex(sites)  ]
        return MPO_new(os, sites ; basis_op_cache_vec = opCacheVec)
    end

    path = "nm_$(nm)_tmp/"
    mkpath(path)
\end{lstlisting}
After specifying the terms for the Hamiltonian \lstinline|tms_hmt| and the QNDiags \lstinline|qnd| like in the previous section, the sites and Hamiltonian MPO can be generated with the function \lstinline|GetMPOSites|.
\begin{lstlisting}
    hmt, sites = GetMPOSites("hmt", tms_hmt, qnd ; path, mpo_method = MyMPO)
\end{lstlisting}
To generate the initial MPS that respects the $\BZ_2$ symmetry, we can use a direct product state.
\begin{lstlisting}
    cfi_p = [ [1, 0][f] for m = 1 : nm for f = 1 : nf ]
    sti_p = MPS(sites, string.(cfi_p))
    cfi_m = [ m == 1 ? [0, 1][f] : [1, 0][f] for m = 1 : nm for f = 1 : nf ]
    sti_m = MPS(sites, string.(cfi_m))
\end{lstlisting}
The lowest eigenenergies and the eigenstate MPSs $|0\rangle,|\sigma\rangle,|\epsilon\rangle$ can be easily generated by the function \lstinline|EasySweep|.
\begin{lstlisting}
    E0, st0 = EasySweep("0", hmt, sti_p ; path)
    Ee, ste = EasySweep("e", hmt, sti_p ; path, proj = ["0"])
    Es, sts = EasySweep("s", hmt, sti_m ; path)
\end{lstlisting}
To measure the angular momentum $L^2$ of the ground state, we generate the MPO for $L^2$.
\begin{lstlisting}
    tms_l2 = GetL2Terms(nm, 2)
    l2 = GetMPO("l2", tms_l2, sites ; path, mpo_method = MyMPO)
    val_l20 = inner(st0', l2, st0)
\end{lstlisting}
Similarly, to measure the OPE coefficient $f_{\sigma\sigma\epsilon}=\langle \sigma|n^x_{00}|\epsilon\rangle/\langle \sigma|n^x_{00}|0\rangle$
\begin{lstlisting}
    obs_nx = GetDensityObs(nm, 2, sgx)
    tms_nx00 = SimplifyTerms(GetComponent(obs_nx, 0.0, 0.0))
    nx00 = GetMPO("nx00", tms_nx00, sites ; path, mpo_method = MyMPO)
    f_sse = abs(inner(sts', nx00, ste) / inner(sts', nx00, st0))
\end{lstlisting}

\section{Practical Examples}
\label{sec:examples}

FuzzifiED can help reproduce almost all the ED and DMRG results in the fuzzy-sphere works. We offer a series of such examples to help the users get started with the package. The code can be found in the \lstinline|examples| directory of the source code repository.

\newcommand{\linkexample}[1]{\href{https://github.com/FuzzifiED/FuzzifiED.jl/blob/main/examples/#1}{\lstinline|#1|}}

\begin{enumerate}
    \item \linkexample{ising\_spectrum.jl} calculates the spectrum of the 3D Ising model on the fuzzy sphere at $N_m = 12$. For each $(\cP,\cZ,\cR)$ sector, 20 states are calculated. This example reproduces Table~I and Figure~4 in Ref.~\cite{Zhu2022}.
    \item \linkexample{ising\_phase\_diagram.jl} calculates the phase diagram of the fuzzy sphere Ising model by calculating the order parameter $\langle M^2\rangle$. This example reproduces Figure~3 in Ref.~\cite{Zhu2022}.
    \item \linkexample{ising\_ope.jl} calculates various OPE coefficients at $N_m = 12$ by taking overlaps between CFT states and density operators and composite. This example reproduces Figure~2 and Table~I in Ref.~\cite{Hu2023Mar}.
    \item \linkexample{ising\_correlator.jl} calculates the $\sigma\sigma$ 2-pt function on sphere and the $\sigma\sigma\sigma\sigma$ 4-pt function of 0 and $\infty$ and two points on sphere. This example reproduces Figures 1c and 2a in Ref.~\cite{Han2023Jun}.
    \item \linkexample{ising\_optimisation.jl} defines a cost function as the square sum of the deviations of descendants and stress tensor to evaluate the conformal symmetry for the Ising model and minimises this cost function to find the best parameter.
    \item \linkexample{ising\_full\_spectrum.jl} calculates the full spectrum of the 3D Ising model on the fuzzy sphere at $N_m = 10$ for sector $(\cP,\cZ,\cR) = (+,+,+)$.
    \item \linkexample{ising\_space\_entangle.jl} calculates the entanglement entropy of the Ising ground state along the real space cut of $\theta = 0.500\pi$ and $0.499\pi$ respectively, and use these two data to extract finite-size $F$-function without subtracting the IQHE contribution. This example reproduces Figure 3 in Ref.~\cite{Hu2024}.
    \item \linkexample{ising\_orbital\_entangle.jl} calculates the entanglement entropy of the Ising ground state along the orbital space cut at $m = 0$, and also the entanglement spectrum in the half-filled $l_z = 0,1$ and both $\BZ_2$ sectors.
    \item \linkexample{ising\_generator.jl} examines the quality of conformal symmetry at $N_m = 12$ by calculating the matrix elements of conformal generators $P^z + K^z$ and compare the states $(P^z + K^z)|\Phi\rangle$ with the CFT expectations. This example reproduces Figure~7 in Ref.~\cite{Fardelli2024}.
    \item \linkexample{defect\_spectrum.jl} calculates the spectrum of pinning-field defect in the 3D Ising model in $l_z=0,\cP=\pm 1$ and $l_z = 1$ sectors, calibrated by bulk $T$. This example reproduces Table~I in Ref.~\cite{Hu2023Aug}.
    \item \linkexample{defect\_correlator.jl} calculates the 1-pt function $\sigma$ and 2-pt function $\sigma\hat{\phi}$ of pinning-field defect in the 3D Ising model. The normalisation of the correlators requires extra bulk data. This example reproduces Figure~4 in Ref.~\cite{Hu2023Aug}.
    \item \linkexample{defect\_changing.jl} calculates the spectrum of the defect-creation and changing operators of the pinning-field defect in the 3D Ising model. This example reproduces Table~2 and Figure~5 in Ref.~\cite{Zhou2024Jan}.
    \item \linkexample{defect\_overlap.jl} calculates the $g$-function of pinning-field defect in the 3D Ising model using the ovelaps between the bulk, defect ground state and the lowest defect-creation state. This example reproduces Figure~6 in Ref.~\cite{Zhou2024Jan}.
    \item \linkexample{cusp\_dim.jl} calculates the scaling dimension of the cusp of the pinning-field defect in the 3D Ising model as a function of the angle $\theta$. This example reproduces Table~2, upper panel in Ref.~\cite{Cuomo2024}.
    \item \linkexample{surface\_ordinary\_spectrum.jl} calculates the spectrum of ordinary surface CFT in the 3D Ising model calibrated by surface displacement operator $D$ in the orbital boundary scheme. This example reproduces Figures 3 and 4 in Ref.~\cite{Zhou2024Jul}.
    \item \linkexample{surface\_normal\_spectrum.jl} calculates the spectrum of normal surface CFT in the 3D Ising model calibrated by surface displacement operator $D$ in the orbital boundary scheme. This example reproduces Figure~5 in Ref.~\cite{Zhou2024Jul}.
    \item \linkexample{hsb\_2l\_spectrum.jl} calculates the spectrum of $\rO(3)$ Wilson-Fisher CFT using the bilayer Heisenberg model. This example reproduces Table~I and Figure~2 in Ref.~\cite{Han2023Dec}.
    \item \linkexample{so5\_spectrum.jl} calculates the spectrum of $\SO(5)$ DQCP on the fuzzy sphere. This example reproduces Table~II in Ref.~\cite{Zhou2023}.
    \item \linkexample{sp3\_spectrum.jl} calculates the spectrum of $\Sp(3)$ CFT on the fuzzy sphere. This example reproduces Table~I in Ref.~\cite{Zhou2024Oct}.
    \item \linkexample{ising\_frac\_fermion.jl} calculates the spectrum of the 3D Ising model on the fuzzy sphere for fermions at fractional filling $\nu = 1/3$. This example reproduces Figure~10 in Ref.~\cite{Voinea2024}.
    \item \linkexample{ising\_frac\_boson.jl} calculates the spectrum of the 3D Ising model on the fuzzy sphere for bosons at fractional filling $\nu = 1/2$ with the module Fuzzifino. This example reproduces Figure 12a, b in Ref.~\cite{Voinea2024}.
    \item \linkexample{potts\_spectrum.jl} calculates the spectrum of 3D Potts model on the fuzzy sphere. This example reproduces Table I and Figure 4 in Ref.~\cite{Yang2025Jan}.
    \item \linkexample{circle\_ising.jl} calculates the spectrum of 2D Ising CFT on a fuzzy thin torus. This example reproduces Figure 4 and Tables I--III in Ref.~\cite{Han2025}.
    \item \linkexample{yang\_lee\_spectrum.jl} calculates the spectrum of 3D non-unitary Yang-Lee CFT on the fuzzy sphere. This example reproduces Figure 11 in Ref.~\cite{ArguelloCruz2025}.
    \item \linkexample{free\_scalar\_spectrum.jl} calculates the spectrum of free real scalar. This example reproduces Figure 2 in Ref.~\cite{He2025Jun}.
    \item \linkexample{free\_scalar\_correlator.jl} calculates the correlator of $n^i$ for $\phi$, $\phi^2$ and $\pi$ of free real scalar. This example reproduces Figure 3 in Ref.~\cite{He2025Jun}.
    \item \linkexample{o4\_dqcp\_spectrum.jl} calculates the spectrum of $\rO(4)$ DQCP on the fuzzy sphere. This example reproduces Figure 4 and Table I in Ref.~\cite{Yang2025Jul}.
    \item \linkexample{su2\_1\_scal\_spectrum.jl} calculates the spectrum of the $\rU(1)_2$ coupled to a complex scalar Chern-Simons matter CFT on the fuzzy sphere. This example reproduces Figures 2d and 3 in Ref.~\cite{Zhou2025Jul}.
    \item \linkexample{free\_majorana\_spectrum.jl} calculates the spectrum of free Majorana fermion. This example reproduces Figure 5 in Ref.~\cite{Zhou2025Sep}.
    \item \linkexample{free\_majorana\_correlator.jl} calculates the 2-pt correlator of free Majorana fermion. This example reproduces Figure 7 in Ref.~\cite{Zhou2025Sep}.
    \item \linkexample{bpf\_220\_spectrum.jl} calculates the spectrum of the transition between bosonic Pfaffian and Halperin 220 described by gauged Majorana fermion CFT. This example reproduces Figure 4a in Ref.~\cite{Voinea2025}.
    \item \linkexample{o3\_wf\_spectrum.jl} calculates the spectrum of $\rO(3)$ Wilson-Fisher CFT. This example takes the model from Ref.~\cite{Dey2025} and reproduces partly Tables I and II, Figures 1 and 2.
    \item \linkexample{o2\_wf\_spectrum.jl} calculates the spectrum of $\rO(2)$ Wilson-Fisher CFT. This example reproduces Figure 3, upper panel in Ref.~\cite{Guo2025}.
    \item \linkexample{o2\_free\_spectrum.jl} calculates the spectrum of $\rO(2)$ free scalar CFT. This example reproduces Figure 4, upper-left panel in Ref.~\cite{Guo2025}.
    \item \linkexample{o4\_wf\_spectrum.jl} calculates the spectrum of $\rO(4)$ Wilson-Fisher CFT and demonstrates the general set-up for O(N) vector models. This example reproduces Figure 3, lower panel in Ref.~\cite{Guo2025}.
    \item \linkexample{ising\_spectrum\_krylov.jl} calculates the spectrum of the 3D Ising model on the fuzzy sphere by calling the \lstinline|eigsolve| function in \lstinline|KrylovKit.jl| instead of Arpack.
    \item \linkexample{ising\_spectrum\_cuda.jl} calculates the spectrum of the 3D Ising model on the fuzzy sphere for one sector by performing the sparse matrix multiplication on CUDA.
\end{enumerate}

\clearpage
\acknowledgments

We extend our deepest gratitude to Yin-Chen He, who provided meticulous testing, unwavering support, and invaluable guidance throughout the development of the package. We also thank Andrew Liam Fitzpatrick, Wenhan Guo, Liangdong Hu, Xiang Li, Christian Voinea, Kay J\"org Wiese, Shuai Yang, and Wei Zhu for their suggestions and contributions to the package.

Z.~Z.~acknowledges support from the Natural Sciences and Engineering Research Council of Canada (NSERC) through Discovery Grants. Research at Perimeter Institute is supported in part by the Government of Canada through the Department of Innovation, Science and Industry Canada and by the Province of Ontario through the Ministry of Colleges and Universities.

\cleardoublepage
\appendix
\addcontentsline{toc}{part}{Appendix}

\section{Data Structures in Exact Diagonalisation}
\label{app:data}

In this appendix, we describe several data structures used in the exact diagonalisation (ED) process, \textit{viz.}~the Lin table for the reverse look-up process that returns the index from the binary string, the compressed sparse column (CSC) format to store the sparse matrices, and the indexing of boson configurations.

\subsection{Construction of Lin Table}
\label{app:data_lin}

\newcommand{\conf}{\mathit{conf}}
\newcommand{\lid}{\mathit{lid}}
\newcommand{\rid}{\mathit{rid}}
\newcommand{\colptr}{\mathit{colptr}}
\newcommand{\rowid}{\mathit{rowid}}
\newcommand{\elval}{\mathit{elval}}
\newcommand{\id}{\mathit{id}}

In this appendix, we describe the construction of the Lin table used for a reverse look-up process that returns the index $i$ from the binary string $\conf_i$ mentioned in Section~\ref{sec:ed_confs}.

Each binary string $\conf_i$ is divided into a left part $l(\conf_i)$ from the $N_{or}$-th bit to the $(N_o-1)$-th bit and a right part $r(\conf_i)$ from the $0$-th bit to the $(N_{or}-1)$-th bit from the right. The cut $N_{or}$ is typically taken as $N_o/2$. Two arrays $\lid$ and $\rid$ with length $2^{N_o-N_{or}}$ and $2^{N_{or}}$ are constructed, such that
\begin{equation}
    i=\lid_{l(\conf_i)}+\rid_{r(\conf_i)}.
\end{equation}

\begin{table}[hbtp]
    \centering
    \begin{tabular}{ccc|cccccc|cccccc}
        \hline\hline
        $c$&$\lid_c$&$\rid_c$&$i$&$\conf_i$&$l_i$&$r_i$&$\lid_{l_i}$&$\rid_{r_i}$&$i$&$\conf_i$&$l_i$&$r_i$&$\lid_{l_i}$&$\rid_{r_i}$\\
        \hline
        0& 1&0& 1&000111&0&7&1&0&11&100011&4&3&11&0\\
        1& 2&0& 2&001011&1&3&2&0&12&100101&4&5&11&1\\
        2& 5&1& 3&001101&1&5&2&1&13&100110&4&6&11&2\\
        3& 8&0& 4&001110&1&6&2&2&14&101001&5&1&14&0\\
        4&11&2& 5&010011&2&3&5&0&15&101010&5&2&14&1\\
        5&14&1& 6&010101&2&5&5&1&16&101100&5&4&14&2\\
        6&17&2& 7&010110&2&6&5&2&17&110001&6&1&17&0\\
        7&20&0& 8&011001&3&1&8&0&18&110010&6&2&17&1\\
         &  & & 9&011010&3&2&8&1&19&110100&6&4&17&2\\
         &  & &10&011100&3&4&8&2&20&111000&7&0&20&0\\
        \hline\hline
    \end{tabular}
    \caption{An example of explicit construction of the Lin table in the configurations on 6 sites filled with 3 particles. Here we use a shorthand notation $l_i=l(\conf_i)$ and $r_i=r(\conf_i)$}
    \label{tbl:lin_eg}
\end{table}

As a simple example, consider the configurations on 6 sites filled with 3 particles. The construction is listed in Table~\ref{tbl:lin_eg}.

\FloatBarrier

\subsection{Compressed Sparse Column (CSC) Sparse Matrix}
\label{app:data_csc}

In this appendix, we describe the storage of sparse matrices in the format of compressed sparse column (CSC) mentioned in Section~\ref{sec:ed_opmat}.

In the CSC format, the matrix is stored in three arrays. We first index elements from $1$ to the number of elements $N_\el$ in the ascending order from the column to column, and in each column from row to row.
\begin{enumerate}
    \item The array $\colptr$ records the column index of each element. It is of length $N+1$ where $N$ is the number of columns. The first $N$ elements record the index of the first element in each column~; the last element records $N_\el+1$, so that the elements that belong to the $i$-th column have indices from $\colptr_i$ to $\colptr_{i+1}-1$.
    \item The array $\rowid$ records the row index of each element and is of length $N_\el$.
    \item The array $\elval$ records the value of each element and is of length $N_\el$.
\end{enumerate}
We give a simple example in Table~\ref{tbl:csc_eg}.

\begin{table}[htbp]
    \centering
    \begin{equation*}
        M=\begin{pmatrix}
            10&40&0&0\\
            20&0&70&80\\
            0&0&0&90\\
            30&50&0&100\\
            0&60&0&0\\
        \end{pmatrix}
    \end{equation*}
    \begin{tabular}{c|cccccccccc}
        \hline\hline
        $i$&1&2&3&4&5&6&7&8&9&10\\
        \hline
        $\colptr_i$&1&4&7&8&11\\
        $\rowid_i$&1&2&4&1&4&5&2&2&3&4\\
        $\elval_i$&10&20&30&40&50&60&70&80&90&100\\
        \hline\hline
    \end{tabular}
    \caption{An example of the storing a $4\times 5$ matrix into the CSC format. }
    \label{tbl:csc_eg}
\end{table}

\FloatBarrier

\subsection{Indexing the Boson Configurations}
\label{app:data_boson}

In this appendix, we describe the indexing of the bosonic configurations with $N_o$ sites and a maximal total occupation $N_{\max}$ mentioned in Section~\ref{sec:ed_fuzzifino}. This is equivalent to the common procedure indexing $N_o+1$ sites with a fixed occupation $N_e=N_{\max}$. The boson configurations are indexed in ascending order from the last site to the first site. \textit{E.~g.}, for 3 sites with maximal occupation 2, the configurations for $n_3n_2n_1$ are 000, 001, 002, 010, 011, 020, 100, 101, 110, 200 numbered from 1 to 10.

The total number of configurations is the number of non-negative integer solutions to the Diophantine equation $n_1+n_2+\dots+n_{N_o}+n_{N_o+1}=N_{\max}$, which is
\begin{equation}
    N_\textrm{cf}(N_o,N_{\max})=\frac{(N_o+N_{\max})!}{N_o!N_{\max}!}
\end{equation}
The index of the
\begin{align}
    \id(n_{N_o}n_{N_o-1}\dots n_1;N_{\max})&=N_\textrm{cf}(0)+N_\textrm{cf}(1)+\dots+N_\textrm{cf}(n_{N_o}-1)\nonumber\\
    &+N_\textrm{cf}(n_{N_o}1)+N_\textrm{cf}(n_{N_o}2)+\dots+N_\textrm{cf}(n_{N_o}(n_{N_o-1}-1))\nonumber\\
    &+\dots\nonumber\\
    &+N_\textrm{cf}(n_{N_o}\dots n_20)+N_\textrm{cf}(n_{N_o}\dots n_21)+\dots+N_\textrm{cf}(n_{N_o}\dots n_2n_1),
\end{align}
where $N_\textrm{cf}(n_{N_o}n_{N_o-1}\dots n_{o})$ is the number of configurations that begin with $n_{N_o}n_{N_o-1}\dots n_{o}$
\begin{equation}
    N_\textrm{cf}(n_{N_o}n_{N_o-1}\dots n_{o})=\frac{[(o-1)+(N_{\max}-n_{N_o}-\dots-n_o)]!}{(o-1)!(N_{\max}-n_{N_o}-\dots-n_o)!}.
\end{equation}
Hence, the index of each configuration is
\begin{equation}
    \id(n_{N_o}n_{N_o-1}\dots n_1;N_{\max})=\sum_{o=1}^{N_o}\sum_{i=0}^{n_o}\frac{\left[(o-1)+\left(N_{\max}-\sum_{o'=o}^{N_o}n_{o'}\right)\right]!}{(o-1)!\left(N_{\max}-\sum_{o'=o}^{N_o}n_{o'}\right)!}
\end{equation}

The reverse look-up is realised by a similar Lin-table, with the left part and right part taken as
\begin{equation}
    l(\{n_o\})=\id(n_{N_o}n_{N_o-1}\dots n_{N_{or}+1};N_{\max})\quad\textrm{and}\quad r(\{n_o\})=\id(n_{N_{or}}n_{N_{or}-1}\dots n_{1};N_{\max}).
\end{equation}

\FloatBarrier

\section{Landau Levels and Pseudo-Potentials on the Torus}
\label{app:manifolds}

In this appendix, we describe the construction of Landau levels and pseudo-potentials on the torus~\cite{Yoshioka1983Torus,Haldane1985Torus,Haldane1987Torus}, mentioned in Sections~\ref{sec:construct_pspot} and \ref{sec:ed_manifolds}.

\subsection{Lowest Landau Level}

The torus is nothing more than a $l_x\times l_y$ rectangle with periodic boundary conditions, \textit{i.~e.}~$x\sim x+l_x$ and $y\sim y+l_y$. We adopt the Landau gauge $A_x=0,A_y=xB$ and set the magnetic length as the unit length $l_B=1$. In this way, the number of orbitals is $N_m=l_xl_y/2\pi$. The lowest Landau level wave-functions are labelled by the momenta along $y$ direction $m$
\begin{equation}
    \phi_m(x,y)=\textrm{const.}\times\sum_{n=-\infty}^\infty\exp\left[-\tfrac{1}{2}(x-X_m+nl_x)^2+iX_m(y+kl_y)\right],
\end{equation}
where $m=1,\dots,N_m$, and
\begin{equation*}
    X_m=2\pi m/l_y.
\end{equation*}
Each orbital is localised in the vicinity of a stripe $x=X_m$. We can then similarly project the fermion operators onto the lowest Landau level
\begin{equation}
    \psi_{f}^\dagger(\br)=\phi_m(\br)c^\dagger_{mf},\quad\br=(x,y)
\end{equation}

\subsection{Many-Body Interaction}

The simplest many-body interaction term is the density-density interaction with a potential function. Omitting the flavour index, it can be expressed as
\begin{equation}
    H_\Int=\frac{1}{N_m^2}\int\rd^2\br_1\rd^2\br_2\,U(\br_1-\br_2)n(\br_1)n(\br_2),\qquad n(\br)=\psi^\dagger(\br)\psi(\br)
\end{equation}
The interaction potential can be expanded into Fourier modes
\begin{equation}
    U(\br)=\sum_\bq U_\bq e^{i\bq\cdot\br},
\end{equation}
where the allowed momenta are labelled by two integers $s$ and $t$
\begin{equation}
    \bq=(2\pi s/l_x,2\pi t/l_y)
\end{equation}
After performing the spatial integrals, the Hamiltonian is expressed in the orbital space as\footnote{In many literatures, the summation excludes the origin point $\bq=(0,0)$, but here the interaction of background charge in fact plays certain role.}
\begin{equation}
    H_\Int=\sum_{m_1m_2m_3m_4}\sum_\bq\frac{U_\bq}{2N_m}e^{iq_x(X_{m_1}-X_{m_3})}e^{-q^2/2}\delta'_{t,m_1-m_4}\delta'_{m_1+m_2,m_3+m_4}c^\dagger_{m_1}c^\dagger_{m_2}c_{m_3}c_{m_4},
\end{equation}
where the Kronecker delta is defined in the sense of mod $N_m$
\begin{equation*}
    \delta'_{m,n}=\left\{\begin{aligned}
        &1,&&m=n\mod N_m\\
        &0,&&\textrm{otherwise.}
    \end{aligned}\right.
\end{equation*}
The interactions can alternatively be organised into pseudo-potentials $U_l$, describing the energy cost of a pair of electrons with relative angular momentum $l$.
\begin{equation}
    U_l=\int\frac{\rd^2\bq}{2\pi}V_\bq e^{-q^2}L_l(q^2),
\end{equation}
where $L_l$ is the Laguerre polynomial. Finally, the interaction Hamiltonian reads
\begin{align}
    H_\Int&=\sum_{lm_1m_2m_3m_4}U_lC_{m_1m_2m_3m_4}^l c^\dagger_{m_1}c^\dagger_{m_2}c_{m_3}c_{m_4}\nonumber\\
    C_{m_1m_2m_3m_4}^l&=\frac{1}{N_m}\sum_\textbf{q}e^{iq_x(X_{m_1}-X_{m_3})}e^{-q^2/2}L_l(q^2)\delta'_{t,m_1-m_4}\delta'_{m_1+m_2,m_3+m_4}.
\end{align}

\section{Tutorial Code}
\label{app:code}

In Sections~\ref{sec:ed} and \ref{sec:dmrg}, we demonstrate the usage of FuzzifiED interfaces with an example that
\begin{enumerate}
    \item calculates the lowest eigenstates in the symmetry sector $L^z=0$ and $(\cP,\cZ,\cR)=(+,+,+)$ (for DMRG, $\cZ=+$ only),
    \item measures their total angular momenta (for DMRG, the ground state only), and
    \item calcultes the OPE coefficient $f_{\sigma\sigma\epsilon}=\langle \sigma|n^z_{00}|\epsilon\rangle/\langle \sigma|n^z_{00}|0\rangle$.
\end{enumerate}
In this appendix, we collect the full version of the codes, including
\begin{enumerate}
    \item the ED code that uses only the core functions,
    \item the ED code that uses the built-in models,
    \item the DMRG code that converts the format into ITensor, and
    \item the DMRG code that uses the EasySweep extension.
\end{enumerate}

\subsection{ED Using Core Functions}
\label{app:code_ed1}

\begin{lstlisting}[numbers=left]
    using FuzzifiED
    using WignerSymbols

    let

    nf = 2
    nm = 12
    no = nf * nm
    qnd = [
        QNDiag(fill(1, no)),
        QNDiag([ 2 * m - nm - 1 for m = 1 : nm for f = 1 : nf ])
    ]
    cfs = Confs(no, [nm, 0], qnd)

    qnf = [
        QNOffd([ (m - 1) * nf + [2, 1][f] for m = 1 : nm for f = 1 : nf ], true,
            ComplexF64[ [-1, 1][f] for m = 1 : nm for f = 1 : nf ]),
        QNOffd([ (m - 1) * nf + [2, 1][f] for m = 1 : nm for f = 1 : nf ]),
        QNOffd([ (nm - m) * nf + f for m = 1 : nm for f = 1 : nf],
            ComplexF64[ iseven(m) ? 1 : -1 for m = 1 : nm for f = 1 : nf ])
    ]
    bs = Basis(cfs, [1, 1, 1], qnf)

    ps_pot = [ 4.75, 1.0 ] * 2.
    h = 3.16
    tms_hmt = Term[]
    m = zeros(Int64, 4)
    for m[1] = 0 : nm - 1, m[2] = 0 : nm - 1, m[3] = 0 : nm - 1
        m[4] = m[1] + m[2] - m[3]
        (m[4] < 0 || m[4] >= nm) && continue
        f = [0, 1, 1, 0]
        o = m .* nf .+ f .+ 1
        mr = m .- s

        val = ComplexF64(0)
        for l in eachindex(ps_pot)
            (abs(mr[1] + mr[2]) > nm - l || abs(mr[3] + mr[4]) > nm - l) && break
            val += ps_pot[l] * (2 * nm - 2 * l + 1) * wigner3j(s, s, nm - l, mr[1], mr[2], -mr[1] - mr[2]) * wigner3j(s, s, nm - l, mr[4], mr[3], -mr[3] - mr[4])
        end
        tms_hmt += Terms(val, [1, o[1], 1, o[2], 0, o[3], 0, o[4]])
    end
    for m = 0 : nm - 1
        o = m * nf .+ [1, 2]
        tms_hmt += Terms(-h, [1, o[1], 0, o[2]])
        tms_hmt += Terms(-h, [1, o[2], 0, o[1]])
    end

    hmt = Operator(bs, tms_hmt)
    hmt_mat = OpMat(hmt)
    enrg, st = GetEigensystem(hmt_mat, 10)

    tms_lz = [ Term(m - s - 1, [1, (m - 1) * nf + f, 0, (m - 1) * nf + f]) for m = 1 : nm for f = 1 : nf ]
    tms_lp = [ Term(sqrt((nm - m) * m), [1, m * nf + f, 0, (m - 1) * nf + f]) for m = 1 : nm - 1 for f = 1 : nf ]
    tms_lm = tms_lp'
    tms_l2 = SimplifyTerms(tms_lz * tms_lz - tms_lz + tms_lp * tms_lm)
    l2 = Operator(bs, tms_l2)
    l2_mat = OpMat(l2)
    l2_val = [ st[:, i]' * l2_mat * st[:, i] for i in eachindex(enrg)]

    bs_m = Basis(cfs, [ 1, -1, 1 ], qnf)
    hmt = Operator(bs_m, bs_m, tms_hmt ; red_q = 1, sym_q = 1)
    hmt_mat = OpMat(hmt)
    enrg_m, st_m = GetEigensystem(hmt_mat, 10)
    st0 = st[:, 1]
    ste = st[:, 2]
    sts = st_m[:, 1]

    tms_nz00 = Term[]
    for m = 0 : nm - 1
        o = m * nf .+ [1, 2]
        tms_nz00 += Terms( 1 / nm, [1, o[1], 0, o[1]])
        tms_nz00 += Terms(-1 / nm, [1, o[2], 0, o[2]])
    end
    nz00 = Operator(bs, bs_m, tms_nz00 ; red_q = 1)
    f_sse = abs((sts' * nz00 * ste) / (sts' * nz00 * st0))

    end
\end{lstlisting}

\subsection{ED Using Built-in Models}
\label{app:code_ed2}

\begin{lstlisting}[numbers=left]
    using FuzzifiED
    sg1 = [  1  0 ;  0  0 ]
    sg2 = [  0  0 ;  0  1 ]
    sgx = [  0  1 ;  1  0 ]
    sgz = [  1  0 ;  0 -1 ]

    nm = 12
    qnd = [
        GetNeQNDiag(2 * nm),
        GetLz2QNDiag(nm, 2)
    ]
    cfs = Confs(2 * nm, [nm, 0], qnd)

    qnf = [
        GetParityQNOffd(nm, 2, [2, 1], [-1, 1]),
        GetFlavPermQNOffd(nm, 2, [2, 1]),
        GetRotyQNOffd(nm, 2)
    ]
    bs = Basis(cfs, [1, 1, 1], qnf)

    tms_hmt = SimplifyTerms(
        GetDenIntTerms(nm, nf, 2 .* [4.75, 1.0], sg1, sg2) -
        3.16 * GetPolTerms(nm, 2, sgx)
    )

    hmt = Operator(bs, tms_hmt)
    hmt_mat = OpMat(hmt)
    enrg, st = GetEigensystem(hmt_mat, 10)

    tms_l2 = GetL2Terms(nm, 2)
    l2 = Operator(bs, tms_l2)
    l2_mat = OpMat(l2)
    l2_val = [ st[:, i]' * l2_mat * st[:, i] for i in eachindex(enrg)]

    bs_m = Basis(cfs, [1, -1, 1], qnf)
    hmt = Operator(bs_m, bs_m, tms_hmt ; red_q = 1, sym_q = 1)
    hmt_mat = OpMat(hmt)
    enrg_m, st_m = GetEigensystem(hmt_mat, 10)
    st0 = st[:, 1]
    ste = st[:, 2]
    sts = st_m[:, 1]

    obs_nz = GetDensityObs(nm, 2, sgz)
    tms_nz00 = SimplifyTerms(GetComponent(obs_nz, 0.0, 0.0))
    nz00 = Operator(bs, bs_m, tms_nz00 ; red_q = 1)
    f_sse = abs((sts' * nz00 * ste) / (sts' * nz00 * st0))
\end{lstlisting}

\subsection{DMRG Using Format Conversion into ITensor}

\begin{lstlisting}[numbers=left]
    using FuzzifiED
    using ITensors, ITensorMPS
    FuzzifiED.ElementType = Float64
    sgx = [  0  1 ;  1  0 ]
    sgz = [  1  0 ;  0 -1 ]

    nm = 12
    nf = 2
    no = nm * nf

    sites = GetSites([
        GetNeQNDiag(nm * nf),
        GetLz2QNDiag(nm, nf),
        GetZnfChargeQNDiag(nm, nf)
    ])
    ps_pot = [4.75, 1.0] ./ 2
    tms_hmt = SimplifyTerms(
        GetDenIntTerms(nm, nf, ps_pot) -
        GetDenIntTerms(nm, nf, ps_pot, sgx) -
        3.16 * GetPolTerms(nm, nf, sgz)
    )
    os_hmt = OpSum(tms_hmt)
    hmt = MPO(os_hmt, sites)

    cfi_p = [ [1, 0][f] for m = 1 : nm for f = 1 : nf ]
    sti_p = MPS(sites, string.(cfi_p))
    cfi_m = [ m == 1 ? [0, 1][f] : [1, 0][f] for m = 1 : nm for f = 1 : nf ]
    sti_m = MPS(sites, string.(cfi_m))

    E0, st0 = dmrg(hmt, sti_p ;
        nsweeps = 10,
        maxdim = [10,20,50,100,200,500],
        noise = [1E-4,3E-5,1E-5,3E-6,1E-6,3E-7],
        cutoff = [1E-8])
    Ee, ste = dmrg(hmt, [st0], sti_p ;
        nsweeps = 10,
        maxdim = [10,20,50,100,200,500],
        noise = [1E-4,3E-5,1E-5,3E-6,1E-6,3E-7],
        cutoff = [1E-8],
        weight = 100)
    Es, sts = dmrg(hmt, sti_m ;
        nsweeps = 10,
        maxdim = [10,20,50,100,200,500],
        noise = [1E-4,3E-5,1E-5,3E-6,1E-6,3E-7],
        cutoff = [1E-8])

    tms_l2 = GetL2Terms(nm, 2)
    l2 = MPO(OpSum(tms_l2), sites)
    val_l20 = inner(st0', l2, st0)

    obs_nx = GetDensityObs(nm, 2, sgx)
    tms_nx00 = SimplifyTerms(GetComponent(obs_nx, 0.0, 0.0))
    nx00 = MPO(OpSum(tms_nx00), sites)
    f_sse = abs(inner(sts', nx00, ste) / inner(sts', nx00, st0))
\end{lstlisting}

\subsection{DMRG with Easy-Sweep}

\begin{lstlisting}[numbers=left]
    using FuzzifiED
    using ITensors, ITensorMPS, HDF5
    using ITensorMPOConstruction
    sgx = [  0  1 ;  1  0 ]
    sgz = [  1  0 ;  0 -1 ]

    function MyMPO(os, sites)
        operatorNames = [ "I", "C", "Cdag", "N" ]
        opCacheVec = [ [OpInfo(ITensors.Op(name, n), sites[n]) for name in operatorNames] for n in eachindex(sites)  ]
        return MPO_new(os, sites ; basis_op_cache_vec = opCacheVec)
    end

    nm = 12
    nf = 2
    no = nm * nf

    path = "nm_$(nm)_tmp/"
    mkpath(path)

    ps_pot = [4.75, 1.0] ./ 2
    tms_hmt = SimplifyTerms(
        GetDenIntTerms(nm, nf, ps_pot) -
        GetDenIntTerms(nm, nf, ps_pot, sgx) -
        3.16 * GetPolTerms(nm, 2, sgz)
    )
    qnd = [
        GetNeQNDiag(no),
        GetLz2QNDiag(nm, nf),
        GetZnfChargeQNDiag(nm, nf)
    ]
    hmt, sites = GetMPOSites("hmt", tms_hmt, qnd ; path, mpo_method = MyMPO)

    cfi_p = [ [1, 0][f] for m = 1 : nm for f = 1 : nf ]
    sti_p = MPS(sites, string.(cfi_p))
    cfi_m = [ m == 1 ? [0, 1][f] : [1, 0][f] for m = 1 : nm for f = 1 : nf ]
    sti_m = MPS(sites, string.(cfi_m))

    E0, st0 = EasySweep("0", hmt, sti_p ; path)
    Ee, ste = EasySweep("e", hmt, sti_p ; path, proj = ["0"])
    Es, sts = EasySweep("s", hmt, sti_m ; path)

    tms_l2 = GetL2Terms(nm, 2)
    l2 = GetMPO("l2", tms_l2, sites ; path, mpo_method = MyMPO)
    l2_val = st0' * l2 * stg

    obs_nx = GetDensityObs(nm, 2, sgx)
    tms_nx00 = SimplifyTerms(GetComponent(obs_nx, 0.0, 0.0))
    nx00 = GetMPO("nx00", tms_nx00, sites ; path, mpo_method = MyMPO)
    f_sse = abs(inner(sts', nx00, ste) / inner(sts', nx00, st0))
\end{lstlisting}

\section{Glossary for Interfaces in FuzzifiED}
\label{app:glossary}

\begin{multicols}{3}
    \noindent
    \href{https://docs.fuzzified.world/models/#FuzzifiED.AngModes}{\lstinline|AngModes|}\\
    \href{https://docs.fuzzified.world/core/#FuzzifiED.Basis}{\lstinline|Basis|}\\
    \href{https://docs.fuzzified.world/core/#FuzzifiED.Confs}{\lstinline|Confs|}\\
    \href{https://docs.fuzzified.world/models/#FuzzifiED.ContractMod-Tuple%7BAngModes,%20AngModes,%20Dict%7D}{\lstinline|ContractMod|}\\
    \href{https://docs.fuzzified.world/extension/#CUDA.CUSPARSE.CuSparseMatrixCSC-Tuple{OpMat{ComplexF64}}}{\lstinline|CuSparseMatrixCSC|}\\
    \href{https://docs.fuzzified.world/itensors/#FuzzifiED.EasySweep-Tuple{String,%20MPO,%20MPS}}{\lstinline|EasySweep|}\\
    \href{https://docs.fuzzified.world/core/#FuzzifiED.ElementType}{\lstinline|FuzzifiED.ElementType|}\\
    \href{https://docs.fuzzified.world/models/#FuzzifiED.FilterComponent-Tuple{SphereObs,%20Any}}{\lstinline|FilterComponent|}\\
    \href{https://docs.fuzzified.world/models/#FuzzifiED.FilterL2-Tuple{AngModes,%20Number}}{\lstinline|FilterL2|}\\
    \href{https://docs.fuzzified.world/fuzzifino/#FuzzifiED.Fuzzifino.GetBosonC2STerms}{\lstinline|GetBosonC2STerms|}\\
    \href{https://docs.fuzzified.world/fuzzifino/#FuzzifiED.Fuzzifino.GetBosonDenIntSTerms}{\lstinline|GetBosonDenIntSTerms|}\\
    \href{https://docs.fuzzified.world/fuzzifino/#FuzzifiED.Fuzzifino.GetBosonFlavSQNDiag}{\lstinline|GetBosonFlavSQNDiag|}\\
    \href{https://docs.fuzzified.world/fuzzifino/#FuzzifiED.Fuzzifino.GetBosonFlavPermSQNOffd}{\lstinline|GetBosonFlavPermSQNOffd|}\\
    \href{https://docs.fuzzified.world/fuzzifino/#FuzzifiED.Fuzzifino.GetBosonL2STerms}{\lstinline|GetBosonL2STerms|}\\
    \href{https://docs.fuzzified.world/fuzzifino/#FuzzifiED.Fuzzifino.GetBosonLpLzSTerms}{\lstinline|GetBosonLpLzSTerms|}\\
    \href{https://docs.fuzzified.world/fuzzifino/#FuzzifiED.Fuzzifino.GetBosonLz2SQNDiag}{\lstinline|GetBosonLz2SQNDiag|}\\
    \href{https://docs.fuzzified.world/fuzzifino/#FuzzifiED.Fuzzifino.GetBosonPairIntSTerms}{\lstinline|GetBosonPairIntSTerms|}\\
    \href{https://docs.fuzzified.world/fuzzifino/#FuzzifiED.Fuzzifino.GetBosonPolSTerms}{\lstinline|GetBosonPolSTerms|}\\
    \href{https://docs.fuzzified.world/fuzzifino/#FuzzifiED.Fuzzifino.GetBosonRotySQNOffd}{\lstinline|GetBosonRotySQNOffd|}\\
    \href{https://docs.fuzzified.world/models/#FuzzifiED.GetC2Terms-Tuple{Int64,%20Int64,%20Vector{%3C:AbstractMatrix{%3C:Number}}}}{\lstinline|GetC2Terms|}\\
    \href{https://docs.fuzzified.world/models/#FuzzifiED.GetComponent-Tuple{SphereObs,%20Number,%20Number}}{\lstinline|GetComponent|}\\
    \href{https://docs.fuzzified.world/core/#FuzzifiED.GetConfId-Tuple{Confs,%20Int64}}{\lstinline|GetConfId|}\\
    \href{https://docs.fuzzified.world/core/#FuzzifiED.GetConfWeight-Tuple{Basis,%20Union{Vector{ComplexF64},%20Vector{Float64}},%20Int64}}{\lstinline|GetConfWeight|}\\
    \href{https://docs.fuzzified.world/models/#FuzzifiED.GetDenIntTerms}{\lstinline|GetDenIntTerms|}\\
    \href{https://docs.fuzzified.world/models/#FuzzifiED.GetDensityMod-Tuple{Int64,%20Int64,%20Matrix{%3C:Number}}}{\lstinline|GetDensityMod|}\\
    \href{https://docs.fuzzified.world/models/#FuzzifiED.GetDensityObs-Tuple{Int64,%20Int64,%20Matrix{%3C:Number}}}{\lstinline|GetDensityObs|}\\
    \href{https://docs.fuzzified.world/core/#FuzzifiED.GetEigensystem-Tuple{OpMat{ComplexF64},%20Int64}}{\lstinline|GetEigensystem|}\\
    \href{https://docs.fuzzified.world/extension/#FuzzifiED.GetEigensystemCuda-Tuple{OpMat{ComplexF64},%20Int64}}{\lstinline|GetEigensystemCuda|}\\
    \href{https://docs.fuzzified.world/extension/#FuzzifiED.GetEigensystemKrylov-Tuple{OpMat{ComplexF64},%20Int64}}{\lstinline|GetEigensystemKrylov|}\\
    \href{https://docs.fuzzified.world/models/#FuzzifiED.GetElectronMod-Tuple{Int64,%20Int64,%20Int64}}{\lstinline|GetElectronMod|}\\
    \href{https://docs.fuzzified.world/models/#FuzzifiED.GetElectronObs-Tuple{Int64,%20Int64,%20Int64}}{\lstinline|GetElectronObs|}\\
    \href{https://docs.fuzzified.world/core/#FuzzifiED.GetEntSpec-Tuple{Vector{%3C:Number},%20Basis,%20Vector{Vector{Vector{Int64}}},%20Vector{Vector{Vector{Int64}}}}}{\lstinline|GetEntSpec|}\\
    \href{https://docs.fuzzified.world/models/#FuzzifiED.GetFlavPermQNOffd}{\lstinline|GetFlavPermQNOffd|}\\
    \href{https://docs.fuzzified.world/models/#FuzzifiED.GetFlavQNDiag}{\lstinline|GetFlavQNDiag|}\\
    \href{https://docs.fuzzified.world/models/#FuzzifiED.GetIntegral-Tuple{SphereObs}}{\lstinline|GetIntegral|}\\
    \href{https://docs.fuzzified.world/models/#FuzzifiED.GetIntMatrix-Tuple{Int64,%20Vector{%3C:Number}}}{\lstinline|GetIntMatrix|}\\
    \href{https://docs.fuzzified.world/fuzzifino/#FuzzifiED.Fuzzifino.GetL2STerms-Tuple%7BTuple%7BVector%7BSTerm%7D,%20Vector%7BSTerm%7D%7D%7D}{\lstinline|GetL2STerms|}\\
    \href{https://docs.fuzzified.world/models/#FuzzifiED.GetL2Terms-Tuple{Int64,%20Int64}}{\lstinline|GetL2Terms|}\\
    \href{https://docs.fuzzified.world/models/#FuzzifiED.GetLpLzTerms}{\lstinline|GetLpLzTerms|}\\
    \href{https://docs.fuzzified.world/models/#FuzzifiED.GetLz2QNDiag-Tuple%7BInt64,%20Int64%7D}{\lstinline|GetLz2QNDiag|}\\
    \href{https://docs.fuzzified.world/itensors/#FuzzifiED.GetMPO-Tuple{String,%20Union{Sum{Scaled{ComplexF64,%20Prod{Op}}},%20Vector{Term}},%20Vector{%3C:Index}}}{\lstinline|GetMPO|}\\
    \href{https://docs.fuzzified.world/itensors/#FuzzifiED.GetMPOSites-Tuple{String,%20Union{Sum{Scaled{ComplexF64,%20Prod{Op}}},%20Vector{Term}},%20Vector{QNDiag}}}{\lstinline|GetMPOSites|}\\
    \href{https://docs.fuzzified.world/fuzzifino/#FuzzifiED.Fuzzifino.GetNeSQNDiag}{\lstinline|GetNeSQNDiag|}\\
    \href{https://docs.fuzzified.world/models/#FuzzifiED.GetNeQNDiag-Tuple{Int64}}{\lstinline|GetNeQNDiag|}\\
    \href{https://docs.fuzzified.world/models/#FuzzifiED.GetPairingMod-Tuple{Int64,%20Int64,%20Matrix{%3C:Number}}}{\lstinline|GetPairingMod|}\\
    \href{https://docs.fuzzified.world/models/#FuzzifiED.GetPairingObs-Tuple{Int64,%20Int64,%20Matrix{%3C:Number}}}{\lstinline|GetPairingObs|}\\
    \href{https://docs.fuzzified.world/models/#FuzzifiED.GetPairIntTerms}{\lstinline|GetPairIntTerms|}\\
    \href{https://docs.fuzzified.world/models/#FuzzifiED.GetParityQNOffd}{\lstinline|GetParityQNOffd|}\\
    \href{https://docs.fuzzified.world/models/#FuzzifiED.GetPinOrbQNDiag}{\lstinline|GetPinOrbQNDiag|}\\
    \href{https://docs.fuzzified.world/models/#FuzzifiED.GetPointValue-Tuple{SphereObs,%20Float64,%20Float64}}{\lstinline|GetPointValue|}\\
    \href{https://docs.fuzzified.world/models/#FuzzifiED.GetPolTerms-Tuple{Int64,%20Int64,%20Matrix{%3C:Number}}}{\lstinline|GetPolTerms|}\\
    \href{https://docs.fuzzified.world/itensors/#FuzzifiED.GetQNDiags-Tuple{Vector{Index{Vector{Pair{QN,%20Int64}}}}}}{\lstinline|GetQNDiags|}\\
    \href{https://docs.fuzzified.world/models/#FuzzifiED.GetRotyQNOffd-Tuple{Int64,%20Int64}}{\lstinline|GetRotyQNOffd|}\\
    \href{https://docs.fuzzified.world/itensors/#FuzzifiED.GetSites-Tuple{Vector{QNDiag}}}{\lstinline|GetSites|}\\
    \href{https://docs.fuzzified.world/manifolds/#FuzzifiED.FuzzyManifolds.GetTorusDenIntTerms}{\lstinline|GetTorusDenIntTerms|}\\
    \href{https://docs.fuzzified.world/manifolds/#FuzzifiED.FuzzyManifolds.GetTorusIntMatrix-Tuple{Int64,%20Number,%20Vector{%3C:Number}}}{\lstinline|GetTorusIntMatrix|}\\
    \href{https://docs.fuzzified.world/manifolds/#FuzzifiED.FuzzyManifolds.GetTorusLz2QNDiag-Tuple{Int64,%20Int64}}{\lstinline|GetTorusLz2QNDiag|}\\
    \href{https://docs.fuzzified.world/manifolds/#FuzzifiED.FuzzyManifolds.GetTorusTranslQNOffd-Tuple{Int64,%20Int64}}{\lstinline|GetTorusTranslQNOffd|}\\
    \href{https://docs.fuzzified.world/manifolds/#FuzzifiED.FuzzyManifolds.GetTorusPairIntTerms}{\lstinline|GetTorusPairIntTerms|}\\
    \href{https://docs.fuzzified.world/models/#FuzzifiED.GetZnfChargeQNDiag-Tuple{Int64,%20Int64}}{\lstinline|GetZnfChargeQNDiag|}\\
    \href{https://docs.fuzzified.world/models/#FuzzifiED.Laplacian-Tuple{SphereObs}}{\lstinline|Laplacian|}\\
    \href{https://docs.fuzzified.world/core/#FuzzifiED.Libpath}{\lstinline|FuzzifiED.Libpath|}\\
    \href{https://docs.fuzzified.world/fuzzifino/#FuzzifiED.Fuzzifino.Libpathino}{\lstinline|Fuzzifino.Libpathino|}\\
    \href{https://docs.fuzzified.world/extension/#Base.Matrix-Tuple{OpMat}}{\lstinline|Matrix|}\\
    \href{https://docs.fuzzified.world/core/#FuzzifiED.NormalOrder-Tuple{Term}}{\lstinline|NormalOrder|}\\
    \href{https://docs.fuzzified.world/core/#FuzzifiED.NumThreads}{\lstinline|FuzzifiED.NumThreads|}\\
    \href{https://docs.fuzzified.world/models/#FuzzifiED.ObsMomIncr}{\lstinline|FuzzifiED.ObsMomIncr|}\\
    \href{https://docs.fuzzified.world/models/#FuzzifiED.ObsNormRadSq}{\lstinline|FuzzifiED.ObsNormRadSq|}\\
    \href{https://docs.fuzzified.world/core/#FuzzifiED.OpenHelp!-Tuple{}}{\lstinline|FuzzifiED.OpenHelp!|}\\
    \href{https://docs.fuzzified.world/core/#FuzzifiED.Operator}{\lstinline|Operator|}\\
    \href{https://docs.fuzzified.world/core/#FuzzifiED.OpMat}{\lstinline|OpMat|}\\
    \href{https://docs.fuzzified.world/itensors/#ITensors.Ops.OpSum-Tuple{Vector{Term}}}{\lstinline|OpSum|}\\
    \href{https://docs.fuzzified.world/models/#FuzzifiED.PadAngModes}{\lstinline|PadAngModes|}\\
    \href{https://docs.fuzzified.world/core/#FuzzifiED.PadQNDiag}{\lstinline|PadQNDiag|}\\
    \href{https://docs.fuzzified.world/core/#FuzzifiED.PadQNOffd}{\lstinline|PadQNOffd|}\\
    \href{https://docs.fuzzified.world/models/#FuzzifiED.PadSphereObs}{\lstinline|PadSphereObs|}\\
    \href{https://docs.fuzzified.world/fuzzifino/#FuzzifiED.PadSQNDiag}{\lstinline|PadSQNDiag|}\\
    \href{https://docs.fuzzified.world/fuzzifino/#FuzzifiED.PadSQNOffd}{\lstinline|PadSQNOffd|}\\
    \href{https://docs.fuzzified.world/fuzzifino/#FuzzifiED.PadSTerms}{\lstinline|PadSTerms|}\\
    \href{https://docs.fuzzified.world/core/#FuzzifiED.PadTerms}{\lstinline|PadTerms|}\\
    \href{https://docs.fuzzified.world/core/#FuzzifiED.ParticleHole-Tuple{Vector{Term}}}{\lstinline|ParticleHole|}\\
    \href{https://docs.fuzzified.world/core/#FuzzifiED.QNDiag}{\lstinline|QNDiag|}\\
    \href{https://docs.fuzzified.world/core/#FuzzifiED.QNOffd}{\lstinline|QNOffd|}\\
    \href{https://docs.fuzzified.world/core/#FuzzifiED.RelabelOrbs-Tuple%7BVector%7BTerm%7D,%20Dict%7BInt64,%20Int64%7D%7D}{\lstinline|RelabelOrbs|}\\
    \href{https://docs.fuzzified.world/core/#FuzzifiED.RemoveOrbs-Tuple%7BVector%7BTerm%7D,%20Vector%7BInt64%7D%7D}{\lstinline|RemoveOrbs|}\\
    \href{https://docs.fuzzified.world/fuzzifino/#FuzzifiED.Fuzzifino.SAngModes}{\lstinline|SAngModes|}\\
    \href{https://docs.fuzzified.world/fuzzifino/#FuzzifiED.Fuzzifino.SBasis}{\lstinline|SBasis|}\\
    \href{https://docs.fuzzified.world/fuzzifino/#FuzzifiED.Fuzzifino.SConfs}{\lstinline|SConfs|}\\
    \href{https://docs.fuzzified.world/core/#FuzzifiED.SilentStd}{\lstinline|FuzzifiED.SilentStd|}\\
    \href{https://docs.fuzzified.world/core/#FuzzifiED.SimplifyTerms-Tuple{Vector{Term}}}{\lstinline|SimplifyTerms|}\\
    \href{https://docs.fuzzified.world/fuzzifino/#FuzzifiED.Fuzzifino.SOperator}{\lstinline|SOperator|}\\
    \href{https://docs.fuzzified.world/extension/#SparseArrays.SparseMatrixCSC-Tuple{OpMat}}{\lstinline|SparseMatrixCSC|}\\
    \href{https://docs.fuzzified.world/models/#FuzzifiED.SphereObs}{\lstinline|SphereObs|}\\
    \href{https://docs.fuzzified.world/fuzzifino/#FuzzifiED.Fuzzifino.SQNDiag}{\lstinline|SQNDiag|}\\
    \href{https://docs.fuzzified.world/fuzzifino/#FuzzifiED.Fuzzifino.SQNOffd}{\lstinline|SQNOffd|}\\
    \href{https://docs.fuzzified.world/fuzzifino/#FuzzifiED.Fuzzifino.SSphereObs}{\lstinline|SSphereObs|}\\
    \href{https://docs.fuzzified.world/core/#FuzzifiED.StateDecompMat-Tuple{Vector{%3C:Number},%20Basis,%20Basis,%20Basis,%20Vector{%3C:Number},%20Vector{%3C:Number}}}{\lstinline|StateDecompMat|}\\
    \href{https://docs.fuzzified.world/fuzzifino/#FuzzifiED.Fuzzifino.STerm}{\lstinline|STerm|}\\
    \href{https://docs.fuzzified.world/fuzzifino/#FuzzifiED.Fuzzifino.STerms}{\lstinline|STerms|}\\
    \href{https://docs.fuzzified.world/models/#FuzzifiED.StoreComps-Tuple{SphereObs}}{\lstinline|StoreComps|}\\
    \href{https://docs.fuzzified.world/models/#FuzzifiED.StoreComps!-Tuple{SphereObs}}{\lstinline|StoreComps!|}\\
    \href{https://docs.fuzzified.world/fuzzifino/#FuzzifiED.Fuzzifino.STransf}{\lstinline|STransf|}\\
    \href{https://docs.fuzzified.world/itensors/#FuzzifiED.SweepOne-Tuple{String,%20MPO,%20MPS,%20Int64}}{\lstinline|SweepOne|}\\
    \href{https://docs.fuzzified.world/core/#FuzzifiED.Term}{\lstinline|Term|}\\
    \href{https://docs.fuzzified.world/core/#FuzzifiED.Terms}{\lstinline|Terms|}\\
    \href{https://docs.fuzzified.world/core/#FuzzifiED.Transf}{\lstinline|Transf|}\\
    \href{https://docs.fuzzified.world/itensors/#FuzzifiED.TruncateQNDiag-Tuple{Vector{QNDiag}}}{\lstinline|TruncateQNDiag|}\\
    \subsection*{Operations}
    \noindent
    \href{https://docs.fuzzified.world/models/#Base.:+-Tuple{AngModes,%20AngModes}}{\lstinline|amd + amd|}\footnote{Here each symbol represents a specific type~: \lstinline[basicstyle=\ttfamily\scriptsize]|amd| is a \lstinline[basicstyle=\ttfamily\scriptsize]|AngModes|,  \lstinline[basicstyle=\ttfamily\scriptsize]|obs| is a \lstinline[basicstyle=\ttfamily\scriptsize]|SphereObs|, \lstinline[basicstyle=\ttfamily\scriptsize]|op| is a \lstinline[basicstyle=\ttfamily\scriptsize]|Operator|, \lstinline[basicstyle=\ttfamily\scriptsize]|sop| is a \lstinline[basicstyle=\ttfamily\scriptsize]|SOperator|, \lstinline[basicstyle=\ttfamily\scriptsize]|mat| is a \lstinline[basicstyle=\ttfamily\scriptsize]|OpMat|, \lstinline[basicstyle=\ttfamily\scriptsize]|qnd| is a \lstinline[basicstyle=\ttfamily\scriptsize]|QNDiag|, \lstinline[basicstyle=\ttfamily\scriptsize]|sqnd| is a  \lstinline[basicstyle=\ttfamily\scriptsize]|SQNDiag|, \lstinline[basicstyle=\ttfamily\scriptsize]|qnf| is a \lstinline[basicstyle=\ttfamily\scriptsize]|QNOffd|, \lstinline[basicstyle=\ttfamily\scriptsize]|sqnf| is a  \lstinline[basicstyle=\ttfamily\scriptsize]|SQNOffd|, \lstinline[basicstyle=\ttfamily\scriptsize]|trs| is a \lstinline[basicstyle=\ttfamily\scriptsize]|Transf|, \lstinline[basicstyle=\ttfamily\scriptsize]|strs| is a \lstinline[basicstyle=\ttfamily\scriptsize]|STransf|, \lstinline[basicstyle=\ttfamily\scriptsize]|vec| is a \lstinline[basicstyle=\ttfamily\scriptsize]|Vector|, \lstinline[basicstyle=\ttfamily\scriptsize]|#| is a number. In general, \lstinline[basicstyle=\ttfamily\scriptsize]|a1 - a2| is supported where \lstinline[basicstyle=\ttfamily\scriptsize]|a1 + a2| is supported~; \lstinline[basicstyle=\ttfamily\scriptsize]|a * #|, \lstinline[basicstyle=\ttfamily\scriptsize]|a / #| and \lstinline[basicstyle=\ttfamily\scriptsize]|-a| are supported where \lstinline[basicstyle=\ttfamily\scriptsize]|# * a| is supported.}\\
    \href{https://docs.fuzzified.world/models/#Base.:+-Tuple{SphereObs,%20SphereObs}}{\lstinline|obs + obs|}\\
    \href{https://docs.fuzzified.world/core/#Base.:+-Tuple{QNDiag,%20QNDiag}}{\lstinline|qnd + qnd|}\\
    \href{https://docs.fuzzified.world/fuzzifino/#Base.:+-Tuple{SQNDiag,%20SQNDiag}}{\lstinline|sqnd + sqnd|}\\
    \href{https://docs.fuzzified.world/core/#Base.:+-Tuple{Vector{Term},%20Vector{Term}}}{\lstinline|tms + tms|}\\
    \href{https://docs.fuzzified.world/fuzzifino/#Base.:+-Tuple{Vector{STerm},%20Vector{STerm}}}{\lstinline|stms + stms|}\\
    \href{https://docs.fuzzified.world/models/#Base.:*-Tuple{Number,%20AngModes}}{\lstinline|# * amd|}\\
    \href{https://docs.fuzzified.world/models/#Base.:*-Tuple{Number,%20SphereObs}}{\lstinline|# * obs|}\\
    \href{https://docs.fuzzified.world/core/#Base.:*-Tuple{Int64,%20QNDiag}}{\lstinline|# * qnd|}\\
    \href{https://docs.fuzzified.world/fuzzifino/#Base.:*-Tuple{Int64,%20SQNDiag}}{\lstinline|# * sqnd|}\\
    \href{https://docs.fuzzified.world/core/#Base.:*-Tuple{Number,%20Vector{Term}}}{\lstinline|# * tms|}\\
    \href{https://docs.fuzzified.world/fuzzifino/#Base.:*-Tuple{Number,%20Vector{STerm}}}{\lstinline|# * stms|}\\
    \href{https://docs.fuzzified.world/models/#Base.:*-Tuple{AngModes,%20AngModes}}{\lstinline|amd * amd|}\\
    \href{https://docs.fuzzified.world/models/#Base.:*-Tuple{SphereObs,%20SphereObs}}{\lstinline|obs * obs|}\\
    \href{https://docs.fuzzified.world/core/#Base.:*-Tuple{QNOffd,%20QNOffd}}{\lstinline|qnf * qnf|}\\
    \href{https://docs.fuzzified.world/fuzzifino/#Base.:*-Tuple{SQNOffd,%20SQNOffd}}{\lstinline|sqnf * sqnf|}\\
    \href{https://docs.fuzzified.world/core/#Base.:*-Tuple{Vector{Term},%20Vector{Term}}}{\lstinline|tms * tms|}\\
    \href{https://docs.fuzzified.world/fuzzifino/#Base.:*-Tuple{Vector{STerm},%20Vector{STerm}}}{\lstinline|stms * stms|}\\
    \href{https://docs.fuzzified.world/core/#Base.:*-Tuple{OpMat{ComplexF64},%20Vector{ComplexF64}}}{\lstinline|mat * vec|}\\
    \href{https://docs.fuzzified.world/core/#Base.:*-Tuple{Operator,%20Vector{ComplexF64}}}{\lstinline|op * vec|}\\
    \href{https://docs.fuzzified.world/fuzzifino/#Base.:*-Tuple{SOperator,%20Vector{ComplexF64}}}{\lstinline|sop * vec|}\\
    \href{https://docs.fuzzified.world/core/#Base.:*-Tuple{Transf,%20Vector{ComplexF64}}}{\lstinline|trs * vec|}\\
    \href{https://docs.fuzzified.world/fuzzifino/#Base.:*-Tuple{STransf,%20Vector{ComplexF64}}}{\lstinline|strs * vec|}\\
    \href{https://docs.fuzzified.world/fuzzifino/#Base.:*-Tuple{Adjoint{ComplexF64,%20Vector{ComplexF64}},%20OpMat{ComplexF64},%20Vector{ComplexF64}}}{\lstinline|vec' * mat * vec|}\\
    \href{https://docs.fuzzified.world/core/#Base.:*-Tuple{Adjoint{ComplexF64,%20Vector{ComplexF64}},%20Operator,%20Vector{ComplexF64}}}{\lstinline|vec' * op * vec|}\\
    \href{https://docs.fuzzified.world/fuzzifino/#Base.:*-Tuple{Adjoint{ComplexF64,%20Vector{ComplexF64}},%20SOperator,%20Vector{ComplexF64}}}{\lstinline|vec' * sop * vec|}\\
    \href{https://docs.fuzzified.world/models/#Base.adjoint-Tuple{AngModes}}{\lstinline|amd'|}\\
    \href{https://docs.fuzzified.world/models/#Base.adjoint-Tuple{SphereObs}}{\lstinline|obs'|}\\
    \href{https://docs.fuzzified.world/core/#Base.adjoint-Tuple{Vector{Term}}}{\lstinline|tms'|}\\
    \href{https://docs.fuzzified.world/fuzzifino/#Base.adjoint-Tuple{Vector{STerm}}}{\lstinline|stms'|}\\
    \href{https://docs.fuzzified.world/core/#Base.:*-Tuple{Vector{Term},%20Vector{Term}}}{\lstinline|tms ^ #|}\\
    \href{https://docs.fuzzified.world/fuzzifino/#Base.:*-Tuple{Vector{STerm},%20Vector{STerm}}}{\lstinline|stms ^ #|}\\
\end{multicols}

\cleardoublepage

\providecommand{\href}[2]{#2}\begingroup\raggedright\endgroup


\begin{thebibliography}{100}

\bibitem{Zhu2022}
W.~Zhu, C.~Han, E.~Huffman, J.~S.~Hofmann and Y.-C.~He, \emph{Uncovering
  conformal symmetry in the 3d {Ising} transition: State-operator
  correspondence from a quantum fuzzy sphere regularization},
  \href{https://doi.org/10.1103/PhysRevX.13.021009}{\emph{Phys. Rev. X}
  {\bfseries 13} (2023) 021009}
  [\href{https://arxiv.org/abs/2210.13482}{{arXiv:2210.13482}}].

\bibitem{Hu2023Mar}
L.~Hu, Y.-C.~He and W.~Zhu, \emph{Operator product expansion coefficients of
  the 3d {Ising} criticality via quantum fuzzy spheres},
  \href{https://doi.org/10.1103/PhysRevLett.131.031601}{\emph{Phys. Rev. Lett.}
  {\bfseries 131} (2023) 031601}
  [\href{https://arxiv.org/abs/2303.08844}{{arXiv:2303.08844}}].

\bibitem{Han2023Jun}
C.~Han, L.~Hu, W.~Zhu and Y.-C.~He, \emph{Conformal four-point correlators of
  the three-dimensional {Ising} transition via the quantum fuzzy sphere},
  \href{https://doi.org/10.1103/PhysRevB.108.235123}{\emph{Phys. Rev. B}
  {\bfseries 108} (2023) 235123}
  [\href{https://arxiv.org/abs/2306.04681}{{arXiv:2306.04681}}].

\bibitem{Zhou2023}
Z.~Zhou, L.~Hu, W.~Zhu and Y.-C.~He, \emph{$\mathrm{SO}(5)$ deconfined phase
  transition under the fuzzy-sphere microscope: Approximate conformal symmetry,
  pseudo-criticality, and operator spectrum},
  \href{https://doi.org/10.1103/PhysRevX.14.021044}{\emph{Phys. Rev. X}
  {\bfseries 14} (2024) 021044}
  [\href{https://arxiv.org/abs/2306.16435}{{arXiv:2306.16435}}].

\bibitem{Lao2023}
B.-X.~Lao and S.~Rychkov, \emph{3d {Ising} {CFT} and exact diagonalization on
  icosahedron: The power of conformal perturbation theory},
  \href{https://doi.org/10.21468/SciPostPhys.15.6.243}{\emph{SciPost Phys.}
  {\bfseries 15} (2023) 243}
  [\href{https://arxiv.org/abs/2307.02540}{{arXiv:2307.02540}}].

\bibitem{Hu2023Aug}
L.~Hu, Y.-C.~He and W.~Zhu, \emph{Solving conformal defects in 3d conformal
  field theory using fuzzy sphere regularization},
  \href{https://doi.org/10.1038/s41467-024-47978-y}{\emph{Nature Commun.}
  {\bfseries 15} (2024) 9013}
  [\href{https://arxiv.org/abs/2308.01903}{{arXiv:2308.01903}}].

\bibitem{Hofmann2023}
J.~S.~Hofmann, F.~Goth, W.~Zhu, Y.-C.~He and E.~Huffman, \emph{Quantum {Monte
  Carlo} simulation of the 3d {Ising} transition on the fuzzy sphere},
  \href{https://doi.org/10.21468/SciPostPhysCore.7.2.028}{\emph{SciPost Phys.
  Core} {\bfseries 7} (2024) 028}
  [\href{https://arxiv.org/abs/2310.19880}{{arXiv:2310.19880}}].

\bibitem{Han2023Dec}
C.~Han, L.~Hu and W.~Zhu, \emph{Conformal operator content of the
  {Wilson-Fisher} transition on fuzzy sphere bilayers},
  \href{https://doi.org/10.1103/PhysRevB.110.115113}{\emph{Phys. Rev. B}
  {\bfseries 110} (2024) 115113}
  [\href{https://arxiv.org/abs/2312.04047}{{arXiv:2312.04047}}].

\bibitem{Zhou2024Jan}
Z.~Zhou, D.~Gaiotto, Y.-C.~He and Y.~Zou, \emph{The $g$-function and defect
  changing operators from wavefunction overlap on a fuzzy sphere},
  \href{https://doi.org/10.21468/SciPostPhys.17.1.021}{\emph{SciPost Phys.}
  {\bfseries 17} (2024) 021}
  [\href{https://arxiv.org/abs/2401.00039}{{arXiv:2401.00039}}].

\bibitem{Hu2024}
L.~Hu, W.~Zhu and Y.-C.~He, \emph{Entropic {$F$}-function of 3d {Ising}
  conformal field theory via the fuzzy sphere regularization},
  \href{https://doi.org/10.1103/PhysRevB.111.155151}{\emph{Phys. Rev. B}
  {\bfseries 111} (2025) 155151}
  [\href{https://arxiv.org/abs/2401.17362}{{arXiv:2401.17362}}].

\bibitem{Cuomo2024}
G.~Cuomo, Y.-C.~He and Z.~Komargodski, \emph{Impurities with a cusp: general
  theory and 3d {Ising}},
  \href{https://doi.org/10.1007/JHEP11(2024)061}{\emph{JHEP} {\bfseries 11}
  (2024) 061} [\href{https://arxiv.org/abs/2406.10186}{{arXiv:2406.10186}}].

\bibitem{Zhou2024Jul}
Z.~Zhou and Y.~Zou, \emph{Studying the 3d {Ising} surface {CFTs} on the fuzzy
  sphere}, \href{https://doi.org/10.21468/SciPostPhys.18.1.031}{\emph{SciPost
  Phys.} {\bfseries 18} (2025) 031}
  [\href{https://arxiv.org/abs/2407.15914}{{arXiv:2407.15914}}].

\bibitem{Dedushenko2024}
M.~Dedushenko, \emph{{Ising} {BCFT} from fuzzy hemisphere},
  \href{https://arxiv.org/abs/2407.15948}{{arXiv:2407.15948}}.

\bibitem{Fardelli2024}
G.~Fardelli, A.~L.~Fitzpatrick and E.~Katz, \emph{Constructing the infrared
  conformal generators on the fuzzy sphere},
  \href{https://doi.org/10.21468/SciPostPhys.18.3.086}{\emph{SciPost Phys.}
  {\bfseries 18} (2025) 086}
  [\href{https://arxiv.org/abs/2409.02998}{{arXiv:2409.02998}}].

\bibitem{Fan2024}
R.~Fan, \emph{Note on explicit construction of conformal generators on the
  fuzzy sphere},  \href{https://arxiv.org/abs/2409.08257}{{arXiv:2409.08257}}.

\bibitem{Zhou2024Oct}
Z.~Zhou and Y.-C.~He, \emph{3d conformal field theories with $\mathrm{Sp}({N})$
  global symmetry on a fuzzy sphere},
  \href{https://doi.org/10.1103/xstj-xvcy}{\emph{Phys. Rev. Lett.} {\bfseries
  135} (2025) 026504}
  [\href{https://arxiv.org/abs/2410.00087}{{arXiv:2410.00087}}].

\bibitem{Voinea2024}
C.~Voinea, R.~Fan, N.~Regnault and Z.~Papi{\'c}, \emph{Regularizing 3d
  conformal field theories via anyons on the fuzzy sphere},
  \href{https://doi.org/10.1103/bf4k-phl9}{\emph{Phys. Rev. X} {\bfseries 15}
  (2025) 031007} [\href{https://arxiv.org/abs/2411.15299}{{arXiv:2411.15299}}].

\bibitem{Yang2025Jan}
S.~Yang, Y.-G.~Yue, Y.~Tang, C.~Han, W.~Zhu and Y.~Chen, \emph{Microscopic
  study of 3d {Potts} phase transition via fuzzy sphere regularization},
  \href{https://arxiv.org/abs/2501.14320}{{arXiv:2501.14320}}.

\bibitem{Han2025}
C.~Han and W.~Zhu, \emph{Quantum phase transitions on the noncommutative
  circle}, \href{https://doi.org/10.1103/PhysRevB.111.085113}{\emph{Phys. Rev.
  B} {\bfseries 111} (2025) 085113}.

\bibitem{Laeuchli2025}
A.~M.~L{\"a}uchli, L.~Herviou, P.~H.~Wilhelm and S.~Rychkov, \emph{Exact
  diagonalization, matrix product states and conformal perturbation theory
  study of a 3d {Ising} fuzzy sphere model},
  \href{https://doi.org/10.21468/SciPostPhys.19.3.076}{\emph{SciPost Phys.}
  {\bfseries 19} (2025) 076}
  [\href{https://arxiv.org/abs/2504.00842}{{arXiv:2504.00842}}].

\bibitem{Fan2025}
R.~Fan, J.~Dong and A.~Vishwanath, \emph{Simulating the non-unitary
  {Yang}-{Lee} conformal field theory on the fuzzy sphere},
  \href{https://arxiv.org/abs/2505.06342}{{arXiv:2505.06342}}.

\bibitem{ArguelloCruz2025}
E.~Arguello~Cruz, I.~R.~Klebanov, G.~Tarnopolsky and Y.~Xin, \emph{{Yang}-{Lee}
  quantum criticality in various dimensions},
  \href{https://arxiv.org/abs/2505.06369}{{arXiv:2505.06369}}.

\bibitem{EliasMiro2025}
J.~Elias~Mir{\'o} and O.~Delouche, \emph{Flowing from the {Ising} model on the
  fuzzy sphere to the 3d {Lee}-{Yang} {CFT}},
  \href{https://doi.org/10.1007/JHEP10(2025)037}{\emph{JHEP} {\bfseries 10}
  (2025) 037} [\href{https://arxiv.org/abs/2505.07655}{{arXiv:2505.07655}}].

\bibitem{He2025Jun}
Y.-C.~He, \emph{Free real scalar {CFT} on fuzzy sphere: spectrum, algebra and
  wavefunction ansatz},
  \href{https://arxiv.org/abs/2506.14904}{{arXiv:2506.14904}}.

\bibitem{Taylor2025}
J.~Taylor, C.~Voinea, Z.~Papi{\'c} and R.~Fan, \emph{Conformal scalar field
  theory from {Ising} tricriticality on the fuzzy sphere},
  \href{https://arxiv.org/abs/2506.22539}{{arXiv:2506.22539}}.

\bibitem{Yang2025Jul}
S.~Yang, L.-D.~Hu, C.~Han, W.~Zhu and Y.~Chen, \emph{Conformal operator flows
  of the deconfined quantum criticality from $\mathrm{SO}(5)$ to
  $\mathrm{O}(4)$},
  \href{https://arxiv.org/abs/2507.01322}{{arXiv:2507.01322}}.

\bibitem{Zhou2025Jul}
Z.~Zhou, C.~Wang and Y.-C.~He, \emph{{Chern}-{Simons}-matter conformal field
  theory on fuzzy sphere: {Confinement} transition of {Kalmeyer}-{Laughlin}
  chiral spin liquid},
  \href{https://arxiv.org/abs/2507.19580}{{arXiv:2507.19580}}.

\bibitem{Dong2025}
J.-M.~Dong, Y.~Zhang, K.-W.~Huang, H.-H.~Tu and Y.-H.~Wu, \emph{Numerical
  extraction of crosscap coefficients in microscopic models for $(2+1)${D}
  conformal field theory},
  \href{https://arxiv.org/abs/2507.20005}{{arXiv:2507.20005}}.

\bibitem{Zhou2025Sep}
Z.~Zhou, D.~Gaiotto and Y.-C.~He, \emph{Free {Majorana} fermion meets gauged
  {Ising} conformal field theory on the fuzzy sphere},
  \href{https://arxiv.org/abs/2509.08038}{{arXiv:2509.08038}}.

\bibitem{Voinea2025}
C.~Voinea, W.~Zhu, N.~Regnault and Z.~Papi{\'c}, \emph{Critical {Majorana}
  fermion at a topological quantum {Hall} bilayer transition},
  \href{https://arxiv.org/abs/2509.08036}{{arXiv:2509.08036}}.

\bibitem{Wiese2025}
K.~J.~Wiese, \emph{Locating the {Ising} {CFT} via the ground-state energy on
  the fuzzy sphere},
  \href{https://arxiv.org/abs/2510.09482}{{arXiv:2510.09482}}.

\bibitem{Dey2025}
A.~Dey, L.~Herviou, C.~Mudry and A.~M.~L{\"a}uchli, \emph{Conformal data for
  the $\mathrm{O}(3)$ {Wilson}-{Fisher} {CFT} from fuzzy sphere realization of
  quantum rotor model},
  \href{https://arxiv.org/abs/2510.09755}{{arXiv:2510.09755}}.

\bibitem{Guo2025}
W.~Guo, Z.~Zhou, T.-C.~Wei and Y.-C.~He, \emph{The ${O}({N})$ free-scalar and
  {Wilson}-{Fisher} conformal field theories on the fuzzy sphere}, {\emph{to
  appear} (2025) }.

\bibitem{Huffman2025}
E.~Huffman, Z.~Zhou, Y.-C.~He and J.~S.~Hofmann, \emph{Generalizing deconfined
  criticality to 3d ${N}$-flavor $\mathrm{SU}(2)$ quantum chromodynamics on the
  fuzzy sphere}, {\emph{to appear} (2025) }.

\bibitem{ITensor}
M.~Fishman, S.~R.~White and E.~M.~Stoudenmire, \emph{The {ITensor} software
  library for tensor network calculations},
  \href{https://doi.org/10.21468/SciPostPhysCodeb.4}{\emph{SciPost Phys.
  Codeb.} (2022) 4}
  [\href{https://arxiv.org/abs/2007.14822}{{arXiv:2007.14822}}].

\bibitem{Julia}
J.~Bezanson, A.~Edelman, S.~Karpinski and V.~B.~Shah, \emph{Julia: A fresh
  approach to numerical computing},
  \href{https://doi.org/10.1137/141000671}{\emph{SIAM Review} {\bfseries 59}
  (2017) 65}.

\bibitem{Rychkov2016CFT}
S.~Rychkov, \emph{{EPFL} Lectures on Conformal Field Theory in ${D}\ge 3$
  Dimensions}, Briefs in Physics, Springer (1, 2016),
  \href{https://doi.org/10.1007/978-3-319-43626-5}{10.1007/978-3-319-43626-5},
  [\href{https://arxiv.org/abs/1601.05000}{{arXiv:1601.05000}}].

\bibitem{SimmonsDuffin2016CFT}
D.~Simmons-Duffin, \emph{The conformal bootstrap},  in \emph{Theoretical
  Advanced Study Institute in Elementary Particle Physics: New Frontiers in
  Fields and Strings}, pp.~1--74, 2017,
  \href{https://doi.org/10.1142/9789813149441_0001}{DOI}
  [\href{https://arxiv.org/abs/1602.07982}{{arXiv:1602.07982}}].

\bibitem{Polyakov1970Conformal}
A.~M.~Polyakov, \emph{Conformal symmetry of critical fluctuations}, {\emph{JETP
  Lett.} {\bfseries 12} (1970) 381}.

\bibitem{Cardy1996Scaling}
J.~L.~Cardy, \emph{Scaling and Renormalization in Statistical Physics},
  Cambridge Lecture Notes in Physics, Cambridge University Press (1996).

\bibitem{Sachdev2011Quantum}
S.~Sachdev, \emph{Quantum Phase Transitions}, Cambridge University Press, 2~ed.
  (2011),
  \href{https://doi.org/10.1017/CBO9780511973765}{10.1017/CBO9780511973765}.

\bibitem{Polchinski1998String}
J.~Polchinski, \emph{{String theory. Vol. 1: An introduction to the bosonic
  string}}, Cambridge Monographs on Mathematical Physics, Cambridge University
  Press (12, 2007),
  \href{https://doi.org/10.1017/CBO9780511816079}{10.1017/CBO9780511816079}.

\bibitem{Maldacena1998AdSCFT}
J.~M.~Maldacena, \emph{The large-${N}$ limit of superconformal field theories
  and supergravity}, {\emph{Adv. Theor. Math. Phys.} {\bfseries 2} (1998) 231}.

\bibitem{Zamolodchikov1986Irreversibility}
A.~B.~Zamolodchikov, \emph{Irreversibility of the flux of the renormalization
  group in a {2D} field theory}, {\emph{JETP Lett.} {\bfseries 43} (1986) 730}.

\bibitem{DiFrancesco1997CFT}
P.~Di~Francesco, P.~Mathieu and D.~S\'en\'echal, \emph{Conformal field theory},
  Graduate texts in contemporary physics, Springer, New York, NY (1997),
  \href{https://doi.org/10.1007/978-1-4612-2256-9}{10.1007/978-1-4612-2256-9}.

\bibitem{Ginsparg1988CFT}
P.~H.~Ginsparg, \emph{Applied conformal field theory},  in \emph{Les Houches
  Summer School in Theoretical Physics: Fields, Strings, Critical Phenomena},
  9, 1988
  [\href{https://arxiv.org/abs/hep-th/9108028}{{arXiv:hep-th/9108028}}].

\bibitem{Belavin1984BPZ}
A.~A.~Belavin, A.~M.~Polyakov and A.~B.~Zamolodchikov, \emph{Infinite conformal
  symmetry in two-dimensional quantum field theory},
  \href{https://doi.org/https://doi.org/10.1016/0550-3213(84)90052-X}{\emph{Nucl.
  Phys. B} {\bfseries 241} (1984) 333}.

\bibitem{Wess1971WZW}
J.~Wess and B.~Zumino, \emph{Consequences of anomalous {Ward} identities},
  \href{https://doi.org/10.1016/0370-2693(71)90582-X}{\emph{Phys. Lett. B}
  {\bfseries 37} (1971) 95}.

\bibitem{Witten1983WZW}
E.~Witten, \emph{Nonabelian bosonization in two-dimensions},
  \href{https://doi.org/10.1007/BF01215276}{\emph{Commun. Math. Phys.}
  {\bfseries 92} (1984) 455}.

\bibitem{Poland2018Bootstrap}
D.~Poland, S.~Rychkov and A.~Vichi, \emph{The conformal bootstrap: Theory,
  numerical techniques, and applications},
  \href{https://doi.org/10.1103/RevModPhys.91.015002}{\emph{Rev. Mod. Phys.}
  {\bfseries 91} (2019) 015002}
  [\href{https://arxiv.org/abs/1805.04405}{{arXiv:1805.04405}}].

\bibitem{Rychkov2023Bootstrap}
S.~Rychkov and N.~Su, \emph{New developments in the numerical conformal
  bootstrap}, \href{https://doi.org/10.1103/RevModPhys.96.045004}{\emph{Rev.
  Mod. Phys.} {\bfseries 96} (2024) 045004}
  [\href{https://arxiv.org/abs/2311.15844}{{arXiv:2311.15844}}].

\bibitem{ElShowk2012Ising}
S.~El-Showk, M.~F.~Paulos, D.~Poland, S.~Rychkov, D.~Simmons-Duffin and
  A.~Vichi, \emph{Solving the 3d {Ising} model with the conformal bootstrap},
  \href{https://doi.org/10.1103/PhysRevD.86.025022}{\emph{Phys. Rev. D}
  {\bfseries 86} (2012) 025022}
  [\href{https://arxiv.org/abs/1203.6064}{{arXiv:1203.6064}}].

\bibitem{Kos2016Ising}
F.~Kos, D.~Poland, D.~Simmons-Duffin and A.~Vichi, \emph{Precision islands in
  the {Ising} and $\mathrm{O}({N})$ models},
  \href{https://doi.org/10.1007/JHEP08(2016)036}{\emph{JHEP} {\bfseries 08}
  (2016) 036} [\href{https://arxiv.org/abs/1603.04436}{{arXiv:1603.04436}}].

\bibitem{Chester2019O2}
S.~M.~Chester, W.~Landry, J.~Liu, D.~Poland, D.~Simmons-Duffin, N.~Su et~al.,
  \emph{Carving out {OPE} space and precise $o(2)$ model critical exponents},
  \href{https://doi.org/10.1007/JHEP06(2020)142}{\emph{JHEP} {\bfseries 06}
  (2020) 142} [\href{https://arxiv.org/abs/1912.03324}{{arXiv:1912.03324}}].

\bibitem{Chester2020O3}
S.~M.~Chester, W.~Landry, J.~Liu, D.~Poland, D.~Simmons-Duffin, N.~Su et~al.,
  \emph{Bootstrapping {Heisenberg} magnets and their cubic instability},
  \href{https://doi.org/10.1103/PhysRevD.104.105013}{\emph{Phys. Rev. D}
  {\bfseries 104} (2021) 105013}
  [\href{https://arxiv.org/abs/2011.14647}{{arXiv:2011.14647}}].

\bibitem{Iliesiu2015GNY}
L.~Iliesiu, F.~Kos, D.~Poland, S.~S.~Pufu, D.~Simmons-Duffin and R.~Yacoby,
  \emph{Bootstrapping 3d fermions},
  \href{https://doi.org/10.1007/JHEP03(2016)120}{\emph{JHEP} {\bfseries 03}
  (2016) 120} [\href{https://arxiv.org/abs/1508.00012}{{arXiv:1508.00012}}].

\bibitem{Ferrenberg2018IsingMC}
A.~M.~Ferrenberg, J.~Xu and D.~P.~Landau, \emph{Pushing the limits of {Monte}
  {Carlo} simulations for the three-dimensional {Ising} model},
  \href{https://doi.org/10.1103/PhysRevE.97.043301}{\emph{Phys. Rev. E}
  {\bfseries 97} (2018) 043301}
  [\href{https://arxiv.org/abs/1806.03558}{{arXiv:1806.03558}}].

\bibitem{Sandvik2010FSS}
A.~W.~Sandvik, \emph{Computational studies of quantum spin systems},
  \href{https://doi.org/10.1063/1.3518900}{\emph{AIP Conf. Proc.} {\bfseries
  1297} (2010) 135}
  [\href{https://arxiv.org/abs/1101.3281}{{arXiv:1101.3281}}].

\bibitem{Fisher1972FSS}
M.~E.~Fisher and M.~N.~Barber, \emph{Scaling theory for finite-size effects in
  the critical region},
  \href{https://doi.org/10.1103/PhysRevLett.28.1516}{\emph{Phys. Rev. Lett.}
  {\bfseries 28} (1972) 1516}.

\bibitem{Pappadopulo2012Radial}
D.~Pappadopulo, S.~Rychkov, J.~Espin and R.~Rattazzi, \emph{{OPE} convergence
  in conformal field theory},
  \href{https://doi.org/10.1103/PhysRevD.86.105043}{\emph{Phys. Rev. D}
  {\bfseries 86} (2012) 105043}
  [\href{https://arxiv.org/abs/1208.6449}{{arXiv:1208.6449}}].

\bibitem{Brower2024Sphere}
R.~C.~Brower and E.~K.~Owen, \emph{The {Ising} model on $\mathbb{S}^2$},
  \href{https://arxiv.org/abs/2407.00459}{{arXiv:2407.00459}}.

\bibitem{Madore1991Fuzzy}
J.~Madore, \emph{The fuzzy sphere},
  \href{https://doi.org/10.1088/0264-9381/9/1/008}{\emph{Class. Quant. Grav.}
  {\bfseries 9} (1992) 69}.

\bibitem{Haldane1983LLL}
F.~D.~M.~Haldane, \emph{Fractional quantization of the {Hall} effect: A
  hierarchy of incompressible quantum fluid states},
  \href{https://doi.org/10.1103/PhysRevLett.51.605}{\emph{Phys. Rev. Lett.}
  {\bfseries 51} (1983) 605}.

\bibitem{Wu1976LLL}
T.~T.~Wu and C.~N.~Yang, \emph{Dirac monopole without strings: Monopole
  harmonics}, \href{https://doi.org/10.1016/0550-3213(76)90143-7}{\emph{Nucl.
  Phys. B} {\bfseries 107} (1976) 365}.

\bibitem{Greiter2011LLL}
M.~Greiter, \emph{Landau level quantization on the sphere},
  \href{https://doi.org/10.1103/PhysRevB.83.115129}{\emph{Phys. Rev. B}
  {\bfseries 83} (2011) 115129}
  [\href{https://arxiv.org/abs/1101.3943}{{arXiv:1101.3943}}].

\bibitem{Hasebe2010LLL}
K.~Hasebe, \emph{Hopf maps, lowest {Landau} level, and fuzzy spheres},
  \href{https://doi.org/10.3842/SIGMA.2010.071}{\emph{SIGMA} {\bfseries 6}
  (2010) 071} [\href{https://arxiv.org/abs/1009.1192}{{arXiv:1009.1192}}].

\bibitem{Pasquier2000HallFM}
V.~Pasquier, \emph{Skyrmions in the quantum {Hall} effect and noncommutative
  solitons}, \href{https://doi.org/10.1016/S0370-2693(00)00965-5}{\emph{Phys.
  Lett. B} {\bfseries 490} (2000) 258}
  [\href{https://arxiv.org/abs/hep-th/0007176}{{arXiv:hep-th/0007176}}].

\bibitem{Girvin2010HallFM}
S.~M.~Girvin, \emph{Spin and isospin: Exotic order in quantum {Hall}
  ferromagnets},  in \emph{{The multifaceted skyrmion}}, G.~E.~Brown and
  M.~Rho, eds., pp.~217--231 (2010),
  \href{https://doi.org/10.1142/9789814280709_0009}{DOI}.

\bibitem{Myers2010Fthm}
R.~C.~Myers and A.~Sinha, \emph{{Seeing a $c$-theorem with holography}},
  \href{https://doi.org/10.1103/PhysRevD.82.046006}{\emph{Phys. Rev. D}
  {\bfseries 82} (2010) 046006}
  [\href{https://arxiv.org/abs/1006.1263}{{arXiv:1006.1263}}].

\bibitem{Casini2011Fthm}
H.~Casini, M.~Huerta and R.~C.~Myers, \emph{Towards a derivation of holographic
  entanglement entropy},
  \href{https://doi.org/10.1007/JHEP05(2011)036}{\emph{JHEP} {\bfseries 05}
  (2011) 036} [\href{https://arxiv.org/abs/1102.0440}{{arXiv:1102.0440}}].

\bibitem{Jafferis2011Fthm}
D.~L.~Jafferis, I.~R.~Klebanov, S.~S.~Pufu and B.~R.~Safdi, \emph{Towards the
  ${F}$-theorem: $\mathcal{N}=2$ field theories on the three-sphere},
  \href{https://doi.org/10.1007/JHEP06(2011)102}{\emph{JHEP} {\bfseries 06}
  (2011) 102} [\href{https://arxiv.org/abs/1103.1181}{{arXiv:1103.1181}}].

\bibitem{Klebanov2011Fthm}
I.~R.~Klebanov, S.~S.~Pufu and B.~R.~Safdi, \emph{${F}$-theorem without
  supersymmetry}, \href{https://doi.org/10.1007/JHEP10(2011)038}{\emph{JHEP}
  {\bfseries 10} (2011) 038}
  [\href{https://arxiv.org/abs/1105.4598}{{arXiv:1105.4598}}].

\bibitem{Casini2012Fthm}
H.~Casini and M.~Huerta, \emph{On the {RG} running of the entanglement entropy
  of a circle}, \href{https://doi.org/10.1103/PhysRevD.85.125016}{\emph{Phys.
  Rev. D} {\bfseries 85} (2012) 125016}
  [\href{https://arxiv.org/abs/1202.5650}{{arXiv:1202.5650}}].

\bibitem{Sterdyniak2011RealEnt}
A.~Sterdyniak, A.~Chandran, N.~Regnault, B.~A.~Bernevig and P.~Bonderson,
  \emph{Real-space entanglement spectrum of quantum {Hall} states},
  \href{https://doi.org/10.1103/PhysRevB.85.125308}{\emph{Phys. Rev. B}
  {\bfseries 85} (2012) 125308}
  [\href{https://arxiv.org/abs/1110.2810}{{arXiv:1110.2810}}].

\bibitem{Dubail2011RealEnt}
J.~Dubail, N.~Read and E.~H.~Rezayi, \emph{Real-space entanglement spectrum of
  quantum {Hall} systems},
  \href{https://doi.org/10.1103/PhysRevB.85.115321}{\emph{Phys. Rev. B}
  {\bfseries 85} (2012) 115321}
  [\href{https://arxiv.org/abs/1111.2811}{{arXiv:1111.2811}}].

\bibitem{Zaletel2012RealEnt}
M.~P.~Zaletel and R.~S.~K.~Mong, \emph{Exact matrix product states for quantum
  {Hall} wave functions},
  \href{https://doi.org/10.1103/PhysRevB.86.245305}{\emph{Phys. Rev. B}
  {\bfseries 86} (2012) 245305}
  [\href{https://arxiv.org/abs/1208.4862}{{arXiv:1208.4862}}].

\bibitem{Rodriguez2011RealEnt}
I.~D.~Rodriguez, S.~H.~Simon and J.~K.~Slingerland, \emph{Evaluation of ranks
  of real space and particle entanglement spectra for large systems},
  \href{https://doi.org/10.1103/PhysRevLett.108.256806}{\emph{Phys. Rev. Lett.}
  {\bfseries 108} (2012) 256806}
  [\href{https://arxiv.org/abs/1111.3634}{{arXiv:1111.3634}}].

\bibitem{Reinicke1987Perturbation1}
P.~Reinicke, \emph{Finite size scaling functions and conformal invariance},
  \href{https://doi.org/10.1088/0305-4470/20/13/048}{\emph{J. Phys. A}
  {\bfseries 20} (1987) 4501}.

\bibitem{Reinicke1987Perturbation2}
P.~Reinicke, \emph{Analytical and non-analytical corrections to finite-size
  scaling}, \href{https://doi.org/10.1088/0305-4470/20/15/044}{\emph{J. Phys.
  A} {\bfseries 20} (1987) 5325}.

\bibitem{Dotsenko1984Potts}
V.~S.~Dotsenko, \emph{{Critical Behavior and Associated Conformal Algebra of
  the {$Z(3)$} {Potts} Model}},
  \href{https://doi.org/10.1016/0550-3213(84)90148-2}{\emph{Nucl. Phys. B}
  {\bfseries 235} (1984) 54}.

\bibitem{Tang2024Potts}
Y.~Tang, H.~Ma, Q.~Tang, Y.-C.~He and W.~Zhu, \emph{Reclaiming the lost
  conformality in a non-{Hermitian} quantum 5-state {Potts} model},
  \href{https://doi.org/10.1103/PhysRevLett.133.076504}{\emph{Phys. Rev. Lett.}
  {\bfseries 133} (2024) 076504}
  [\href{https://arxiv.org/abs/2403.00852}{{arXiv:2403.00852}}].

\bibitem{Barkema1991Potts}
G.~Barkema and J.~de~Boer, \emph{Numerical study of phase transitions in potts
  models}, \href{https://doi.org/10.1103/PhysRevA.44.8000}{\emph{Phys. Rev. A}
  {\bfseries 44} (1991) 8000}.

\bibitem{Chester2022Potts}
S.~M.~Chester and N.~Su, \emph{Upper critical dimension of the 3-state {Potts}
  model},  \href{https://arxiv.org/abs/2210.09091}{{arXiv:2210.09091}}.

\bibitem{Yang1952Singularity}
C.~N.~Yang and T.~D.~Lee, \emph{Statistical theory of equations of state and
  phase transitions. i. theory of condensation},
  \href{https://doi.org/10.1103/PhysRev.87.404}{\emph{Phys. Rev.} {\bfseries
  87} (1952) 404}.

\bibitem{Lee1952Singularity}
T.~D.~Lee and C.~N.~Yang, \emph{Statistical theory of equations of state and
  phase transitions. ii. lattice gas and {Ising} model},
  \href{https://doi.org/10.1103/PhysRev.87.410}{\emph{Phys. Rev.} {\bfseries
  87} (1952) 410}.

\bibitem{Cardy2023Singularity}
J.~Cardy, \emph{The {Yang}-{Lee} edge singularity and related problems},  in
  \emph{50 Years of the Renormalization Group}, 5, 2023
  [\href{https://arxiv.org/abs/2305.13288}{{arXiv:2305.13288}}].

\bibitem{Wilson1971WF}
K.~G.~Wilson and M.~E.~Fisher, \emph{{Critical exponents in 3.99 dimensions}},
  \href{https://doi.org/10.1103/PhysRevLett.28.240}{\emph{Phys. Rev. Lett.}
  {\bfseries 28} (1972) 240}.

\bibitem{Lipa2003SF}
J.~A.~Lipa, J.~A.~Nissen, D.~A.~Stricker, D.~R.~Swanson and T.~C.~P.~Chui,
  \emph{Specific heat of liquid helium in zero gravity very near the lambda
  point}, \href{https://doi.org/10.1103/PhysRevB.68.174518}{\emph{Phys. Rev. B}
  {\bfseries 68} (2003) 174518}
  [\href{https://arxiv.org/abs/cond-mat/0310163}{{arXiv:cond-mat/0310163}}].

\bibitem{Girvin1986GMP}
S.~M.~Girvin, A.~H.~MacDonald and P.~M.~Platzman, \emph{Magneto-roton theory of
  collective excitations in the fractional quantum {Hall} effect},
  \href{https://doi.org/10.1103/PhysRevB.33.2481}{\emph{Phys. Rev. B}
  {\bfseries 33} (1986) 2481}.

\bibitem{Lee2014WZW}
J.~Lee and S.~Sachdev, \emph{{Wess-Zumino-Witten} terms in graphene {Landau}
  levels}, \href{https://doi.org/10.1103/PhysRevLett.114.226801}{\emph{Phys.
  Rev. Lett.} {\bfseries 114} (2015) 226801}
  [\href{https://arxiv.org/abs/1411.5684}{{arXiv:1411.5684}}].

\bibitem{Komargodski2017QCD}
Z.~Komargodski and N.~Seiberg, \emph{A symmetry breaking scenario for
  {QCD}$_{3}$}, \href{https://doi.org/10.1007/JHEP01(2018)109}{\emph{JHEP}
  {\bfseries 01} (2018) 109}
  [\href{https://arxiv.org/abs/1706.08755}{{arXiv:1706.08755}}].

\bibitem{Senthil2003DQCP}
T.~Senthil, A.~Vishwanath, L.~Balents, S.~Sachdev and M.~P.~A.~Fisher,
  \emph{Deconfined quantum critical points},
  \href{https://doi.org/10.1126/science.1091806}{\emph{Science} {\bfseries 303}
  (2004) 1490}
  [\href{https://arxiv.org/abs/cond-mat/0311326}{{arXiv:cond-mat/0311326}}].

\bibitem{Senthil2004DQCP}
T.~Senthil, L.~Balents, S.~Sachdev, A.~Vishwanath and M.~P.~A.~Fisher,
  \emph{Quantum criticality beyond the {Landau-Ginzburg-Wilson} paradigm},
  \href{https://doi.org/10.1103/PhysRevB.70.144407}{\emph{Phys. Rev. B}
  {\bfseries 70} (2004) 144407}
  [\href{https://arxiv.org/abs/cond-mat/0312617}{{arXiv:cond-mat/0312617}}].

\bibitem{Senthil2023DQCP}
T.~Senthil, \emph{{Deconfined quantum critical points: a review}},
  \href{https://arxiv.org/abs/2306.12638}{{arXiv:2306.12638}}.

\bibitem{Nahum2015DQCP}
A.~Nahum, P.~Serna, J.~T.~Chalker, M.~Ortu\~no and A.~M.~Somoza, \emph{Emergent
  $\mathrm{SO}(5)$ symmetry at the {N\'eel} to valence-bond-solid transition},
  \href{https://doi.org/10.1103/PhysRevLett.115.267203}{\emph{Phys. Rev. Lett.}
  {\bfseries 115} (2015) 267203}
  [\href{https://arxiv.org/abs/1508.06668}{{arXiv:1508.06668}}].

\bibitem{Wang2017DQCP}
C.~Wang, A.~Nahum, M.~A.~Metlitski, C.~Xu and T.~Senthil, \emph{{Deconfined
  quantum critical points: symmetries and dualities}},
  \href{https://doi.org/10.1103/PhysRevX.7.031051}{\emph{Phys. Rev. X}
  {\bfseries 7} (2017) 031051}
  [\href{https://arxiv.org/abs/1703.02426}{{arXiv:1703.02426}}].

\bibitem{Gorbenko2018Complex}
V.~Gorbenko, S.~Rychkov and B.~Zan, \emph{Walking, weak first-order
  transitions, and complex {CFTs}},
  \href{https://doi.org/10.1007/JHEP10(2018)108}{\emph{JHEP} {\bfseries 10}
  (2018) 108} [\href{https://arxiv.org/abs/1807.11512}{{arXiv:1807.11512}}].

\bibitem{Ippoliti2018DQCP}
M.~Ippoliti, R.~S.~K.~Mong, F.~F.~Assaad and M.~P.~Zaletel, \emph{Half-filled
  {Landau} levels: A continuum and sign-free regularization for
  three-dimensional quantum critical points},
  \href{https://doi.org/10.1103/PhysRevB.98.235108}{\emph{Phys. Rev. B}
  {\bfseries 98} (2018) 235108}
  [\href{https://arxiv.org/abs/1810.00009}{{arXiv:1810.00009}}].

\bibitem{Wang2020DQCP}
Z.~Wang, M.~P.~Zaletel, R.~S.~K.~Mong and F.~F.~Assaad, \emph{Phases of the
  $(2+1)$ dimensional $\mathrm{SO}(5)$ nonlinear sigma model with topological
  term}, \href{https://doi.org/10.1103/PhysRevLett.126.045701}{\emph{Phys. Rev.
  Lett.} {\bfseries 126} (2021) 045701}
  [\href{https://arxiv.org/abs/2003.08368}{{arXiv:2003.08368}}].

\bibitem{Chen2023WZW}
B.-B.~Chen, X.~Zhang, Y.~Wang, K.~Sun and Z.~Y.~Meng, \emph{Phases of $(2+1)$d
  $\mathrm{SO}(5)$ nonlinear sigma model with a topological term on a sphere:
  Multicritical point and disorder phase},
  \href{https://doi.org/10.1103/PhysRevLett.132.246503}{\emph{Phys. Rev. Lett.}
  {\bfseries 132} (2024) 246503}
  [\href{https://arxiv.org/abs/2307.05307}{{arXiv:2307.05307}}].

\bibitem{Chen2024WZW}
B.-B.~Chen, X.~Zhang and Z.~Yang~Meng, \emph{Emergent conformal symmetry at the
  multicritical point of $(2+1)$d $\mathrm{SO}(5)$ model with
  {Wess-Zumino-Witten} term on a sphere},
  \href{https://doi.org/10.1103/PhysRevB.110.125153}{\emph{Phys. Rev. B}
  {\bfseries 110} (2024) 125153}
  [\href{https://arxiv.org/abs/2405.04470}{{arXiv:2405.04470}}].

\bibitem{Laughlin1983Anomalous}
R.~B.~Laughlin, \emph{Anomalous quantum {Hall} effect: An incompressible
  quantum fluid with fractionally charged excitations},
  \href{https://doi.org/10.1103/PhysRevLett.50.1395}{\emph{Phys. Rev. Lett.}
  {\bfseries 50} (1983) 1395}.

\bibitem{Jain1989Composite}
J.~K.~Jain, \emph{Composite fermion approach for the fractional quantum {Hall}
  effect}, \href{https://doi.org/10.1103/PhysRevLett.63.199}{\emph{Phys. Rev.
  Lett.} {\bfseries 63} (1989) 199}.

\bibitem{Read1998Rezayi}
N.~Read and E.~Rezayi, \emph{{Beyond paired quantum {Hall} states: Parafermions
  and incompressible states in the first excited {Landau} level}},
  \href{https://doi.org/10.1103/PhysRevB.59.8084}{\emph{Phys. Rev. B}
  {\bfseries 59} (1999) 8084}
  [\href{https://arxiv.org/abs/cond-mat/9809384}{{arXiv:cond-mat/9809384}}].

\bibitem{Moore1991Pfaffian}
G.~W.~Moore and N.~Read, \emph{Nonabelions in the fractional quantum {Hall}
  effect}, \href{https://doi.org/10.1016/0550-3213(91)90407-O}{\emph{Nucl.
  Phys. B} {\bfseries 360} (1991) 362}.

\bibitem{Dasgupta1981Duality}
C.~Dasgupta and B.~I.~Halperin, \emph{Phase transition in a lattice model of
  superconductivity},
  \href{https://doi.org/10.1103/PhysRevLett.47.1556}{\emph{Phys. Rev. Lett.}
  {\bfseries 47} (1981) 1556}.

\bibitem{Metlitski2015Duality}
M.~A.~Metlitski and A.~Vishwanath, \emph{Particle-vortex duality of
  two-dimensional {Dirac} fermion from electric-magnetic duality of
  three-dimensional topological insulators},
  \href{https://doi.org/10.1103/PhysRevB.93.245151}{\emph{Phys. Rev. B}
  {\bfseries 93} (2016) 245151}
  [\href{https://arxiv.org/abs/1505.05142}{{arXiv:1505.05142}}].

\bibitem{Seiberg2016Duality}
N.~Seiberg, T.~Senthil, C.~Wang and E.~Witten, \emph{A duality web in $2+1$
  dimensions and condensed matter physics},
  \href{https://doi.org/10.1016/j.aop.2016.08.007}{\emph{Annals Phys.}
  {\bfseries 374} (2016) 395}
  [\href{https://arxiv.org/abs/1606.01989}{{arXiv:1606.01989}}].

\bibitem{Cai2023Moire}
J.~Cai, E.~Anderson, C.~Wang, X.~Zhang, X.~Liu, W.~Holtzmann et~al.,
  \emph{Signatures of fractional quantum anomalous {Hall} states in twisted
  {MoTe}$_2$}, \href{https://doi.org/10.1038/s41586-023-06289-w}{\emph{Nature}
  {\bfseries 622} (2023) 63}
  [\href{https://arxiv.org/abs/1809.09091}{{arXiv:1809.09091}}].

\bibitem{Zeng2023Moire}
Y.~Zeng, Z.~Xia, K.~Kang, J.~Zhu, P.~Kn{\"u}ppel, C.~Vaswani et~al.,
  \emph{{Thermodynamic evidence of fractional Chern insulator in moir{\'e}
  MoTe$_{2}$}}, \href{https://doi.org/10.1038/s41586-023-06452-3}{\emph{Nature}
  {\bfseries 622} (2023) 69}
  [\href{https://arxiv.org/abs/2305.00973}{{arXiv:2305.00973}}].

\bibitem{Park2023Moire}
H.~Park, J.~Cai, E.~Anderson, Y.~Zhang, J.~Zhu, X.~Liu et~al.,
  \emph{Observation of fractionally quantized anomalous hall effect},
  \href{https://doi.org/10.1038/s41586-023-06536-0}{\emph{Nature} {\bfseries
  622} (2023) 74}
  [\href{https://arxiv.org/abs/2308.02657}{{arXiv:2308.02657}}].

\bibitem{Laughlin1983FQHE}
R.~B.~Laughlin, \emph{Anomalous quantum {Hall} effect: An incompressible
  quantum fluid with fractionally charged excitations},
  \href{https://doi.org/10.1103/PhysRevLett.50.1395}{\emph{Phys. Rev. Lett.}
  {\bfseries 50} (1983) 1395}.

\bibitem{Hsin2016LevelRank}
P.-S.~Hsin and N.~Seiberg, \emph{Level/rank duality and {Chern}-{Simons}-matter
  theories}, \href{https://doi.org/10.1007/JHEP09(2016)095}{\emph{JHEP}
  {\bfseries 09} (2016) 095}
  [\href{https://arxiv.org/abs/1607.07457}{{arXiv:1607.07457}}].

\bibitem{Benini2017Duality}
F.~Benini, P.-S.~Hsin and N.~Seiberg, \emph{Comments on global symmetries,
  anomalies, and duality in $(2 + 1)d$},
  \href{https://doi.org/10.1007/JHEP04(2017)135}{\emph{JHEP} {\bfseries 04}
  (2017) 135} [\href{https://arxiv.org/abs/1702.07035}{{arXiv:1702.07035}}].

\bibitem{Kalmeyer1987CSL}
V.~Kalmeyer and R.~B.~Laughlin, \emph{Equivalence of the resonating valence
  bond and fractional quantum {Hall} states},
  \href{https://doi.org/10.1103/PhysRevLett.59.2095}{\emph{Phys. Rev. Lett.}
  {\bfseries 59} (1987) 2095}.

\bibitem{Kalmeyer1989CSL}
V.~Kalmeyer and R.~B.~Laughlin, \emph{Theory of the spin liquid state of the
  {Heisenberg} antiferromagnet},
  \href{https://doi.org/10.1103/PhysRevB.39.11879}{\emph{Phys. Rev. B}
  {\bfseries 39} (1989) 11879}.

\bibitem{Divic2025AnyonSC}
S.~Divic, V.~Cr{\'e}pel, T.~Soejima, X.-Y.~Song, A.~J.~Millis, M.~P.~Zaletel
  et~al., \emph{Anyon superconductivity from topological criticality in a
  hofstadter-hubbard model},
  \href{https://doi.org/10.1073/pnas.2426680122}{\emph{Proc. Nat. Acad. Sci.}
  {\bfseries 122} (2025) e2426680122}
  [\href{https://arxiv.org/abs/2410.18175}{{arXiv:2410.18175}}].

\bibitem{Kim2024AnyonSC}
M.~Kim, A.~Timmel, L.~Ju and X.-G.~Wen, \emph{Topological chiral
  superconductivity beyond pairing in a {Fermi} liquid},
  \href{https://doi.org/10.1103/PhysRevB.111.014508}{\emph{Phys. Rev. B}
  {\bfseries 111} (2025) 014508}
  [\href{https://arxiv.org/abs/2409.18067}{{arXiv:2409.18067}}].

\bibitem{Shi2024AnyonSC}
Z.~D.~Shi and T.~Senthil, \emph{{Doping a Fractional Quantum Anomalous Hall
  Insulator}}, \href{https://doi.org/10.1103/kcm5-hx56}{\emph{Phys. Rev. X}
  {\bfseries 15} (2025) 031069}
  [\href{https://arxiv.org/abs/2409.20567}{{arXiv:2409.20567}}].

\bibitem{Yang2008Feshbach}
K.~Yang and H.~Zhai, \emph{Quantum {Hall} transition near a fermion {Feshbach}
  resonance in a rotating trap},
  \href{https://doi.org/10.1103/PhysRevLett.100.030404}{\emph{Phys. Rev. Lett.}
  {\bfseries 100} (2008) 030404}
  [\href{https://arxiv.org/abs/0709.2934}{{arXiv:0709.2934}}].

\bibitem{Liou2018Feshbach}
S.-F.~Liou, Z.-X.~Hu and K.~Yang, \emph{Topological phase transition in a
  two-species fermion system: Effects of a rotating trap potential or a
  synthetic gauge field},
  \href{https://doi.org/10.1103/PhysRevB.97.245140}{\emph{Phys. Rev. B}
  {\bfseries 97} (2018) 245140}
  [\href{https://arxiv.org/abs/1802.10553}{{arXiv:1802.10553}}].

\bibitem{Halperin1983}
B.~I.~Halperin, \emph{Theory of the quantized hall conductance}, {\emph{Helv.
  Phys. Acta} {\bfseries 56} (1983) 75}.

\bibitem{Wen2000Majorana}
X.-G.~Wen, \emph{Continuous topological phase transitions between clean quantum
  hall states}, \href{https://doi.org/10.1103/physrevlett.84.3950}{\emph{Phys.
  Rev. Lett.} {\bfseries 84} (2000) 3950}
  [\href{https://arxiv.org/abs/cond-mat/9908394}{{arXiv:cond-mat/9908394}}].

\bibitem{Billo2013Defect}
M.~Bill\'o, M.~Caselle, D.~Gaiotto, F.~Gliozzi, M.~Meineri and R.~Pellegrini,
  \emph{Line defects in the 3d {Ising} model},
  \href{https://doi.org/10.1007/JHEP07(2013)055}{\emph{JHEP} {\bfseries 07}
  (2013) 055} [\href{https://arxiv.org/abs/1304.4110}{{arXiv:1304.4110}}].

\bibitem{Billo2016Defect}
M.~Bill\`o, V.~Gon\c{c}alves, E.~Lauria and M.~Meineri, \emph{Defects in
  conformal field theory},
  \href{https://doi.org/10.1007/JHEP04(2016)091}{\emph{JHEP} {\bfseries 04}
  (2016) 091} [\href{https://arxiv.org/abs/1601.02883}{{arXiv:1601.02883}}].

\bibitem{Andreas2000MagLine}
A.~Hanke, \emph{Critical adsorption on defects in ising magnets and binary
  alloys}, \href{https://doi.org/10.1103/PhysRevLett.84.2180}{\emph{Phys. Rev.
  Lett.} {\bfseries 84} (2000) 2180}.

\bibitem{Allais2014MagLine}
A.~Allais, \emph{Magnetic defect line in a critical {Ising} bath},
  \href{https://arxiv.org/abs/1412.3449}{{arXiv:1412.3449}}.

\bibitem{Allais2013MagLine}
A.~Allais and S.~Sachdev, \emph{Spectral function of a localized fermion
  coupled to the {Wilson-Fisher} conformal field theory},
  \href{https://doi.org/10.1103/PhysRevB.90.035131}{\emph{Phys. Rev. B}
  {\bfseries 90} (2014) 035131}
  [\href{https://arxiv.org/abs/1406.3022}{{arXiv:1406.3022}}].

\bibitem{Pannell2023MagLine}
W.~H.~Pannell and A.~Stergiou, \emph{{Line defect {RG} flows in the $\epsilon$
  expansion}}, \href{https://doi.org/10.1007/JHEP06(2023)186}{\emph{JHEP}
  {\bfseries 06} (2023) 186}
  [\href{https://arxiv.org/abs/2302.14069}{{arXiv:2302.14069}}].

\bibitem{Cuomo2021gfn}
G.~Cuomo, Z.~Komargodski and A.~Raviv-Moshe, \emph{Renormalization group flows
  on line defects},
  \href{https://doi.org/10.1103/PhysRevLett.128.021603}{\emph{Phys. Rev. Lett.}
  {\bfseries 128} (2022) 021603}
  [\href{https://arxiv.org/abs/2108.01117}{{arXiv:2108.01117}}].

\bibitem{Casini2022gfn}
H.~Casini, I.~Salazar~Landea and G.~Torroba, \emph{Entropic $g$ theorem in
  general spacetime dimensions},
  \href{https://doi.org/10.1103/PhysRevLett.130.111603}{\emph{Phys. Rev. Lett.}
  {\bfseries 130} (2023) 111603}
  [\href{https://arxiv.org/abs/2212.10575}{{arXiv:2212.10575}}].

\bibitem{Affleck1994DefCh}
I.~Affleck and A.~W.~W.~Ludwig, \emph{The fermi edge singularity and boundary
  condition changing operators},
  \href{https://doi.org/10.1088/0305-4470/27/16/007}{\emph{J. Phys. A}
  {\bfseries 27} (1994) 5375}
  [\href{https://arxiv.org/abs/cond-mat/9405057}{{arXiv:cond-mat/9405057}}].

\bibitem{Affleck1996DefCh}
I.~Affleck, \emph{{Boundary condition changing operators in conformal field
  theory and condensed matter physics}},
  \href{https://doi.org/10.1016/S0920-5632(97)00411-8}{\emph{Nucl. Phys. B
  Proc. Suppl.} {\bfseries 58} (1997) 35}
  [\href{https://arxiv.org/abs/hep-th/9611064}{{arXiv:hep-th/9611064}}].

\bibitem{Metlitski2020IsingBd}
M.~A.~Metlitski, \emph{Boundary criticality of the $\mathrm{O}({N})$ model in
  $d=3$ critically revisited},
  \href{https://doi.org/10.21468/SciPostPhys.12.4.131}{\emph{SciPost Phys.}
  {\bfseries 12} (2022) 131}
  [\href{https://arxiv.org/abs/2009.05119}{{arXiv:2009.05119}}].

\bibitem{Krishnan2023IsingBd}
A.~Krishnan and M.~A.~Metlitski, \emph{A plane defect in the 3d
  $\mathrm{O}({N})$ model},
  \href{https://doi.org/10.21468/SciPostPhys.15.3.090}{\emph{SciPost Phys.}
  {\bfseries 15} (2023) 090}
  [\href{https://arxiv.org/abs/2301.05728}{{arXiv:2301.05728}}].

\bibitem{Giombi2023IsingBd}
S.~Giombi and B.~Liu, \emph{Notes on a surface defect in the $\mathrm{O}({N})$
  model}, \href{https://doi.org/10.1007/JHEP12(2023)004}{\emph{JHEP} {\bfseries
  12} (2023) 004}
  [\href{https://arxiv.org/abs/2305.11402}{{arXiv:2305.11402}}].

\bibitem{Hasebe2020Landau}
K.~Hasebe, \emph{{$\mathrm{SO}(5)$} {Landau} models and nested {Nambu} matrix
  geometry}, \href{https://doi.org/10.1016/j.nuclphysb.2020.115012}{\emph{Nucl.
  Phys. B} {\bfseries 956} (2020) 115012}
  [\href{https://arxiv.org/abs/2002.05010}{{arXiv:2002.05010}}].

\bibitem{Wu1977Monopole}
T.~T.~Wu and C.~N.~Yang, \emph{Some properties of monopole harmonics},
  \href{https://doi.org/10.1103/PhysRevD.16.1018}{\emph{Phys. Rev. D}
  {\bfseries 16} (1977) 1018}.

\bibitem{Shnir2005Monopole}
Y.~M.~Shnir, \emph{Magnetic Monopoles}, Text and Monographs in Physics,
  Springer, Berlin/Heidelberg (2005),
  \href{https://doi.org/10.1007/3-540-29082-6}{10.1007/3-540-29082-6}.

\bibitem{Biedenharn1984Angular}
L.~C.~Biedenharn and J.~D.~Louck, \emph{Angular Momentum in Quantum Physics:
  Theory and Application}, Encyclopedia of Mathematics and its Applications,
  Cambridge University Press (1984).

\bibitem{Trugman1985Pseudo}
S.~A.~Trugman and S.~Kivelson, \emph{Exact results for the fractional quantum
  hall effect with general interactions},
  \href{https://doi.org/10.1103/PhysRevB.31.5280}{\emph{Phys. Rev. B}
  {\bfseries 31} (1985) 5280}.

\bibitem{Yoshioka1983Torus}
D.~Yoshioka, B.~I.~Halperin and P.~A.~Lee, \emph{Ground state of
  two-dimensional electrons in strong magnetic fields and $\frac{1}{3}$
  quantized {Hall} effect},
  \href{https://doi.org/10.1103/PhysRevLett.50.1219}{\emph{Phys. Rev. Lett.}
  {\bfseries 50} (1983) 1219}.

\bibitem{Haldane1985Torus}
F.~D.~M.~Haldane, \emph{Many-particle translational symmetries of
  two-dimensional electrons at rational {Landau}-level filling},
  \href{https://doi.org/10.1103/PhysRevLett.55.2095}{\emph{Phys. Rev. Lett.}
  {\bfseries 55} (1985) 2095}.

\bibitem{Haldane1987Torus}
F.~D.~M.~Haldane, \emph{The hierarchy of fractional states and numerical
  studies},  in \emph{The Quantum {Hall} Effect}, R.~E.~Prange and
  S.~M.~Girvin, eds., (New York, NY), pp.~303--352, Springer US (1987),
  \href{https://doi.org/10.1007/978-1-4684-0499-9_8}{DOI}.

\bibitem{Zou2019Operator2d}
Y.~Zou, A.~Milsted and G.~Vidal, \emph{Conformal fields and operator product
  expansion in critical quantum spin chains},
  \href{https://doi.org/10.1103/PhysRevLett.124.040604}{\emph{Phys. Rev. Lett.}
  {\bfseries 124} (2020) 040604}
  [\href{https://arxiv.org/abs/1901.06439}{{arXiv:1901.06439}}].

\bibitem{Arnoldi1951}
W.~E.~Arnoldi, \emph{The principle of minimized iterations in the solution of
  the matrix eigenvalue problem}, {\emph{Quart. Appl. Math.} {\bfseries 9}
  (1951) 17}.

\bibitem{Arpack1998}
R.~B.~Lehoucq, D.~C.~Sorensen and C.~Yang, \emph{ARPACK Users' Guide}, Society
  for Industrial and Applied Mathematics (1998),
  \href{https://doi.org/10.1137/1.9780898719628}{10.1137/1.9780898719628}.

\bibitem{White1992DMRG}
S.~R.~White, \emph{Density matrix formulation for quantum renormalization
  groups}, \href{https://doi.org/10.1103/PhysRevLett.69.2863}{\emph{Phys. Rev.
  Lett.} {\bfseries 69} (1992) 2863}.

\bibitem{Schollwock2005DMRG}
U.~Schollwock, \emph{{The density-matrix renormalization group}},
  \href{https://doi.org/10.1103/RevModPhys.77.259}{\emph{Rev. Mod. Phys.}
  {\bfseries 77} (2005) 259}
  [\href{https://arxiv.org/abs/cond-mat/0409292}{{arXiv:cond-mat/0409292}}].

\bibitem{Schollwoeck2010DMRG}
U.~Schollwoeck, \emph{{The density-matrix renormalization group in the age of
  matrix product states}},
  \href{https://doi.org/10.1016/j.aop.2010.09.012}{\emph{Annals Phys.}
  {\bfseries 326} (2011) 96}
  [\href{https://arxiv.org/abs/1008.3477}{{arXiv:1008.3477}}].

\end{thebibliography}
\end{document}